\documentclass[11pt]{article}
\usepackage{jheppub}
\usepackage{graphicx}

\usepackage{bbm}
\def\beq{\begin{equation}}
\def\eeq{\end{equation}}
\def\bea{\begin{eqnarray}}
\def\eea{\end{eqnarray}}
\def\half{\frac{1}{2}}

\def\tr{ \, \textrm{tr} \,}
\def\Re{ \, \textrm{Re} \,}
\def\Im{ \, \textrm{Im} \,}
\def\min{ \, \textrm{min} \,}

\def\d{\partial}
\def\cA{\mathcal{A}}
\def\cH{\mathcal{H}}
\def\cW{\mathcal{W}}
\def\cV{\mathcal{V}}
\def\cM{\mathcal{M}}
\def\cP{\mathcal{P}}
\def\cT{\mathcal{T}}

\def\cO{\mathcal{O}}

\def\bb{\bar{b}}

\def\Re{ \, \textrm{Re} \, }
\def\MeV{ \, \textrm{MeV} \, }

\def\tm{\tilde{m}}
\def\tsigma{\tilde{\sigma}}

\def\sgn{ \,\textrm{sgn} }

\title{Successes and failures of a more comprehensive hard wall AdS/QCD}
\author[a]{S.~K.~Domokos,}
\author[b]{J.~A.~Harvey,}
\author[c]{A.~B.~Royston}

\affiliation[a]{Department of Particle Physics and Astrophysics\\
Weizmann Institute of Science, Rehovot 76100, Israel}
\affiliation[b]{Enrico Fermi Institute and Department of Physics
\\5640 Ellis Ave., Chicago IL 60637, USA} \affiliation[c]{NHETC and
Department of Physics and Astronomy, Rutgers University \\ 126
Frelinghuysen Rd., Piscataway NJ 08855, USA}

\emailAdd{sophia.domokos@weizmann.ac.il}
\emailAdd{j-harvey@uchicago.edu}
\emailAdd{aroyston@physics.rutgers.edu}

\abstract{We explore corrections to the ``hard wall'' gravity dual
of QCD with two-form tensor fields $b_{MN}$. These fields correspond
to the quark bilinear  $O^T=\bar q \sigma^{\mu \nu} q$ of
QCD, which generates states with quantum numbers $1^{--}$ and
$1^{+-}$, or $\omega/\rho$-like and $h_1/b_1$-like mesons,
respectively. We include new interaction terms, which render the model complete
up to dimension six. 
We find that breaking chiral symmetry induces modifications to the spectrum, by mixing 
the vector current $J^V$ and $O^T$, and
breaking the degeneracy  of the $1^{--}$ and $1^{+-}$ spectra. We
also compute the $b_1\rightarrow\omega\pi$ decay rate and $D/S$
ratio and the couplings of excited $\rho$ states to two pions, performing a comprehensive new fit to 
the QCD data.  We obtain a value for the magnetic susceptibility of the  QCD quark condensate which is consistent with
phenomenological estimates. These results  allow us to evaluate the success of dimension counting and UV matching in this model.}


\begin{document}
\begin{flushright} WIS/15/12-SEPT-DPPA \\ EFI-12-26 \\ RUNHETC-2012-20 \end{flushright}

\maketitle

\section{Introduction}

Bottom-up  AdS/QCD (pioneered in \cite{Erlich:2005qh,Da Rold:2005zs} following earlier work on deconstruction \cite{Son:2003et}) successfully
describes many features of
low energy hadron physics. Bottom-up models do not descend directly from
string theory (as do the top-down setups of \cite{Sakai:2004cn,Kuperstein:2008cq}), but rather posit the existence of a
gravity dual for low-energy QCD, and study the dynamics of
fields dual to QCD operators in a simple confining
background.

Though these frameworks are admittedly more ad hoc than top-down
models, they  remain essential tools for gaining a firmer grasp on the
five-dimensional (5d) dual to QCD. For example, despite their greater predictive power
and more ``fundamental'' origin, top-down models working in the
supergravity limit represent a point in parameter space that is
clearly distinct from the infrared (IR) limit of QCD.\footnote{Some attempts to go beyond this limit in top-down duals are explored
in \cite{Imoto:2010ef}.}  This supergravity (or $\alpha'\rightarrow 0$) limit
relies on a large separation of scales between massless and massive
string excitations. However, experimentally observed slopes of mesonic Regge
trajectories suggest that no such separation exists:  massive worldsheet excitations and
massless ones all weigh in at $\sim 1~ \textrm{GeV}$. Bottom-up
models are not restricted to this $\alpha'=l_s^2\rightarrow 0$ limit, as they
admit would-be massive string modes as new bulk fields.\footnote{Relaxing the truncation at
massless string modes and allowing the string length to be finite might also lead one to question whether 
the theory is  local. In what follows we assume this to be the case.} Bottom-up frameworks 
act as a simple testing ground for approximations made in
top-down duals. Assuming that any bottom-up model we consider can be embedded into string theory, these simple frameworks
 may also shed light on the massive
string sectors of top-down models. For example, as described in
\cite{Domokos:2011dn}, the Lagrangian for the two-tensor field
representing a leading massive string mode must  be first
order, a result which may shed light on the effective
Lagrangian one would expect to derive from string field theory.

Bottom-up frameworks have a significant disadvantage, however. Their Lagrangians are constructed term by term, instead
of descending from a known string theory action: they contain a large number of undetermined parameters as a result.
There is some disagreement in the literature about how to fix these free
parameters (coupling constants, mass scales and operator scaling
dimensions). For instance, \cite{Erlich:2005qh, Cappiello:2010tu, Alvares:2011wb} advocate fixing as
many parameters as possible to the values they assume in
perturbative QCD. The argument is that this procedure
maximizes the predictive power of the model, leaving the fewest
possible constants to be determined in the IR.

We argue here that the above approach is somewhat counterproductive,
as it might cause one to discard an otherwise viable framework
simply because it undergoes renormalization group (RG) flow from the ultraviolet (UV)
to the IR, and gives a poor matching to the low-energy spectrum evaluated using UV-valued parameters.
It is important to remember that while the hard wall
model lives on a space which is asymptotically $AdS_5$, and thus has
a conformal UV limit, the coupling constant is never small. The UV
limit of the hard wall model is not the UV limit of QCD.
One tacitly assumes that the UV boundary of the hard wall model
corresponds to an energy
scale large compared to the scale of IR physics, but where the
supergravity approximation of AdS/CFT is valid and the QCD
coupling is already large. The hard wall framework thus approximates these
intermediate energies with a ``conformal window.'' As suggested by its original proponents
\cite{Erlich:2005qh,Da Rold:2005zs}, though naive, this approximation works reasonably well for
some quantities. Matching the UV behavior of this gravity dual to
the UV of perturbative QCD presumes that correlation functions and
scaling dimensions are not (or little-) affected by RG flow.
There are certain operators -- such as conserved currents -- for which this should indeed hold true, but
these represent only a fraction of the operators needed to describe
the low-energy meson spectrum. We should treat all
parameters not directly associated with conserved
currents as free parameters in the
bottom-up framework, and fit their values using low energy IR data. 

Note for instance that the extended hard wall framework
suggested in \cite{Domokos:2011dn} was subsequently analyzed in \cite{Alvares:2011wb}. The authors determined
all free parameters introduced with the two-form $b_{MN}$ 
by matching to perturbative calculations.  The extension thus introduced no new free parameters. The holographic correlation functions of the
vector and tensor operators were  furthermore found to reproduce the large $Q^2$ behavior expected from QCD. The
resulting spectrum matched experimental data very poorly, however.
Here, we leave as free parameters all quantities which are
not explicitly  protected under RG, and fit them to the low-energy data
available. 

Even with a larger number of free parameters, AdS/QCD provides an
appealing structure for modeling low-energy hadron physics that is
motivated by top-down as well as phenomenological reasons, and which
naturally incorporates many previously-proposed organizing
principles, like Vector Meson Dominance and Hidden Local Symmetry.
Furthermore, by allowing these parameters to flow, one might even
gain insight into the nature of the QCD renormalization group flow.

In this work we continue our exploration  of the hard wall model
\cite{Da Rold:2005zs,Erlich:2005qh} extended to include a two-tensor field dual to
the QCD operator $\bar{q}\sigma^{\mu\nu}T^aq$. With this extension, the hard wall framework includes all naive
dimension three QCD operators. We focus our study on the 
vector ($1^{--}$)  and axial vector ($1^{+-}$) modes generated by
this operator, which should appear as massive worldsheet fields in a string description. We incorporate chiral symmetry breaking effects
expected to appear at $\cO(N_c^{-1/2}$). We also study the leading
$\cO(N_{c}^{-1})$ modification to the mass of the $b_1$-like modes, performing a global fit to all free parameters in the model, including those of
the original hard wall framework. For consistency, we add all interaction terms up to 5d dimension six.

Our results  shed light on the reliability of QCD dimension counting, the large $\lambda$ approximation, and more.
We find the interesting result that quantities associated with the massive string excitations tend to run significantly under RG, while most quantities related to
massless excitations do not.

The paper is organized as follows. In Section \ref{sec:setup} we
review the hard wall model and its extension to include all naive
dimension three operators. We also describe the five-dimensional
interaction terms that are responsible for chiral symmetry-breaking
and leading corrections of $b_1$ and $\rho$ spectra. In
Section \ref{sec:computation} we review the requisite  calculational techniques,
relegating most details to the appendix. In Section \ref{sec:results} we
present our results, including the deviation from perturbative QCD
expectations. We conclude in Section \ref{sec:conclusions}.

\section{Model set-up}\label{sec:setup}

We will now describe in detail the bottom-up hard wall model of \cite{Erlich:2005qh} (and the extension proposed in \cite{Domokos:2011dn}) that forms the basis
of this work. 
Briefly summarized, the low-energy regime of QCD is characterized by  the confinement of quarks and gluons to color singlet states, and by the breaking of the
$U(N_f)_L\times U(N_f)_R$ flavor symmetry of QCD to an approximate vector (isospin) subgroup. These elements are realized holographically via a gravitational background
that induces confinement, and a non-trivial vacuum configuration of the quark mass operator $\bar{q}q$. The other fields living
in this background are dual to all dimension 3 quark bilinears, which generate mesonic states from the vacuum. On the gravity side
 the corresponding fields can be expanded in terms of an infinite (but discrete) tower of normalizable modes solving the equations of motion with different
values of the 4d four-momentum (or meson mass). 
The vacuum state of the field dual to $\bar{q}q$ modifies the classical equations of motion
for these fields, lifting the degeneracy between  vector or pseudo-vector states.

\subsection{The extended hard wall model: background and fields}

The hard wall model lives on
a slice of $AdS_5$ truncated so as to induce confinement. We use the Poincar\'e metric,
\begin{equation}\label{AdS}
ds^2 = \frac{\ell^2}{z^2} (\eta_{\mu\nu} dx^\mu dx^\nu - dz^2 )~\qquad \varepsilon < z < z_0~,
\end{equation}
where $\ell$ is the $AdS_5$ radius.
The rough equivalence between inverse radial coordinate $z$ and energy scale
allows one to identify
 $\varepsilon^{-1}$ with a UV cutoff which we eventually take to infinity (i.e. $\varepsilon\rightarrow 0$). Meanwhile $z_0^{-1}$ corresponds to the scale of
the effective theory we generate: the masses of the mesonic excitations are all proportional to $z_0^{-1}$. This somewhat artificial IR cutoff stands in for some more complicated
geometry inducing confinement via a potential that goes gradually to infinity as $z\rightarrow\infty$ (see e.g. \cite{Karch:2006pv}). However, it is still an important toy model:
in fact, some top-down duals, such as the Sakai-Sugimoto model, feature spacetimes that also end at a finite radial value. In addition to simplifying
computations and making some of the IR characteristics of the theory more transparent,  the hard wall model may thus shed light on some aspects of these top-down duals.
Given this truncated geometry, we will have to make careful selections for the boundary conditions on the fields at the IR wall. We will discuss this in detail below.

The fields living in this background are those dual to all of the naive dimension 3 quark bilinears of QCD as summarized in Table (\ref{operators}). These consist of the
bifundamental scalar $\bar{q}q$ which corresponds to the complex scalar $X$, 
the left- and right-handed flavor currents dual to left- and right-handed gauge fields of the gauge group $U(N_f)_L\times U(N_f)_R$, and finally a two-tensor in the bifundamental corresponding to the complex two-form $b_{MN}$.  In these expressions  $\sigma_{\mu \nu}= i (\gamma_\mu \gamma_\nu - \gamma_\nu
\gamma_\mu)/2$, the adjoint index $a=1,2,3$ for $SU(2)$ flavor symmetry or
$a=0,1,2,3$ for $U(2)$, and the fundamental index $i = 1,2$.

\begin{table}[htbp]
\centering
\begin{tabular}{@{}cc @ {}}
\hline
operator & dual AdS field \\
\hline\hline
$\bar{q}_i \half (1 - \gamma^5) q_j$ ~ ($O_{ij}^S = \bar{q}_i q_j$)  & $X_{ij}$ ($\Re(X_{ij})$)\\
$O_{\mu}^{L,a} = \bar q(x)t^a \gamma_\mu \half (1+\gamma^5) q(x)$ & $A_{L,M}^a$\\
$ O_{\mu}^{R,a} = \bar q(x)t^a \gamma_\mu \half (1-\gamma^5)q(x)$ & $A_{R,M}^a$\\
$ O_{\mu\nu}^{T,a} = \bar q(x) t^a\sigma_{\mu \nu} q(x)$ & $b_{MN}^a$
\end{tabular}
\caption{The operator/field correspondence in the extended hard wall model.}
\label{operators}
\end{table}

It will prove most instructive and convenient to work
with the conserved vector and broken axial-vector
linear combinations of chiral currents,
\begin{equation}\label{OVOA}
J_{\mu}^{V,a} \equiv O_{\mu}^{V,a} \equiv O_{\mu}^{L,a} + O_{\mu}^{R,a} = \bar{q} t^a \gamma_\mu q~, \qquad O_{\mu}^{A,a} \equiv O_{\mu}^{L,a} - O_{\mu}^{R,a} = \bar{q} t^a \gamma_\mu \gamma_5 q~,
\end{equation}
whose dual fields are given by
\begin{equation}\label{VAdef}
V = \half (A_L + A_R)~, \qquad A = \half (A_L - A_R)~,
\end{equation}
as these generate mass eigenstate resonances from the
vacuum when chiral symmetry is broken.  For instance, the vector
current generates the tower of $J^{PC}=1^{--}$ or
$\rho$-meson states, the divergence-free part of the axial current gives rise to a
tower of $1^{++}$ resonances, the $a_1$ states, and its divergence generates the $J^P = 0^-$ pion states. 

These operators and the scalar $X$ have benefitted from extensive study in recent years. Our primary focus will therefore be on
 the tensor operator $O^T=\bar q(x)\sigma_{\mu\nu} q(x)$, continuing the description of its spectrum and interactions
initiated in \cite{Domokos:2011dn}. As can be deduced from QCD, this
operator encapsulates six physical degrees of freedom, and gives
rise to two towers of resonances, one $1^{--}$ and one $1^{+-}$. The
latter are the $b_1$ resonances, nearly degenerate with the
$a_1$. As we will see below, these states also influence the
existing hard wall spectrum as the ``tensor $\rho$s'' mix with the
``vector $\rho$s'' generated by the vector current when chiral
symmetry is broken. We should note at this point that this model is
totally insensitive to isospin: the spectra of isospin-1 and
isospin-0 pairs (like the $\omega$ and $\rho$) are degenerate with
each other. Because isospin-0 modes may undergo mixing with glueball states, which we neglect entirely here, we
will focus our attention (except for a few special cases) on the isotriplet mesons.

\subsection{Action}

The action of our model supplements the original action for the hard wall model with a gauge-invariant first order action for the
two-form field $b_{MN}$:
\begin{equation} \label{Snot}
S = S_{\rm hw}+ S_{\rm CS}+S_{\rm sd}  +S_{\rm int}~,
\end{equation}
where
\begin{equation} \label{Shw}
S_{\rm hw}= \int d^5x \sqrt{g} \tr \left\{ |DX|^2 - \frac{m_{X}^2}{\ell^2} |X|^2 - \frac{1}{4 \ell g_5^2} (F_L^2 +F_R^2) \right\}
\end{equation}
\begin{equation} \label{Scs}
S_{\rm CS} =  \frac{N_c}{24 \pi^2 } \int_{\mathcal{M}} \biggl( \omega_5(A_L) - \omega_5(A_R) \biggr)~,
\end{equation}
\begin{align} \label{Ssd}
S_{\rm sd} = -  \frac{i \sgn(\mu)}{2\ell g_{b}^2}\int_{\cM} \tr \ \left[
\bar{b}\left( D-i \frac{\mu}{\ell} \star\right)b -b\left(
D+i\frac{\mu}{\ell}\star\right)\bar{b}\right]  - \frac{1 }{4\ell g_b^2}\int_{ \d \cM}\tr \ b_{\mu\nu}
b^{\mu\nu}~.
\end{align}
Here $\tr F^3 = d \omega_5$. The additional interaction terms for $b_{MN}$  are described below.

The action is fully invariant under the chiral gauge symmetry corresponding to the
flavor symmetry of QCD;
the bar denotes conjugation of the Lie algebra representation.
The fields $X,b$ transform in the bifundamental of $U(N_f)_L \times U(N_f)_R$.  The field strengths and covariant derivatives are
\begin{align}\label{gaugeconventions}
& (F_{L,R})_{MN} = \d_M (A_{L,R})_N - \d_N (A_{L,R})_M - i [ (A_{L,R})_M , (A_{L,R})_N ]~, \cr
& D_M X = \d_M X - i (A_L)_M X + i X (A_R)_M~, \cr
& D_M b_{NP} = \d_M b_{NP} - i (A_L)_M b_{NP} + i b_{NP} (A_R)_M~.
\end{align}
We work with Hermitian generators and real structure constants normalized such that
\begin{equation}\label{Liealgebra}
\tr \left( t^a t^b \right) = \half \delta^{ab}~, \qquad [t^a,t^b] = i f^{abc} t^c~.
\end{equation}

Chiral symmetry breaking arises when we expand around a non-trivial vacuum of the complex scalar $X$,
\begin{equation}\label{Xvev}
\langle X \rangle \equiv X_0(z) = \frac{1}{2\ell^{3/2}} \left( 2m_q \ell^{\Delta -3} z^{4-\Delta} + \frac{\sigma}{4(\Delta-2)} \ell^{3-\Delta} z^\Delta \right) \mathbbm{1} \equiv \frac{v(z)}{2\ell^{3/2}} \mathbbm{1}~,
\end{equation}
which breaks $U(N_f)_L \times U(N_f)_R$ to the diagonal subgroup.
Here $\Delta = 2 + \sqrt{4 + m_{X}^2}$.   
We choose the factors in front of the non-normalizable and normalizable modes in \eqref{Xvev} such that $m_q$ and $\sigma$ correspond to the quark mass and condensate, respectively.  Since this result differs slightly from \cite{Erlich:2005qh,arXiv:0812.5105}, and since also the precise way in which the quark condensate $\sigma \equiv \langle \bar{q} q \rangle$ is embedded in $X$ will be important for us later, we briefly review the derivation of \eqref{Xvev} in Appendix \ref{Appendix:Xvev}.  The parameterization of $X_0(z)$ in terms of the function $v(z)$ is for later convenience.  The original formulation of the hard wall model fixed the 5d mass of $X$ such that $\Delta=3$, the bare scaling dimension of $\bar{q}q$.
This parameter should in fact run with RG, so we leave it as a free parameter of the model to be determined by comparison with low energy data. It is straightforward to see that evaluating $X$ on
its vev Higgs-es the axial-vector states, lifting their degeneracy with the vector combination of currents.

The action for $b_{MN}$ above is written in terms of differential forms, with
\begin{equation}\label{Dbdef}
(Db)_{MNP} = 3 D_{[M} b_{NP]}~.
\end{equation}
The fact that the action $S_{\rm sd}$ is first order in derivatives is essential for constraining the number of physical degrees of freedom in $b_{MN}$ to match the
six degrees of freedom contained in the dual QCD operator, while maintaining gauge invariance under the flavor symmetry. (See \cite{Domokos:2011dn} for details). 
Naively, a massive complex two-form with a second-order action has twenty degrees of freedom in 5d. Eight of these, the $b_{\mu z}$ fields, are Lagrange multipliers.  If the
equations of motion are first order, only six of the twelve remaining components are independent degrees of freedom.

First order actions on spaces with boundaries  require the addition of boundary terms to ensure a consistent variational principle \cite{Henneaux:1998ch}.
This was first noted by \cite{Arutyunov:1998xt}
for  free two-forms in $AdS_5$.  These boundary terms play a significant role, as without them the two-point function of the fields would vanish. In the hard wall model,
there is also a boundary in the IR which requires a boundary term as well. However, it turns out that for our choice of boundary condition the on-shell value of this term vanishes.  As there
are no symmetries to guide us, choosing the appropriate IR boundary condition for $b_{MN}$ may seem somewhat arbitrary, but in fact the 
choice is unique up to terms higher order in the fields, which are $1/N_c$ suppressed.  In  \cite{Domokos:2011dn} we showed explicitly that of the two possible boundary conditions consistent with the variational principle, only one leads to physically reasonable behavior in the $O^T$ two-point function in the IR.

Note the factor of $\sgn(\mu)$ multiplying the bulk action in
(\ref{Ssd}): as we allow $\mu$ to range over positive and negative
values, this coefficient guarantees that the physical components of
$b_{MN}$ have kinetic terms of the appropriate sign. While this
factor is irrelevant to computations in the free 5d theory, the relative sign
between the tensor kinetic term and the vector kinetic term has
physical meaning once we couple these two sectors.

In summary: the action described so far simply consists of gauge-invariant kinetic and mass terms for the fields in play. We have several free parameters: 
the scales $\ell$ and $z_0$, the couplings $g_5$ and $g_b$, and the AdS masses $m_X$ and $\mu$. The only possible parameter not listed is a mass term for the flavor fields, which we
can set unequivocally to zero
as it is  related to the scaling dimension of the conserved vector current.

\subsection{Interaction terms}

Now let us turn to the interaction terms.
Given that we work in a bottom-up framework, there are an infinite number of interactions we could include. We are most interested
in exploring the effect of the new field $b_{MN}$, particularly the effect of chiral symmetry-breaking in the quadratic order action (for which the most experimental data exists). In order to 
limit the number of new interaction terms, we include all terms only up to dimension six according to 5d dimension counting. 
Once the various fields are rescaled to have canonically normalized
kinetic terms we find that they have mass dimension
\begin{align}
[X] = \frac{3}{2}\qquad [L_M]=[R_M]=\frac{3}{2} \qquad [b_{MN}] =\frac{3}{2}~.
\end{align} 
The canonically normalized versions of the gauge fields have somewhat unconventional dimensions due to the fact that the Yang-Mills coupling has dimension $\text{mass}^{-1/2}$ in 5d. 
Using this dimension counting, we find that there are a total of three interaction terms needed to complete the framework up to dimension six, the
highest dimension present in the original hard-wall model.\footnote{More precisely, these are the highest dimension terms present in the 5d Yang-Mills action. Canonically normalizing the
fields in the Chern-Simons action, however, one would find terms up to dimension $15/2$. Most of these do not contribute to quantities of interest for us, however.}  

First let us describe the terms which affect the quadratic order action for $b_{MN}$ when chiral symmetry is broken.
We begin by writing a term which produces the observed mixing between the vector meson states created by the vector current, and the
vector mesons created by the tensor $O^T$ when chiral symmetry is broken. On the gravity side, this means that we should include a three-point interaction that involves the vector gauge field, $b_{MN}$,
and the scalar $X$. When we expand around the non-trivial vev $X=X_0$, this contributes a quadratic coupling between $V_M$ and $b_{MN}$. As discussed in detail in \cite{Domokos:2011dn}, there 
is only one such term allowed by
the discrete symmetries of the theory with dimension less than six. This is a term of dimension $11/2$, which we label with the coupling  constant $g_1$:
\begin{align}
\label{Sg1} & S_{g_1} =  g_1\int d^5x \sqrt{g} \textrm{tr}\left\{ b_{MN} F_{R}^{MN} \overline{X} + \overline{X} F_{L}^{MN} b_{MN} + X F_{R}^{MN} \bar{b}_{MN} + \bar{b}_{MN} F_{L}^{MN} X \right\}~.
\end{align}
Chiral symmetry-breaking should also lift the degeneracy between the two towers of states ($1^{--}$ and $1^{+-}$)
created by the tensor $O^T$. There is a unique dimension six term which serves this purpose. We label it with the coupling constant $g_2$:
\begin{align}
\label{Sg2} & S_{g_2} = g_2 \ell \int d^5 x \sqrt{g} \tr \left\{ b_{MN} \overline{X} b^{MN} \overline{X} + \bar{b}_{MN} X \bar{b}^{MN} X \right\}~.
\end{align}
We have now described the interaction terms at or below dimension six, which contribute to the quadratic order action when chiral symmetry is broken. There is
one additional term of dimension $11/2$ not included in the gauge-covariant $b_{MN}$ action or the Yang-Mills action so far. We label it with $g_3$
\begin{align}
S_{g_3} =ig_3\int d^5x\sqrt{g}\tr\left\{ \bb^M_{\phantom{M}N}F^N_{(L)P}b^P_{\phantom{P}M} + b^M_{\phantom{M}N}F^N_{(R)P}\bb^P_{\phantom{P}M}\right\} 
\end{align}
and will find that it contributes significantly to the $b_1\rightarrow\omega\pi$ coupling.

Overall, we add three interaction terms, and three free parameters to the model:
\begin{equation}\label{Sint}
S_{\rm int} = S_{g_1} + S_{g_2} + S_{g_3}~.
\end{equation}

Though we do not use $N_c$ counting as our primary organizing principle for the interaction terms, especially from the perspective
of the 4d theory, it is crucial to keep the $N_c$ counting in mind. 
On the gravity side, working order-by-order in $N_c$ essentially corresponds to working order-by-order in fields, as we now explain.
Recall the standard large $N_c$ counting rules for correlation functions of quark bilinears, which can include an infinite number of gauge fields. By counting factors of
$N_c$ that appear in planar correlators of $r$ such mesonic operators $O$ we find
\begin{align}
\langle O_1O_2 \cdots O_r  \rangle\sim N_c^{1-r}~.
\end{align}
We want bilinear operators to create mesons from the vacuum with amplitudes of order 1. 
We should then rescale these currents by factors of $\sqrt{N_c}$ such that
\begin{align}
\langle  (\sqrt{N_c}O_1)(\sqrt{N_c}O_2)\cdots (\sqrt{N_c}O_r)  \rangle \sim N_{c}^{1-r/2}~.
\end{align}
Reasoning analogously on the gravity side, we want to impose the requirement that the two-point functions of quark bilinears be order 1. This forces one to redefine
the fields $A_{L,M}, \ A_{R,M}, \ b_{MN}$ by absorbing appropriate factors of the couplings $g_5$ and $g_b$.  With these rescaled fields and the requirement
that three-point functions should be order $N_c^{-1/2}$, one can quickly deduce, for example, that $g_5\sim\cO (N_c^{-1/2})$.\footnote{While clearly the order 1 coefficients of these terms should run
with RG, we assume that their leading order in $N_c$ remains unchanged between UV and IR.} Furthermore, after these rescalings we know the $N_c$ order of
any coefficient of an $r$-point interaction in the Lagrangian and can arrange terms in $N_{c}^{-1}$ going order by order in fields.
According to $N_c$ counting with our rescaled fields, then, we have $g_1g_bg_5\sim N_c^{-1/2}$,   $g_3g_b^2g_5\sim N_c^{-1/2}$, and  $g_2g_b^2\sim N_c^{-1}$.
We allow these couplings to vary freely, fixing them by fitting to IR data, and checking whether their $N_c$ scaling approximately matches expectations.  Note that modifying any of the terms we have written
above by adding derivative operators would certainly increase their 5d dimension, but would leave them at the same order in $N_c$: it is truncating at dimension six that allows us to remain with a finite number of terms.

\subsection{Gauge fixing}

Some brief remarks about gauge fixing are warranted, as we choose a different gauge from \cite{Erlich:2005qh}.
The pion and sigma fields, with the pions being the Nambu--Goldstone bosons of chiral symmetry breaking, are naturally encoded in the fluctuations in phase and magnitude of $X$, respectively.  While the vacuum breaks the $U(N_f)_L \times U(N_f)_R$ gauge symmetry, the theory itself still possesses the full symmetry.  The simplest way to determine the physical spectrum is to make gauge choices that fully fix the gauge symmetry.  Note that we work with the 5d theory on shell and in the leading saddle point approximation: we are thus dealing with \textit{classical} gauge fixing.

In our model we only consider pion fluctuations, and not sigma modes.  Thus, $X$ takes the restricted form
\begin{equation}\label{pioninX}
X(x,z) = X_0 e^{2i \pi(x,z)}~,
\end{equation}
where $\pi(x,z) = \pi^a(x,z) T^a$, with $\pi^a$ real.  We can use part of the gauge symmetry to eliminate the pion fluctuations from $X$.  Since $X$ transforms as a
bifundamental, the action of the gauge group is
\begin{equation}\label{Xgt}
X \to X' =  e^{-i \Lambda_L} X e^{i\Lambda_R}~.
\end{equation}
By taking $\Lambda_L = - \Lambda_R = \pi$, we have that
\begin{equation}\label{gf1}
X' = X_0~.
\end{equation}
This is our first gauge fixing condition. We will drop the prime on $X$ from now on.

Since $X \propto \mathbbm{1}$, the gauge transformations leaving $X$ invariant are the diagonal $U(N_f)_V$ transformations generated by $\Lambda_L = \Lambda_R \equiv \Lambda_V$.  Under these the axial vector field $A$ transforms as a charged matter field, while the vector gauge field $V$ transforms (not surprisingly!) as a gauge field.
We will use the remaining gauge freedom to impose conditions of $V$, while $A$ is allowed to be a general 5-vector, which we parameterize as
\begin{equation}\label{Aparam}
A_M = (A_{\mu}^\perp + \d_\mu \varphi, A_z )~.
\end{equation}
Here $\d^\mu A_{\mu}^\perp = 0$, and we may assume $\d^\mu \d_\mu \varphi \neq 0$, as it would otherwise have been included in $A_{\mu}^\perp$.  The modes of $A_{\mu}^\perp$ are dual to the tower of axial vector mesons, while $\varphi,A_z$ contain the modes dual to the pion tower.
  
A gauge transformation can always be found that brings us to $V_z = 0$ gauge.  The remaining gauge freedom is to make $U(N_f)_V$ transformations that preserve this condition; these correspond to $z$-independent gauge transformations.  Writing $V_\mu = V_{\mu}^\perp + \d_\mu \phi_V$, where $\d^\mu V_{\mu}^\perp = 0$ and $\d^2 \phi_V \neq 0$, one finds that the equation of motion for $\phi_V$ following from \eqref{Snot} is $\d^2 \d_z \phi_V + \cO(\Phi^2) = 0$, where $\Phi$ represents any field.  Working perturbatively, this means that $\phi_V$ has a leading order $z$-independent piece plus corrections that are quadratic in the fields.  We may use our final bit of residual gauge symmetry to set the leading order piece to zero.  It can be shown that the quadratic order terms in $\phi_V$ do not contribute to any three-point couplings when plugged back into \eqref{Snot}.  Hence for the purposes of this paper we may safely take
\begin{equation}\label{Vparam}
V_M = (V_\mu,0)~, \qquad \d^\mu V_\mu = 0~.
\end{equation}
%

\section{Computational techniques}\label{sec:computation}

We wish to understand the effect of the chiral symmetry breaking
interaction terms on low energy observables such as masses of the
mesonic excitations, their decay constants, and some interactions
that mediate observed decays. In this section we describe in general terms how to
extract this information from the extended hard wall model. 
Details are left to the appendices.

We begin by analyzing the quadratic part of \eqref{Snot}, which we denote $S^{(2)}$.
The linearized equations of motion and boundary conditions that
result from imposing $\delta S^{(2)} = 0$ will determine
the bulk-to-boundary propagators and spectrum of normalizable modes.
From these we derive the two-point functions, which contain the
meson masses and decay constants.  Having properly identified the
physical fluctuations, we then compute a selection of three-point
couplings.  We apply some of these results to obtain the tree-level
contribution to the decay rate and $D/S$ wave amplitude ratio for
$b_1 \to \omega  \pi$.

\subsection{Equations of motion and boundary conditions} \label{subsec:eomandbcs}

We would like to fix $X$ to its vev (\ref{Xvev}), and study the action
\eqref{Snot} on this background as a functional of the 5d fields
$V,A,b$.  One should immediately raise objections to this program since our slice of AdS space together with the scalar vev do not constitute a solution to the coupled Einstein-matter equations of motion.  However, provided that the mass of the scalar is in a certain range, there will exist a solution to the coupled equations with asymptotically AdS boundary conditions \cite{Hertog:2004ns}, (see also \cite{Hertog:2006rr}), and our treatment serves as a reasonable approximation to that exact solution.  The mass range is $-4 < m_{X}^2 < 0$, corresponding to $2 < \Delta < 4$.  The lower limit on the mass-squared  is the Breitenlohner-Freedman bound, below which an asymptotically AdS solution becomes perturbatively unstable\footnote{We could extend the range of $\Delta$ down to the unitarity bound $\Delta >1$ by choosing $\Delta$ to be the smaller of the two roots of $\Delta (\Delta -4) = m_{X}^2$, which is permissible when $-4 < m_{X}^2 < -3$.}.  The upper limit is easier to understand from the dual field theory point of view.  $\Delta > 4$ corresponds to an irrelevant operator.  Deforming the field theory by an irrelevant operator renders it ill-defined in the UV; the corresponding gravity statement is that asymptotically AdS boundary conditions can no longer be maintained when $m_{X}^2 > 0$.  This conclusion is also backed up by a perturbative analysis of the coupled Einstein-matter equations along the lines of \cite{de Haro:2000xn}.  Henceforth we will restrict ourselves to the range $2 < \Delta < 4$ and work in the approximation where backreaction of the scalar vev is neglected.

Variation of the action at quadratic order results in the
linearized equations of motion
\begin{align}\label{5deomb}
0 =&~ i db + \frac{\mu}{\ell} \star b - \frac{2 g_{1} g_{b}^2 \sgn(\mu)}{\sqrt{\ell}} v(z) \star dV - \frac{g_2 g_{b}^2 \sgn(\mu)}{\ell} v(z)^2 \star \bar{b}~, \\ \label{5deomV}
0 =&~ d \star \left( dV - \frac{g_1 g_{5}^2}{\sqrt{\ell}} v(z) (b + \bar{b}) \right)~, \\ \label{5deomA}
0=&~ d \star dA - \frac{g_{5}^2}{\ell^2} v(z)^2 \star A~.
\end{align}
The variation reduces to a set of boundary terms when evaluated on a solution to these equations, namely,
\begin{align}\label{deltaS2}
\delta S^{(2)} =&~ \int d^4 x \tr \displaystyle\biggl\{ \frac{i \sgn(\mu)}{4 \ell g_{b}^2} \epsilon^{\mu\nu\rho\sigma} \left( b_{\mu\nu}^{\mp} \overline{\delta b_{\rho\sigma}^{\pm}} - \overline{b_{\mu\nu}^{\mp}} \delta b_{\rho\sigma}^{\pm} \right) + \frac{2}{g_{5}^2 z} (\d_z A_\mu - \d_\mu A_z) \delta A^{\mu} + \cr
&~ \qquad \qquad \quad + \left[ \frac{2}{g_{5}^2 z} \d_z V_\mu + \frac{2 g_{1} v(z)}{\sqrt{\ell} z} (b_{\mu z} + \bar{b}_{\mu z} ) \right] \delta V^\mu \displaystyle\biggr\} \displaystyle\bigg|_{\varepsilon}^{z_0}~.
\end{align}
Here $b_{\mu\nu}^{\pm}$ is the self-dual $(+)$ or anti-self-dual $(-)$ part of $b_{\mu\nu}$, which can be obtained via projection,
\begin{equation}\label{sdproj}
b_{\mu\nu}^{\pm} = \half \left( \delta_{\mu}^{[\rho} \delta_{\nu}^{\sigma]} \pm \frac{i}{2} \epsilon_{\mu\nu}^{\phantom{\mu\nu}\rho\sigma} \right) b_{\rho\sigma}~.
\end{equation}
The top sign in \eqref{deltaS2} is chosen when $\mu > 0$ and the
bottom sign when $\mu < 0$. See \cite{Domokos:2011dn} for details.

Boundary conditions should be chosen such that the UV boundary
terms at $z = \varepsilon$ and the IR ones at $z = z_0$ in
\eqref{deltaS2} vanish: this guarantees that we are in fact
expanding around a stationary point of the action functional.  We
discuss each of these in turn, beginning with the UV boundary terms,
for which we will require some results concerning the asymptotic
behavior of solutions to \eqref{5deomb}-\eqref{5deomA}.

The equations of motion for $V,A$ are second order. $b$, meanwhile,
has two first order equations involving both the real and imaginary
parts, which we can convert into a single second order equation by
acting with the appropriate differential operator. As usual, the
general solution for each field will be a sum of two types of
solution with differing behavior as $z = \varepsilon \to 0$ conventionally referred to as
normalizable and non-normalizable.  The leading scaling behaviors
are
\begin{align}\label{scaling}
\textrm{non-normalizable:} \quad &  b_{\mu\nu} \sim z^{-|\mu|}~, \quad  V_\mu , A_{\mu}^\perp \sim z^0~, \quad  \varphi \sim z^0~, \cr
\textrm{normalizable:} \quad & b_{\mu\nu} \sim z^{|\mu|}~, \quad  V_\mu, A_{\mu}^\perp \sim z^2~, \quad  \varphi \sim z^{ \min \{ 2, 2\Delta - 4 \} }~.
\end{align}
These results follow from a detailed analysis of
\eqref{5deomb}-\eqref{5deomA} given below. Some comments and caveats
should be stated here, however:
\begin{itemize}
\item When $\mu >0$ ($<0$) it is $b_{\mu\nu}^+$ ($b_{\mu\nu}^-$) that contributes the dominant behavior in \eqref{scaling}.
\item The scaling \eqref{scaling} is independent of $g_1,g_2$.  The interaction terms, being proportional to the tachyon vev, give subleading terms in the differential equations as $z \to 0$.  Therefore these results can also be obtained in the $g_1,g_2 \to 0$ limit where we have the original hard wall model and a decoupled free $b$-sector.  Note that these statements are only true provided $\Delta < 4$ which we are assuming.
\item  We assumed a further restriction on parameter space when stating
\eqref{scaling}.  The $g_1$ terms in \eqref{5deomb}, \eqref{5deomV}
couple $V$ with specific components of $b$.  This means that each
solution for $V$ induces one for $b$ and vice-versa. There are thus
a total of four solutions to the coupled $b$-$V$ system, with
correlated asymptotics $b \sim (z^{\mp |\mu|}, z^{4-\Delta},
z^{6-\Delta})$ and $V \sim (z^{\mp |\mu| + 6-\Delta} , z^0, z^2 )$.
In order that the $V$ wavefunction induced from the non-normalizable
$b$-type solution not dominate the wavefunction of the
non-normalizable $V$-type solution, one must assume $|\mu| < 6 -
\Delta$.  The necessity of this condition can be argued from both physical and technical standpoints.  First,
the leading asymptotics of $V$ are directly tied to the dimension of dual vector current, $\bar{q} \gamma_\mu t^a q$.  This is a conserved quantity; its dimension is protected and given by the classical value $\Delta_V = 3$.  On the gravity side this requires the leading asymptotics $V \sim z^0$.  Secondly, as we show in Appendix \ref{asymptotic}, allowing $|\mu| > 6-\Delta$ leads to divergences in the on-shell action that cannot be removed by counterterms that are local functionals of the sources; the holographic renormalization procedure of \cite{Henningson:1998gx,de Haro:2000xn} breaks down.
\end{itemize}

The two classes of UV boundary conditions we consider are
differentiated by whether or not we turn on non-normalizable
solutions.  If we are only interested in normalizable modes of the 5d
fields, dual to meson states,  we set the coefficients of the
non-normalizable solutions to zero.  This eliminates the UV boundary
term in \eqref{deltaS2} in the $\varepsilon \to 0$ limit.  If we are
interested in the bulk-to-boundary propagators of the 5d fields, with which we can study correlation functions of QCD operators, we turn on
non-normalizable solutions.  The UV
boundary term in \eqref{deltaS2} is made to vanish by holding the
fields fixed, $\delta \Phi = 0$, at the boundary.  The values they
are fixed to are the classical sources dual to the QCD operators. The
precise boundary conditions, given in 4d momentum space, are
\begin{align}\label{sources}
& \Re{b_{\mu\nu}^{\pm}}(k,\varepsilon) =  \frac{\ell^{|\mu| - 1/2}}{\varepsilon^{|\mu|}} \cT_{\mu\nu}(k)~, \qquad V_\mu(k,\varepsilon) = V_{\mu}^0(k)~,  \cr
&  A_{\mu}^\perp(k,\varepsilon) = A_{\mu}^0(k)~, \qquad \varphi(k,\varepsilon) = \pi(k)~,
\end{align}
where $\Phi(k,z) = \int d^4 x e^{-ik \cdot x} \Phi(x,z)$, and the self-dual or anti-self-dual projection on $b$ is again correlated with the sign of $\mu$.  All other components of 5d fields are determined in terms of these
asymptotics via the equations of motion.  The sources $\cT_{\mu\nu}(k)$, $V_{\mu}^0(k)$, and $A_{\mu}^0(k) + i k_\mu \pi(k)$
are dual to the (classical) dimension-three tensor, vector, and axial-vector quark bilinears,
respectively.  In particular, $\pi(k)$ is dual to the divergence of the axial current, which contains the same
information as the scalar quark bilinear.

Next, let us consider the IR contribution to \eqref{deltaS2}.  The boundary term is a sum of terms of the
form $P_{\Phi} \cdot \delta \Phi$, where $P_\Phi$ is some function of the fields.  Naively, the two options are a
 Dirichlet condition that holds $\Phi$ fixed, or a Neumann condition that sets $P_{\Phi} = 0$.  In the case
 of $\delta V^\mu$ and $\delta A^\mu$ we can use $U(N_f)_L \times U(N_f)_R$ gauge invariance to argue in favor
 of the Neumann condition, as the Dirichlet condition would explicitly break this symmetry.  Since $b_{\mu\nu}$ does
 not correspond to a conserved current in QCD, there is no analogous argument to determine its boundary condition. In
 \cite{Domokos:2011dn}, however, we used physical considerations to argue in favor of the Neumann condition.
 In the free limit, $g_1,g_2 \to 0$, choosing the Dirichlet condition results in a pole at $k^2 \to 0$
 in the tensor-tensor two-point function. As
 there are no massless particles in QCD with the appropriate quantum numbers to explain this divergence, it is
 reasonable to conclude that this boundary condition does not yield a good gravity dual to QCD.
 With the Neumann condition there is no such divergence.
 We continue to follow this prescription here, since we will be interested in regions of parameter space
 where $g_1,g_2$ are small.  This results in the following set of IR boundary
 conditions:\footnote{As we discussed in \cite{Domokos:2011dn}, one can modify the IR
 boundary conditions by adding terms to the action \eqref{Snot} that are localized on the IR boundary.
 Such terms, beyond the ones already considered in \eqref{Snot}, are subleading in $1/N_c$, and are on the same footing
 as additional $1/N_c$-suppressed terms that we could add to the bulk.}
\begin{align}\label{IRbcs}
& b_{\mu\nu}^{\mp}(k,z_0) = 0~, \qquad \d_z V_\mu + \frac{g_1 g_{5}^2}{\sqrt{\ell}} v(z) (b_{\mu z} + \bar{b}_{\mu z}) \displaystyle\bigg|_{z = z_0} = 0~, \cr
& \d_z A_{\mu}^\perp |_{z = z_0} = 0~, \qquad (\d_z \varphi - A_z) |_{z = z_0} = 0~.
\end{align}
These conditions apply whether we are restricting to normalizable
modes or determining the bulk-to-boundary propagators.

While the 5d-covariant notation of \eqref{5deomb}-\eqref{5deomA}
provides a relatively simple presentation in which symmetries are
manifest, the dynamics are obscured by the redundancy of the
description.  For example, $b_{MN}$ contains twenty real components,
but only six of these correspond to physical degrees of freedom.
Since we will only be using on-shell quantities in the 5d theory, we
are free to use equations of motion to eliminate extraneous fields.
This process is summarized here; see appendix \ref{S2simp} for
further details.

A minimal set of fields may be chosen as follows.  First, $V_\mu,
A_{\mu}^\perp$ contain all degrees of freedom associated with the
vector-like $\omega/\rho$ and axial $f_1/a_1$ mesons respectively.
Below we will often use rescaled versions of these fields,
\begin{equation}\label{calV}
V_\mu(k,z) = g_5 \cV_\mu(k,z)~, \qquad A_{\mu}^\perp(k,z) = g_5 \cA_{\mu}(k,z)~.
\end{equation}
The scalars $(\varphi, A_z)$ contain all $\eta/\pi$ fluctuations.  The linearized equation of motion for $A_z$ has no $z$-derivatives; thus $A_z$ can be solved for algebraically, leaving a second-order equation in $\d_z$ for $\varphi(k,z)$.  We will find it more convenient, however, to introduce a different field variable in the pion sector.  We define $\hat{A}(k,z)$ such that
\begin{equation}\label{Ahat}
\varphi(k,z) = - \frac{z^3}{k z_{0}^2 v(z)^2} \d_z \hat{A}(k,z)~, \qquad A_z(k,z) = \frac{k z^3}{z_{0}^2 v(z)^2} \hat{A}(k,z)~.
\end{equation}
The relation between $\varphi,A_z$ implied by these identifications is consistent with their equations of motion.  Here $k \equiv \sqrt{k^2}$ is always real and positive since we we will only be using these expressions on shell.

Of the twenty real components of $b_{MN}$, eight are eliminated by a Proca-like condition.  Half of the remaining twelve have the interpretation of momenta and half coordinates.  Using the equations of motion one can show that $b_{MN}$ may be parameterized in terms of two 4d-transverse vectors, $\cH_\mu(k,z)$ and $\cW_\mu(k,z)$, as follows:
\begin{align}\label{bmunudecomp}
b_{\mu\nu}(k,z) =&~ -\frac{i g_b \sqrt{\ell}}{k z_0} \displaystyle\biggl\{ \epsilon_{\mu\nu}^{\phantom{\mu\nu}\rho\sigma} k_\rho \left( \cH_\sigma + i \cW_\sigma \right) + \frac{2i \sgn(\mu) z}{|\mu|+u_2} \d_z k_{[\mu} \cH_{\nu]} + \cr
& \qquad \qquad \qquad \qquad \qquad - \frac{2 \sgn(\mu) z}{|\mu|-u_2} \d_z k_{[\mu} \cW_{\nu]} + \frac{2 (k z_0) u_1}{|\mu| - u_2} k_{[\mu} \cV_{\nu]} \displaystyle\biggr\}~, \\ \label{bmuzdecomp}
b_{\mu z}(k,z) =&~ - g_b \sqrt{\ell} \displaystyle\biggl\{ \frac{\sgn(\mu) k z}{z_0 (|\mu| - u_2)} \cW_\mu + \frac{u_1}{|\mu| - u_2} \d_z \cV_\mu - \frac{i \sgn(\mu) k z}{z_0 (|\mu| + u_2)} \cH_\mu \displaystyle\biggr\}~.
\end{align}
Observe that these expressions also have dependence on $\cV_\mu$.  This comes from mixing terms in auxiliary equations of motion that have been used.  We have introduced notation for combinations of parameters that appear often in the following:
\begin{equation}\label{udefs}
u_1(z) \equiv 2g_1 g_b g_5 v(z)~, \qquad u_2(z) \equiv g_2 g_{b}^2 v(z)^2~.
\end{equation}
As is clear from \eqref{bmunudecomp}, $\cH$ parameterizes the real-transverse components of $b_{\mu\nu}$, while $\cW$ parameterizes the imaginary-transverse components.  The transformation properties of $b$ under charge conjugation and parity identify $\cH$ with the $h_1/b_1$ mesons and $\cW$ with the tensor-like $\omega/\rho$'s.

The factors of $(|\mu| \pm u_2)^{-1}$ arise from
eliminating auxiliary components (such as $b_{\mu z}$) and plugging
back into the equations of motion. As we will discuss soon, we must
restrict parameter ranges such that $|\mu| > |u_2(z)|$, $\forall z
\in [0,z_0]$, otherwise we will not necessarily obtain a positive
definite spectrum. Intuitively, the $g_2$ couplings act like
off-diagonal terms in the mass-squared matrix for $b$ once $X$ is
evaluated on its vev.  If these off-diagonal terms become too large,
the eigenvalues may become negative.  Since $v(z)$ is a
monotonically increasing function, the constraint is equivalent to
$|\mu| > |u_2(z_0)|$. Note that this restriction also makes sense
from the perspective of the large $N_c$ expansion. The combination
of couplings $g_2g_b^2$ that appears in $u_2$ should be
$\cO(N_c^{-1})$ compared to the diagonal terms of the mass-squared
matrix, which, by construction, are $\cO(1)$.

In appendix \ref{S2simp} we show that the quadratic part of \eqref{Snot} is equivalent, on shell, to the following action,
written in terms of fields $\{\hat{A}, \cA_\mu, \cH_\mu, R_\mu \equiv (\cW_\mu, \cV_\mu)^T \}$:
\begin{align}\label{Snot2}
S^{(2)} =&~ - \int \frac{d^4 k}{(2\pi)^4} \int_{\varepsilon}^{z_0} dz \tr \displaystyle\biggl\{ w_{(\pi)} \bar{\hat{A}} (k^2 - O^{(\pi)}) \hat{A} + w_{(a)} \bar{\cA}_\mu (k^2 - O^{(a)} ) \cA^\mu  + \cr
& \qquad  \quad + w_{(b)} \bar{\cH}_\mu (k^2 - O^{(b)} ) \cH^\mu + \bar{R}_{\mu} {\bf W}_{(\rho)} (k^2 {\bf 1} - {\bf O}^{(\rho)}) R^{\mu} \displaystyle\biggr\} + S_{\rm bndry}~,
 \end{align}
where all operators are of Sturm--Liouville type.  Explicitly,
\begin{align}\label{Opia}
O^{(\pi,a)} =&~ - w_{(\pi,a)}^{-1} \d_z \left[ w_{(\pi,a)} \d_z \right] + \frac{g_{5}^2 v(z)^2}{z^2}~, \\ \label{Ob}
O^{(b)} =&~ - w_{(b)}^{-1} \d_z \left[ w_{(b)} \d_z \right] + \frac{1}{z^2} (\mu^2 - u_{2}^2)~, \\ \label{Orho}
{\bf O}^{(\rho)} =&~  \left( \begin{array}{c c} - w_{(w)}^{-1} \d_z [ w_{(w)} \d_z ] + \frac{1}{z^2}(\mu^2 - u_{2}^2) &\quad  \frac{\sgn(\mu) k}{z_0 w_{(w)}} \d_z \left( \frac{u_1}{|\mu|-u_2} \right) \\   \frac{\sgn(\mu) k}{z_0 w_{(v)}} \d_z \left( \frac{u_1}{|\mu|-u_2} \right) &\quad -  w_{(v)}^{-1} \d_z [ w_{(v)} \d_z ]  \end{array}\right)~.
\end{align}
The weight metric in the $\rho$ sector is
\begin{equation}\label{rhomet}
{\bf W}_{(\rho)} = { \, \textrm{diag} \,}(w_{(w)}, w_{(v)} )~,
\end{equation}
and the  weight functions are
\begin{align}\label{weights}
& w_{(\pi)} = \frac{z^3}{z_{0}^4 v(z)^2}~, \qquad w_{(a)} = z^{-1}~, \qquad w_{(b)} = \frac{z}{z_{0}^2 (|\mu| + u_2)}~, \cr
& w_{(w)} = \frac{z}{z_{0}^2 (|\mu| - u_2)}~, \qquad w_{(v)} = \frac{1}{z} \left( 1 - \frac{u_{1}^2}{|\mu| - u_2} \right)~.
\end{align}
Here we see that positivity of the norm in the $b_1$ and $\rho$
sectors imposes the conditions $|\mu| > |u_2(z_0)|$ and $|\mu| >
u_2(z_0) + u_{1}^2(z_0)$ on the parameters of the model. The last
term in \eqref{Snot2} encompasses all of the boundary terms present
and determines the on-shell action to quadratic order in sources
when we evaluate $S^{(2)}$ on the bulk-to-boundary propagators.  We
will have more to say about this below.

Finally, let us state the boundary conditions satisfied by these
fields.  First we introduce some notation for this purpose that will
come in handy below. In each sector, we define linear functionals,
$Q$ and $P$, which encode the UV and IR boundary conditions,
respectively:
\begin{align}\label{Qs}
& Q_{(\pi)}[f] = -z_{0}^2 w_{(\pi)} \d_z f~, \quad   Q_{(a)}[f] = f~, \quad  Q_{(b)}[f] = \half \left( f - z_{0}^2 w_{(b)} \d_z f \right)~, \cr
& Q_{(\rho)} = (Q_{(w)}, Q_{(v)})^T~, \qquad \textrm{with} \cr
& Q_{(w)}[ (f,g)^T] = \half \left( f - z_{0}^2 w_{(w)} \d_z f + \frac{\sgn(\mu) k z_0 u_1}{|\mu| - u_2} g \right)~, \quad  Q_{(v)}[ (f,g)^T] = g~,
\end{align}
and
\begin{align}\label{Ps}
& P_{(\pi)}[f] = f~, \quad P_{(a)}[f] = z_{0}^2 w_{(a)} \d_z f~, \quad P_{(b)}[f]  = f + z_{0}^2 w_{(b)} \d_z f~,  \cr
& P_{(\rho)} = (P_{(w)}, P_{(v)})^T~, \qquad \textrm{with} \cr
& P_{(w)}[ (f,g)^T] =  f + z_{0}^2 w_{(w)} \d_z f - \frac{\sgn(\mu) k z_0 u_1}{|\mu| - u_2} g ~,  \cr
& P_{(v)}[ (f,g)^T] = z_{0}^2 w_{(v)} \d_z g - \frac{\sgn(\mu) k z_0 u_1}{|\mu| - u_2} f~.
\end{align}
Additionally we define sources
\begin{align}\label{Ss}
& S_{(\pi)} = k \pi(k), \quad   S_{(a)\mu} = \frac{1}{g_5} A_{\mu}^0(k)~, \quad  S_{(b)\mu} =  \frac{i z_0 \ell^{|\mu|-1}}{2 g_b k \varepsilon^{|\mu|}} \epsilon_{\mu}^{\phantom{\mu}\nu\rho\sigma} k_\nu \cT_{\rho\sigma}(k)~, \cr
& S_{(\rho)\mu} = (S_{(w)\mu}, S_{(v)\mu})^T~, \qquad \textrm{with} \cr
& S_{(w)\mu} = \frac{i \sgn(\mu) z_0 \ell^{|\mu|-1}}{g_b k \varepsilon^{|\mu|}} k^\nu \cT_{\nu\mu}(k) ~, \quad  S_{(v)\mu} = \frac{1}{g_5} V_{\mu}^0(k)~.
\end{align}
Letting $\{ \Phi \} = \{ \hat{A}, \cA_\mu, \cH_\mu, R_\mu \}$ denote the collection of field content for all of the different sectors and $O$ the corresponding Sturm-Liouville operators, the bulk-to-boundary propagators are solutions to $(k^2 - O)\Phi = 0$, satisfying the UV and IR boundary conditions
\begin{equation}\label{bcshort}
Q[\Phi](k,\varepsilon) = S~, \qquad P[ \Phi](k,z_0) = 0~.
\end{equation}
Here $Q$, $P$ and $S$ refer to the appropriate boundary conditions or sources for a given field in $\left\{\Phi\right\}$~.
Normalizable modes corresponding to individual meson states are eigenfunctions of the various $O$'s.  In this case we
set $Q[\Phi](k,\varepsilon) = 0$; as $\varepsilon \to 0$ this isolates the normalizable modes.  The IR boundary conditions $P[ \Phi](k,z_0) = 0$ thus lead to quantization of the spectra.

\subsection{Bulk-to-boundary propagators, on-shell action, and two-point functions}\label{Section:nonnormmodes}

Although we must ultimately resort to numerics to obtain explicit
solutions for bulk-to-boundary propagators and normalizable
eigenfunctions, a great deal of progress can be made analytically in
evaluating the on-shell action and two-point functions.  The
on-shell action at quadratic order is given by the boundary terms in
\eqref{Snot2}.  Our analysis in appendix \ref{S2simp} results in
\begin{align}\label{Sbndry}
S_{\rm bndry} =&~ - \int \frac{d^4 k}{(2\pi)^4} \tr \displaystyle\biggl\{ w_{(\pi)} \bar{\hat{A}} \d_z \hat{A} + w_{(a)} \cA_\mu \d_z \bar{\cA}^\mu + \frac{1}{2 z_{0}^2} \left( |\cH_\mu|^2 - z_{0}^4 w_{(b)}^2 | \d_z \cH_\mu|^2 \right)  \cr
&~~~ \qquad \qquad \qquad + \frac{1}{2 z_{0}^2} \left[ | \cW_{\mu} |^2 - \left| z_{0}^2 w_{(w)} \d_z \cW_\mu - \frac{ \sgn(\mu) k z_0 u_1}{|\mu| - u_2} \cV_\mu \right|^2 \right] + \cr
&~~~ \qquad \qquad \qquad + \left[ w_{(v)} \d_z \bar{\cV}^\mu - \frac{ \sgn(\mu) k u_1}{z_0 (|\mu| - u_2)} \bar{\cW}^\mu \right] \cV_\mu \displaystyle\biggr\} \displaystyle\bigg|_{z= \varepsilon} \cr
=&~ - \sum_{\left\{ \Phi\right\}} \frac{1}{z_{0}^2} \int \frac{d^4 k}{(2\pi)^4} \tr \left\{ \overline{P[\Phi]}(k,\varepsilon) \cdot Q[\Phi](k,\varepsilon) \right\}~.
\end{align}
Here we have used the IR boundary conditions to eliminate the boundary terms at $z = z_0$.

We would like to evaluate this action on the bulk-to-boundary propagators.  The $Q$'s can simply be replaced with the sources using \eqref{bcshort}, but in order to express the $P$'s in terms of sources we need to find the explicit solutions for the bulk-to-boundary propagators.  Since we will eventually be taking the $\varepsilon \to 0$ limit, it is useful to separate out the non-normalizable and normalizable components of the solution.

\subsubsection{ $\pi$, $a_1$, and $b_1$ sectors}

We begin with the $\pi$, $a_1$, and $b_1$ sectors.  The non-normalizable and normalizable solutions, $f^{\rm n.n.}(k,z)$ and $f^{\rm n.}(k,z)$, are two particular solutions to $(k^2 - O) f = 0$ distinguished by their asymptotic behavior as $z \to 0$.  The general solution is a linear combination $f = c^{\rm n.n.} f^{\rm n.n.} + c^{\rm n.} f^{\rm n.}$.  The coefficients are fixed by the boundary conditions \eqref{bcshort} and this leads to the bulk-to-boundary propagators for $\pi$, $a_1$, and $b_1$ as
\begin{align}\label{btbs}
& \Phi(k,z) = \frac{ f^{\rm n.n.}(k,z) - C(k) f^{\rm n.}(k,z) }{Q^{\rm n.n.}(k,\varepsilon) - C(k) Q^{\rm n.}(k,\varepsilon)} S(k)~, \qquad \textrm{where} \cr
&  C(k) = \frac{P^{\rm n.n.}(k,z_0)}{ P^{\rm n.}(k,z_0)}~.
\end{align}
Here we are using a shorthand $Q^{\rm n.n.} \equiv Q[f^{\rm n.n.}]$, etc.  Acting on an element of $\left\{\Phi\right\}$ with the corresponding $Q$ and evaluating at $z = \varepsilon$ leads to the UV boundary condition, while acting on it with $P$ and evaluating at $z = z_0$ gives zero, thanks to the form of $C(k)$.  Plugging this solution into \eqref{Sbndry} gives a contribution
\begin{equation}\label{Soss}
\overline{P[\Phi]}(k,\varepsilon) \cdot Q[\Phi](k,\varepsilon) = \bar{S}(k) \frac{ P^{\rm n.n.}(k,\varepsilon) - C(k) P^{\rm n.}(k,\varepsilon) }{Q^{\rm n.n.}(k,\varepsilon) - C(k) Q^{\rm n.}(k,\varepsilon)} S(k)~.
\end{equation}
%

\subsubsection{$\rho$ sector}

Essentially the same formula holds for the $\rho$ sector, suitably generalized.  In the $\rho$ sector $(k^2 - {\bf O}) R = 0$ is a coupled second order equation which has a four-dimensional solution space.  We isolate four particular solutions by their behavior near $z = 0$.  They are non-normalizable and normalizable solutions of $w$- and $v$-type.  A solution of ``$w$-type'' can be characterized as one that becomes pure $\cW$, $R = (\cW,0)^T$, in the $g_1 \to 0$ limit where tensor-vector mixing is turned off.  Thus we may also 
denote these as tensor-like and vector-like.  If $i \in \{ {\rm n.n.}(w), {\rm n.n.}(v), {\rm n.}(w), {\rm n.}(v) \}$ is an index running over these four solutions, then a general solution has the form $R = c^i R^i$.

The coefficients are fixed with the four boundary conditions $Q[ R] = S$, $P[R] = 0$.  Let $\alpha,\beta$ be indices running over $\{ w, v \}$.  Define the two-component vectors $c^{\rm n.n.}$, $c^{\rm n.}$ with components $c^{{\rm n.n.}(\beta)}$ and so on. In the $\rho$ sector the boundary conditions are now given by two-by-two matrices ${\bf Q}^{\rm n.n.}$,  ${\bf Q}^{\rm n.}$, ${\bf P}^{\rm n.n.}$,  ${\bf P}^{\rm n.}$ with components $({\bf Q}^{\rm n.n.})_{\alpha\beta} = Q_{(\alpha)}[R^{{\rm n.n.}(\beta)}]$ etc.  Then the boundary conditions take the form
\begin{equation}\label{rhobcs}
{\bf Q}^{\rm n.n.}(k,\varepsilon) c^{\rm n.n.} + {\bf Q}^{\rm n.}(k,\varepsilon) c^{\rm n.} = S~, \qquad {\bf P}^{\rm n.n.}(k,z_0) c^{\rm n.n.} + {\bf P}^{\rm n.}(k,z_0) c^{\rm n.} = 0~,
\end{equation}
the solution to which is
\begin{align}\label{rhobcssolve}
& c^{\rm n.} = - {\bf C}(k) c^{\rm n.n.} ~, \qquad c^{\rm n.n.} = \left[ {\bf Q}^{\rm n.n.}(k,\varepsilon) - {\bf Q}^{\rm n.}(k,\varepsilon) {\bf C}(k) \right]^{-1} S~, \qquad \textrm{with} \cr
& {\bf C}(k) = \left( {\bf P}^{\rm n.}(k,z_0) \right)^{-1} {\bf P}^{\rm n.n.}(k,z_0)~.
\end{align}
This gives the coefficients $c^i$, and hence the bulk-to-boundary propagator $c^i R^i$ in terms of the source.  In order to compute the on-shell action we are interested in $P[R]$, which takes the form
\begin{align}\label{PofR}
P[R](k,\varepsilon) =&~ {\bf P}^{\rm n.n.}(k,\varepsilon) c^{\rm n.n.} + {\bf P}^{\rm n.}(k,\varepsilon) c^{\rm n.}~.
\end{align}
Plugging in the solution for the $c$'s, we find that the $\rho$ sector contribution to \eqref{Sbndry} is
\begin{equation}\label{Sosrho}
\overline{P[R]}(k,\varepsilon) \cdot Q[R](k,\varepsilon) = \bar{S}(k)\frac{ {\bf P}^{\rm n.n.}(k,\varepsilon)^T - {\bf C}(k)^T {\bf P}^{\rm n.}(k,\varepsilon)^T }{[ {\bf Q}^{\rm n.n.}(k,\varepsilon)^T - {\bf C}(k)^T {\bf Q}^{\rm n.}(k,\varepsilon)^T ]} S(k)~,
\end{equation}
using the $\rho$ sector versions of $P$, $Q$ and $S$, as usual.
Here $T$ denotes transpose and the matrix in the denominator is to be understood as
matrix multiplication by the inverse from the left.

\subsubsection{On-shell action and holographic renormalization}

The contributions \eqref{Soss}, \eqref{Sosrho} determine the
on-shell action at finite $\varepsilon$.  In order to compute field
theory correlators via gauge/gravity duality, we need to
take $\varepsilon \to 0$.  Generally there are both divergent and
finite terms in the limit.  The finite terms can be further divided
into those which are entire functions of $k^2$ and those which are
non-analytic in $k^2$. In all cases considered here the divergent terms come from
the $\bar{S} (Q^{\rm n.n.})^{-1} P^{\rm n.n.} S$ piece of
\eqref{Soss}, \eqref{Sosrho}, and are furthermore entire functions
of $k^2$.  Finite terms may come from this piece as well, but
they are also analytic functions of $k^2$.   The non-analytic finite
terms are obtained from the $\bar{S} (Q^{\rm n.n.})^{-1} C(k) P^{\rm
n.} S$ piece.  The $Q^{\rm n.}$ term in the denominator of
\eqref{Soss}, \eqref{Sosrho} never contributes in the $\varepsilon
\to 0$ limit. These results are reviewed in Appendix
\ref{asymptotic}, where we present the necessary asymptotics and, in
particular, give our conventions for the non-normalizable and
normalizable solutions.

The process of holographic renormalization \cite{Henningson:1998gx,de Haro:2000xn} introduces a
counterterm action to cancel the divergences that arise in the
on-shell action. This mechanism is dual to the regularization of UV
divergences in the field theory. The counterterm action should be a
local functional of the sources, as guaranteed by the analyticity in
momentum space of the terms that diverge as $\varepsilon\rightarrow
0$ in the on-shell action. Similarly, the finite terms in the
on-shell action that are entire functions of $k^2$ are
scheme-dependent; they may be shifted to zero by introducing local
finite counterterms. The renormalized action is thus determined by
terms finite in $\varepsilon$ and non-analytic in $k$.  The results
of Appendix \ref{asymptotic} lead to
\begin{align}\label{Sren}
\lim_{\varepsilon \to 0} S_{\rm ren}^{(2)} =&~ \int \frac{d^4 k}{(2\pi)^4} \tr \displaystyle\biggl\{ - \frac{m_{q}^2 k^2 \ell^{2\Delta -6} C_{(\pi)}(k)}{(2\Delta -6) z_{0}^{2\Delta -6}}  \bar{\pi} \pi + \frac{2 C_{(a)}(k)}{g_{5}^2 k^2 z_{0}^2} \bar{A}_{\mu}^0 (k^2 \eta^{\mu\nu} - k^\mu k^\nu) A_{\nu}^0 + \cr
&~ \qquad \qquad \quad  +  \frac{\ell^{2|\mu|-2}}{g_{b}^2 k^2 z_{0}^{2|\mu|}} \bar{\cT}^{\mu\nu} \left[  C_{ww}(k) \cP_{\mu\nu,\rho\sigma}^\parallel - C_{(b)}(k) \cP_{\mu\nu,\rho\sigma}^\perp \right] \cT^{\rho\sigma} + \cr
&~ \qquad \qquad \quad + \frac{2 C_{vv}(k)}{g_{5}^2 k^2 z_{0}^2} \bar{V}_{\mu}^0 (k^2 \eta^{\mu\nu} - k^\mu k^\nu) V_{\nu}^0 + \cr
&~ \qquad \qquad \quad + \frac{2 i \sgn(\mu) \ell^{|\mu|-1}}{g_b g_5 k z_{0}^{|\mu|+1}} \left( C_{wv}(k) + C_{vw}(k) \right) k_\mu \bar{V}_{\nu}^0 \cT^{\mu\nu}  \displaystyle\biggr\}~,
\end{align}
where $C_{\alpha\beta}$ are the components of ${\bf C}_{\rho}$.  This expression is
valid for generic parameter values.  In the special case $\Delta = 3$, the non-normalizable solution in the pion
sector has logarithmic instead of power law behavior for small $z$ and the prefactor in the first term should be
replaced with $(6 - 2\Delta) \to 1$.

\subsubsection{Two-point functions}

This result can be used to compute two-point functions of the QCD quark bilinear operators.  On the one hand,
the sources appear in the generator of QCD correlation functions as
\begin{equation}\label{qcdW}
W = \displaystyle\bigg\langle \exp{ \left( i \int d^4 x \left( V^{0\mu a} O_{\mu}^{V,a} + (A^{0\mu a} + \d^\mu \pi^a) O_{\mu}^{A,a} + \cT^{\mu\nu a} O_{\mu\nu}^{T,a} \right) \right)} \displaystyle\bigg\rangle~,
\end{equation}
where the expectation value is over the QCD path integral.  On the other hand, gauge/gravity duality tells us that $W = Z_{\rm grav}[V^0,A^0,\pi,\cT]$, the gravity partition function with asymptotic boundary conditions given by the sources.

This relationship must be taken with a grain of salt, of course, as we are certainly not working with a well-defined theory of quantum gravity.  We could imagine that our model has an embedding into string theory, and the action \eqref{Snot}, evaluated on shell, gives the leading contribution to $Z_{\rm grav}$ in a $g_s$ and $\alpha'$ expansion, $Z_{\rm grav} = e^{i S} + \cdots$.  These correspond to the $1/N_c$ and low energy expansions in field theory language.  However, the string dual of QCD is not known, and in any event the usefulness of these expansions (in particular the low energy expansion) for QCD is questionable.  Until these issues are improved upon, the approach of AdS/QCD should be understood from a different point of view: we use $Z_{\rm eff} \equiv e^{i S}$ to define the generator of correlators in an effective theory, $W_{\rm eff} = Z_{\rm eff}$, and we see how well this effective theory describes QCD.

The on-shell action \eqref{Sren} is sufficient for determining the two-point functions in our effective theory of QCD.  From the relationship $W_{\rm eff} = e^{i S}$ we schematically have $\langle O_1 O_2 \rangle = - \delta_{\phi_{0}^1} \delta_{\phi_{0}^2} e^{i S} |_{\phi_0 = 0}$, where $\phi_0$ denotes the collection of sources.  The results are conveniently expressed in a set of matrix elements $\Pi$ which are gauge invariant Lorentz scalars, defined in terms of the (momentum space) correlators as follows:
\begin{align}\label{medefs}
\langle O_{\mu}^{A,a}(k) O_{\nu}^{A,b}(-q) \rangle =&~ i (2\pi)^4 \delta^{(4)}(k-q)  \left[ (k^2 \eta_{\mu\nu} - k_\mu k_\nu) \Pi^{A,A}(k) + k_\mu k_\nu \Pi^{\pi,\pi}(k) \right] \delta^{ab}~, \cr
\langle O_{\mu}^{V,a}(k) O_{\nu}^{V,b}(-q) \rangle =&~ i (2\pi)^4 \delta^{(4)}(k-q) (k^2 \eta_{\mu\nu} - k_\mu k_\nu ) \Pi^{V,V}(k) \delta^{ab}~,  \cr
\langle O_{\mu\nu}^{T,a}(k) O_{\rho\sigma}^{T,b}(-q) \rangle  =&~ i (2\pi)^4 \delta^{(4)}(k-q) \left( \cP^{\perp}_{\mu\nu,\rho\sigma} \Pi^{T,T\perp}(k) + \cP^{\parallel}_{\mu\nu,\rho\sigma} \Pi^{T,T\parallel}(k) \right) \delta^{ab}~, \cr
\langle O_{\mu\nu}^{T,a}(k) O_{\lambda}^{V,b}(-q) \rangle =&~  (2\pi)^4 \delta^{(4)}(k-q)  \left( k_{\mu} \eta_{\nu \lambda} - k_\nu \eta_{\mu \lambda} \right) \Pi^{T,V}(k) \delta^{ab}~.
\end{align}
The transverse and longitudinal tensor projectors are defined as
\begin{equation}\label{tensorproj}
(\cP^{\perp})_{\mu\nu}^{\rho\sigma} = k^2 \delta_{[\mu}^{\rho} \delta_{\nu]}^{\sigma} - 2 k_{[\mu} k^{[\rho} \delta_{\nu]}^{\sigma]}~, \qquad (\cP^\parallel)_{\mu\nu}^{\rho\sigma} = 2 k_{[\mu} k^{[\rho} \delta_{\nu]}^{\sigma]}~,
\end{equation}
and satisfy $\cP^2 = k^2 \cP$.  Making use of these definitions, some straightforward computation yields the following set of two-point functions:
\begin{align}\label{2pointers}
& \Pi^{\pi,\pi}(k) = - \frac{m_{q}^2 z_{0}^{6-2\Delta} }{(6 -
2\Delta) k^2 \ell^{6-2\Delta}} C_{(\pi)}(k)~,  &\Pi^{A,A}(k) = -
\frac{2}{g_{5}^2 k^2 z_{0}^2} C_{(a)}(k)~, \cr & \Pi^{T,T \perp} =
\frac{\ell^{2|\mu|-2}}{g_{b}^2 k^2 z_{0}^{2|\mu|}} C_{(b)}(k)~,  &
\Pi^{T,T \parallel} = - \frac{\ell^{2|\mu|-2}}{g_{b}^2 k^2
z_{0}^{2|\mu|}} C_{ww}(k)~, \cr & \Pi^{V,V}(k) = - \frac{2}{g_{5}^2
k^2 z_{0}^2} C_{vv}(k)~, & \Pi^{T,V}(k) = \frac{\sgn(\mu) \ell^{|\mu|-1}}{2 g_b g_5 k z_{0}^{|\mu|+1}} \left(C_{wv}(k) + C_{vw}(k) \right)~.
\end{align}
All of the dynamics is contained in the functions $C(k)$, which must be obtained numerically.  They are given by $C = (P^{\rm n.})^{-1} P^{\rm n.n.}$, evaluated at the IR boundary.  The precise definitions of the normalizable and non-normalizable modes for each sector can be found in Appendix \ref{asymptotic}.

\subsubsection{An interesting one-point function: magnetic susceptibility of the quark condensate}

We can also use \eqref{Sren} to compute one-point functions of operators in the presence of external fields, yielding, for instance, the magnetic susceptibility. 

Let us briefly review how the magnetic susceptibility of the quark condensate is defined \cite{Ioffe:1983ju}.  Consider turning on an external magnetic field---a classical source whose electromagnetic potential we denote $A_{\mu}^{\rm cl}$.  In such a background one expects the tensor current $O_{\mu\nu}^{T,a} =  \bar{q} t^a \sigma_{\mu\nu} q$ to acquire a vev, and the magnetic susceptibility measures the ratio of the tensor current vev to the external field.  More precisely, we have a susceptibility $\chi_i$ for each species of quark $q_i$, defined through
\begin{equation}
\langle \bar{q}_i \sigma_{\mu\nu} q_i \rangle = e_i \chi_i \langle \bar{q}_i q_i \rangle F_{\mu\nu}^{\rm cl}~, \qquad \textrm{(no sum on $i$)}~,
\end{equation}
where $e_i$ is the electric charge of the quark, and $F_{\mu\nu}^{\rm cl} = \d_\mu A_{\nu}^{\rm cl} - \d_\nu A_{\mu}^{\rm cl}$.  To be concrete we will consider the case of the up quark, $e_u = 2e/3$:
\begin{equation}\label{chiupdef}
\langle \bar{u} \sigma_{\mu\nu} u \rangle = \langle O_{\mu\nu}^{T,0} + O_{\mu\nu}^{T,3} \rangle = \frac{2e}{3} \chi_u \langle O^{S,0} + O^{S,3} \rangle F_{\mu\nu}^{\rm cl}~.
\end{equation}

The holographic prescription together with the on-shell action \eqref{Sren} can be used to compute $\chi$.  The electromagnetic current of the quarks
\begin{equation}
J_{\mu}^{\rm EM} = e \bar{q} {\bf Q} \gamma_\mu q~, \qquad {\bf Q} = \left( \begin{array}{c c} 2/3 & 0 \\ 0 & - 1/3 \end{array}\right)~,
\end{equation}
can be expressed in terms of flavor currents $J_{\mu}^{V,a} = \bar{q} t^a \gamma_\mu q$ according to $J_{\mu}^{\rm EM} = e ( \frac{1}{3} J_{\mu}^{V,0} + J_{\mu}^{V,3} )$.  Thus by choosing our vector current sources $V_{\mu}^{0,a}$ to be of the form
\begin{equation}\label{emsources}
V_{\mu}^{0,0} = \frac{e}{3} A_{\mu}^{\rm cl}~, \qquad V_{\mu}^{0,3} = e A_{\mu}^{\rm cl}~, \qquad V_{\mu}^{0,1} = V_{\mu}^{0,2} = 0~,
\end{equation}
we find that the exponent of \eqref{qcdW} contains the term $V_{\mu}^{0,a} \bar{q} t^a \gamma^\mu q = (A^{\rm cl})^\mu J_{\mu}^{\rm EM}$.  This shows that $A_{\mu}^{\rm cl}$ is an electromagnetic background field.  Now we evaluate the tensor one-point functions appearing in \eqref{chiupdef} in the background \eqref{emsources} with all other sources turned off.  To leading order in the electromagnetic field we can use $W = \exp{(i S_{\rm ren}^{(2)})}$, and functionally differentiate with respect to the tensor source.  We find
\begin{equation}
\langle O_{\mu\nu}^{T,a}(k) \rangle = -i \frac{ \delta}{\delta \cT^{\mu\nu,a}(-k)} e^{i S_{\rm ren}^{(2)}} \displaystyle\bigg|_{\cT = A^0 = \pi = 0} = \frac{ \sgn(\mu) }{g_5 \tilde{g}_b k z_{0}^2} \left( C_{wv}(k) + C_{vw}(k) \right) (f_{V})_{\mu\nu}^a~,
\end{equation}
where $\tilde{g}_b = g_b (z_0/\ell)^{|\mu|-1}$ and $(f_{V})_{\mu\nu}^a = -2i k_{[\mu} V_{\nu]}^{0,a}$, leading to
\begin{equation}\label{tensor1pt}
\langle O_{\mu\nu}^{T,0}(k) + O_{\mu\nu}^{T,3}(k) \rangle = \frac{4e \sgn(\mu)}{3 g_5 \tilde{g}_b k z_{0}^2} \left( C_{wv}(k) + C_{vw}(k) \right) F_{\mu\nu}^{\rm cl}~.
\end{equation}
In principle we could study a spacetime-dependent magnetic susceptibility, but in order to keep things simple and compare with the literature later, we will take the $k \to 0$ limit to recover the constant background value.

Notice here that we are working perturbatively in the background field.  To calculate the one-point function in a finite background field one should expand the 5d fields around the constant finite background solution $V_\mu = V_{\mu}^0$, \eqref{emsources}, much  as we expand around the nonzero tachyon vev, \eqref{Xvev}.  This type of calculation was recently carried out in \cite{Gorsky:2012ui} in the context of the model proposed in \cite{Domokos:2011dn}, but with parameters fit to the UV as in \cite{Alvares:2011wb}.  Computing the tensor and scalar one-point functions in this background, the authors of \cite{Gorsky:2012ui}  find a $\chi(A^{\rm cl})$ to all orders in the external field.  We are computing the linear response, $\chi \equiv \chi(0)$.  We will compare with their results as well as other theoretical and lattice results in section \ref{chiresults}.

Since we are computing the linear response to the field it is sufficient to evaluate the scalar vev's on the right-hand side of \eqref{chiupdef} at their leading-order $A_{\mu}^{\rm cl}$-independent value.  This result is given simply by the quark vev:
\begin{equation}\label{scalar1pt}
\langle O^{S,0} \rangle = \half \langle \bar{q} q \rangle = \frac{\sigma}{2} \qquad  \textrm{and} \qquad  \langle O^{S,3} \rangle = 0~.
\end{equation}
(See Appendix \ref{Appendix:Xvev} for details).  Combing the $k \to 0$ limit of \eqref{tensor1pt} with \eqref{scalar1pt} and plugging into \eqref{chiupdef} we find
\begin{equation}\label{chiuptheory}
\chi_u = \frac{4 \sgn(\mu)}{g_5 \tilde{g}_b \sigma z_{0}^2} \cdot \lim_{k \to 0} \frac{1}{k} \left( C_{wv}(k) + C_{vw}(k) \right)~.
\end{equation}
In section \eqref{chiresults} we give numerical estimates for this result based on our best-fit parameter values.

\subsection{Normalizable modes, masses, and decay constants}

The physical quantities we ultimately wish to extract from solving
the linearized equations of motion and from the on-shell action are
of course the masses and decay constants of the resonances. There
are two equivalent procedures for deriving this information: (1)
from the pole structure and residues of the two-point function in
the form given in (\ref{2pointers}), or (2) by writing the two-point
functions explicitly as a sum over momentum space poles whose
residues are given in terms of the tower of normalizable modes
describing the bulk fields (as described in more detail below). It
turns out that the first form makes most apparent such general
properties as the $k^2\rightarrow\infty$ or $k^2 \to 0$ behavior of the
correlator, which the infinite sum form conceals. The second
repackaging, meanwhile, becomes more useful when we are interested
in specific (usually light) resonances and their decay constants,
which, in turn, the first form leaves obscure.

In order to derive the sum-over-poles-type expression, we should
first discuss the towers of normalizable modes.  In each sector define the $L^2$-normalizable Hilbert space of functions,
\begin{equation}\label{L2}
\cH = \left\{ \phi \in L_{w}^2[\varepsilon,z_0]~\big|~ Q[\phi](k,\varepsilon) = P[\phi](k,z_0) = 0 \right\}~.
\end{equation}
One can show that each $O$ is an Hermitian operator on the Hilbert space $\cH$ of a given sector.  In the $\pi, a_1$, and $b_1$ sectors $O$ is furthermore positive definite, while we expect ${\bf O}$ in the
$\rho$ sector to be positive definite for fixed momentum and $g_1$ sufficiently small, though we do not prove it.

\begin{figure}
\begin{center}
\includegraphics{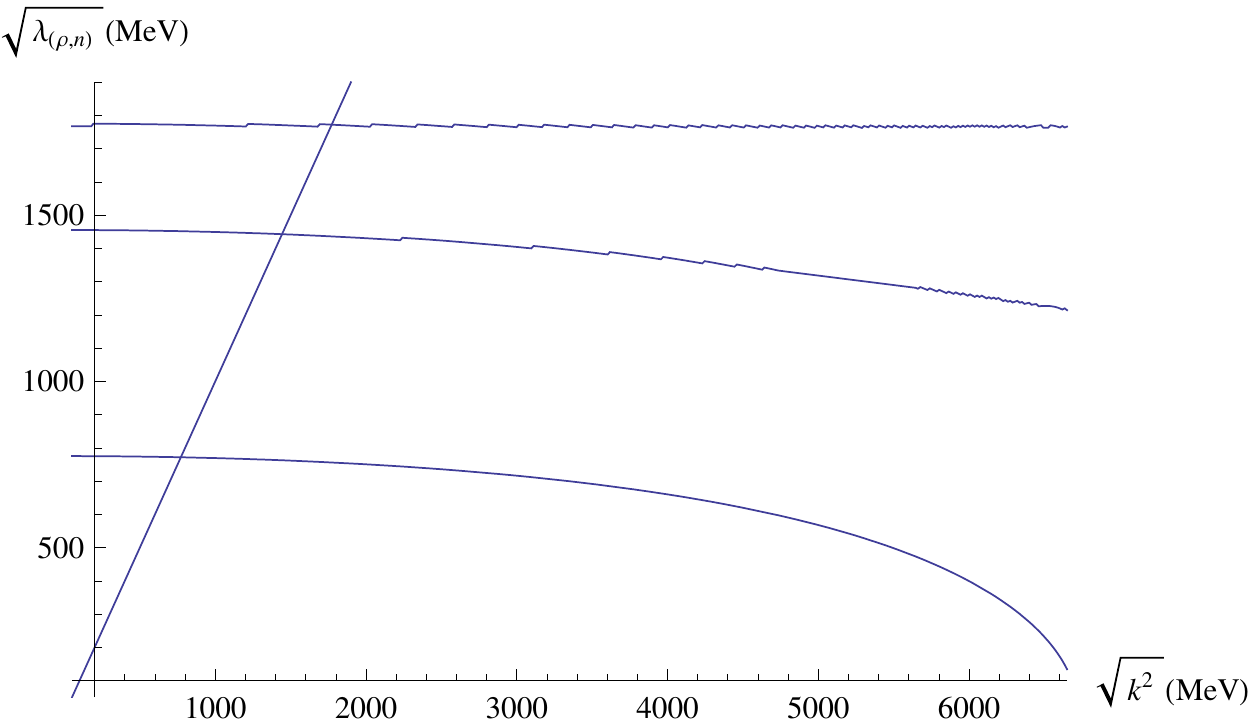}
\caption{The first three eigenvalues of ${\bf O}$ in the $\rho$ sector, as a function of momentum for the best fit parameter values.  (See Table \ref{table:fitparams} below.)  The intercepts with the diagonal give the tower of $\rho$ masses.}
\label{Fig:rhoevalues}
\end{center}
\end{figure}

Given the form of \eqref{Snot2} it is clear that the eigenvalues of the various $O$'s should be interpreted as the mass-squared of normalizable modes.  The $O$'s of the $\pi, a_1$, and $b_1$ sectors are 
independent of the momentum $k^2$, so we get a standard dispersion relation for the 4d states, $k^2 = m_m^2$, where $m=0,1,2,\dots$ labels the eigenvalues.  In the $\rho$ sector we see 
that ${\bf O}$ depends on $k$.  It can be shown that the eigenvalues depend only on the momentum-squared, and thus we get a nontrivial dispersion relation $k^2 = \lambda_{m}(k^2)$.  The $\rho$ masses are the solutions to this equation; we denote them by $k^2 = m_{m}^2$.  In Figure \ref{Fig:rhoevalues} we plot the momentum dependence of the first three eigenvalues $\lambda_{m}$.  The $\rho$ masses are given by the intersection points of the curves with the diagonal.  Notice that when the momentum gets large enough the eigenvalues become negative; the momentum-dependent off-diagonal terms in ${\bf O}$, \eqref{Orho}, are competing with the diagonal terms.  We work with parameter values such that the momentum scale at which this occurs is well beyond the scale where we expect our description to be valid.  Near each mass eigenvalue there is an expansion
\begin{equation}\label{rhoresexp}
\lambda_{m}(k^2) = m_{m}^2 + \frac{d\lambda_{m}}{d(k^2)} \displaystyle\bigg|_{m_m^2} \cdot(k^2 - m_{m}^2) + \cO\left( (k^2 - m_{m}^2)^2 \right)~.
\end{equation}

We have an orthogonal basis of eigenfunctions for the Hilbert space of each sector,  $\cH$, satisfying
\begin{align}\label{eigenfns}
& O \phi_{m}(z) = m_m^2 \phi_m(z)\quad \text{for the}~ \pi, a_1, b_1~\text{sectors} \cr
&  {\bf O} \phi_{m}(k,z) = \lambda_{m}(k^2) \phi_{m}(k,z)\quad\text{for the $\rho$ sector}
\end{align}
and normalized such that
\begin{align}\label{normcon}
& \int_{\varepsilon}^{z_0} dz w \phi_{m} \phi_{l} = \delta_{ml}\quad\text{for the}~ \pi, a_1, b_1~\text{sectors} \cr
&  \int_{\varepsilon}^{z_0} dz \phi_{m}^T {\bf W}_{(\rho)} \phi_{l} = \delta_{ml}~\quad\text{for the $\rho$ sector.}
\end{align}
We have chosen our basis functions to be real-valued for convenience.  Note that $\phi_{m}$ in the $\rho$ sector is a two-component column vector of functions.  When we want to specify its components we write $\phi_{n}^{\alpha}$ where $\alpha,\beta$ run over $w,v$ as usual.  The bulk Green functions for the operators $(k^2 - O)$ can be expressed in terms of the normalized eigenfunctions as
\begin{align}\label{Greenfns}
& G(k,z;z') = \sum_{m} \frac{ \phi_m(z) \phi_{l}(z') }{k^2 - m_{m}^2}~, \quad \text{for the $\pi, a_1, b_1$ sectors} \cr
&  {\bf G}_{\alpha\beta}(k,z;z') = \sum_{m} \frac{\phi_{m}^{\alpha}(k,z) \phi_{l}^{\beta}(k,z') }{k^2 - \lambda_{l}(k^2) } ~\quad\text{for the $\rho$ sector}.
\end{align}
They satisfy, in each sector,
\begin{align}\label{Greendef}
& (k^2 - O) G = w^{-1} \delta(z - z')\quad\text{for the $\pi, a_1, b_1$ sectors} \cr
&   (k^2 - {\bf O}) {\bf G} = {\bf W}^{-1} \delta(z-z')~\quad\text{for the $\rho$ sector.}
\end{align}

We can use these bulk Green functions to obtain convenient expressions for the bulk-to-boundary propagators by generalizing the argument in \cite{Hong:2004sa}.  First observe that the following Green's identities hold, as can be explicitly verified using \eqref{Opia}--\eqref{Orho}, \eqref{Qs}, \eqref{Ps}.  Let $\psi(z),\chi(z)$ be arbitrary functions on $[\varepsilon,z_0]$.  Then
\begin{align}\label{Greenid1}
& \int_{\varepsilon}^{z_0} dz w \left\{\chi (k^2 - O ) \psi - \psi (k^2 - O) \chi \right\} = \cr
& \qquad \qquad  = - z_{0}^{-2} \left( Q[\chi]  P[\psi] - Q[\psi]  P[\chi] \right) \displaystyle\bigg|_{\varepsilon}^{z_0}~, \quad\text{for $\pi,a_1,b_1$.}
\end{align}
For the $\rho$ sector let $\chi_{\alpha}(z), \psi_{\alpha\beta}(z)$ be the components of a two-vector and matrix-valued function.  Then we have the Green's identity
\begin{align}\label{Greenid2}
& \int_{\varepsilon}^{z_0} dz \left\{ \chi_{\alpha} \left( {\bf W} (k^2 - {\bf O}) \right)_{\alpha\gamma} \psi_{\gamma\beta} - \psi_{\gamma\beta}  \left( {\bf W} (k^2 - {\bf O}) \right)_{\gamma\alpha} \chi_{\alpha} \right\} = \cr
& \qquad \qquad =  - z_{0}^{-2} \left(  Q_{\alpha}[\chi] (P_{\alpha}[\psi] )_\beta - (Q_{\alpha}[\psi] )_\beta P_{\alpha}[\chi] \right) \displaystyle\bigg|_{\varepsilon}^{z_0}~.
\end{align}
Here the notation $(P_{(\rho;\alpha)}[\psi] )_\beta$ means that $P$ acts on the first or second column of the matrix $\psi$, with the result indexed by $\beta$.  Now employ these identities as follows.  Take $\chi = \Phi$ to be the bulk-to-boundary propagator, and $\psi = G$ to be the Green function.  On the left-hand sides the second terms vanish because the bulk-to-boundary propagators are solutions to the equations of motion for arbitrary $k^2$, while the first terms integrate to give $\Phi(z')$, using \eqref{Greendef}.  On the right-hand sides both terms vanish at the IR boundary because both $\Phi,G$ satisfy Neumann conditions there, while only the $Q[\Phi] \cdot P[G]$ terms survive on the UV boundary since $Q[G] = 0$ there.  Noting that $Q[\Phi]$ is the source $S$, we find the desired relations:
\begin{align}\label{btbG}
& \Phi(k,z) =  \sum_{m} \frac{ S(k) P[\phi_{m}](\varepsilon) }{z_{0}^2 (k^2 - m_{m}^2) } \phi_{m}(z) \quad\text{for the $\pi,a_1,b_1$ sectors}, \cr
& \Phi(k,z) =  \sum_{m} \frac{ S(k) \cdot P[\phi_{m}] (k,\varepsilon) }{ z_{0}^2(k^2 - \lambda_m(k^2))} \phi_m(k,z)\quad\text{for the $\rho$ sector.} 
\end{align}

With these expressions the boundary action, \eqref{Sbndry}, takes the simple form
\begin{align}\label{Sbndrymodeexp}
 S_{\rm bndry} =&~ - \sum_{ \{ \pi,a_1, b_1 \} } \frac{1}{z_{0}^4} \int \frac{d^4 k}{(2\pi)^4} \tr |S(k)|^2 \sum_{m}  \frac{ \left( P[\phi_{m}](\varepsilon) \right)^2 }{ k^2 - m_{m}^2}  + \cr
&~ -  \frac{1}{z_{0}^4} \int \frac{d^4 k}{(2\pi)^4} \tr  \sum_{m}  \frac{ \left| P[\phi_{m}](k,\varepsilon) \cdot S(k) \right|^2 }{ k^2 - \lambda_{m}(k^2)}\displaystyle\bigg|_{\rho~\text{sector}}~.
\end{align}
This is the sum-over-modes form of the on-shell action, valid at finite $\varepsilon$.  Our next task is to determine the renormalized action in the $\varepsilon \to 0$ limit.  In particular, we want to determine how the $P[\phi]$ for each sector are related to the normalizable modes, $f^{\rm n.}$, which we can compute numerically.

Let's begin with the $\pi$, $a_1$, and $b_1$ sectors.  In each of these sectors, the $\phi_{m}$ are eigenfunctions of $O$  and thus solve the same equation as the bulk-to-boundary propagators, but with $k^2 \to m_{m}^2$.  The general solution is a linear combination of normalizable and non-normalizable modes, $f^{\rm n.}(m_m,z)$ and $f^{\rm n.n.}(m_m,z)$.  Thus UV boundary condition $Q[\phi_m] = 0$ fixes the relative normalization such that
\begin{equation}\label{eigenmodesol}
\phi (m_m,z) = N \left( f^{\rm n.}(m_m,z) - \frac{Q^{\rm n.}(m_m,\varepsilon)}{Q^{\rm n.n.}(m_m,\varepsilon)} f^{\rm n.n.}(m_m,z) \right)~.
\end{equation}
Imposing the IR condition determines the eigenvalues, while $N$ in each case is fixed by the normalization condition \eqref{normcon}.  As $\varepsilon \to 0$, one finds from the asymtotics in Appendix \ref{asymptotic} that the second term vanishes relative to the first for any finite $z$.  Hence, in the limit, the eigenvalues are given by
\begin{equation}\label{evalues}
P[f^{\rm n.}](m_m,z_0) \equiv P^{\rm n.}(m_m,z_0) = 0 \quad \text{for mass eigenvalues $m_m$}~.
 \end{equation}  
This indeed corresponds to the poles of the $C(k)$, \eqref{btbs}, that parameterize the two-point functions \eqref{2pointers}.  Meanwhile the normalization constants are given by
\begin{equation}\label{normconst}
N_{m}^{-2} = \int_{0}^{z_0} w f^{\rm n.}(m_m,z)^2 + \cO(\varepsilon^{>0})~.
\end{equation}
Using the explicit aysmptotics of the $f$'s we can also compute
\begin{align}\label{Pofmodes}
& P_{(\pi)}[\phi_{m}^{(\pi)}](\varepsilon) = N_{(\pi,m)} \left(1 + \cO(\varepsilon^{>0}) \right)~, \quad P_{(a)}[\phi_{m}^{(a)}](\varepsilon) = 2 N_{(a,m)} \left(1 + \cO(\varepsilon^{>0}) \right)~, \cr
& P_{(b)}[\phi_{m}^{(b)}](\varepsilon) = 2 N_{(b,m)} \frac{\varepsilon^{|\mu|}}{z_{0}^{|\mu|}} \left(1 + \cO(\varepsilon^{>0}) \right)~.
\end{align}

The $\rho$ sector requires more care due to the momentum dependence of the eigenvalues.  Eigenfunctions are solutions to the equation $(\lambda - {\bf O}) \phi = 0$.  This equation is slightly different than the equation satisfied by the bulk-to-boundary propagator since $\lambda$ must be treated independently of the momentum $k^2$ appearing in the $\rho$ sector operator ${\bf O}$.  Nevertheless it is easy to show that the leading asymptotics of the four linearly independent solutions are the same so we will continue to use the same notation introduced around \eqref{rhobcs}.  The general solution is $R = \sum_i c^i R^i$, and in this case the UV boundary conditions $Q[R] = 0$ may be written as
\begin{equation}
{\bf Q}^{\rm n.n.}(\lambda,k,\varepsilon) c^{\rm n.n.} + {\bf Q}^{\rm n.}(\lambda,k,\varepsilon) c^{\rm n.} = 0~.
\end{equation}
Using the asymptotics of the ${\bf Q}$ matrices in Appendix \ref{asymptotic}, we find that the $c^{\rm n.n.}$ are linear combinations of the $c^{\rm n.}$ with coefficients that go as positive powers of $\varepsilon$.  Hence for finite $z$ we can neglect the contribution of the non-normalizable modes to the eigenfunctions $\phi^{(\rho)}$.  In particular the IR boundary conditions which determine the eigenvalues become
\begin{equation}\label{rhoevalues}
{\bf P}^{\rm n.}(\lambda,k,z_0) c^{\rm n.} = 0 \quad \Rightarrow \quad \det  {\bf P}^{\rm n.}(\lambda,k,z_0) = 0 \quad \Rightarrow \quad \lambda \in \{ \lambda_{m}(k^2) \}~.
\end{equation}
If we go on shell by setting $\lambda_{m}(k^2) = k^2$ then \eqref{rhoevalues} corresponds to the poles of ${\bf C}_{(\rho)}(k)$, \eqref{rhobcssolve}.  

Notice that for each $\lambda_{m}$ in this sector, the first equation of \eqref{rhoevalues} determines the ratio of the normalizable mode coefficients appearing in $\phi_{m}$.  Let us denote 
these coefficients $c_{m}^{\rm n.} = N_{m} = (N_{(w,m)},N_{(v,m)})^T$ so that we have
\begin{equation}
 {\bf P}^{\rm n.}\left(\lambda_{m}(k^2),k,z_0\right) \cdot N_{m}(k) = 0~.
 \end{equation}
This equation does not determine the $N$'s outright, since the rank of ${\bf P}$ is one.  The overall scale is fixed by the normalization condition which, as $\varepsilon \to 0$, becomes
\begin{align}\label{rhonorm}
1 =&~ \int_{0}^{z_0} dz ( \phi_{m} )^T {\bf W}_{(\rho)} \phi_{m} \cr
=&~  \int_{0}^{z_0} dz \left\{ w_{(w)} \left( N_{(w,m)} f_{(w,m)}^{\rm n.} + N_{(v,m)} f_{(v,m)}^{\rm n.} \right)^2 +  w_{(v)} \left( N_{(w,m)} g_{(w,m)}^{\rm n.} + N_{(v,m)} g_{(v,m)}^{\rm n.} \right)^2 \right\}~, \cr
\end{align}
where in this sector $R^{{\rm n.}(w)}\left( \lambda_{m}(k^2),k,z \right) = (f_{(w,m)}^{\rm n.},g_{(w,m)}^{\rm n.})^T$, etc., and terms that vanish as $\varepsilon \to 0$ have been neglected.  The $N$'s, like the eigenvalues, depend on the momentum.  In order to determine the on-shell action in the $\rho$ sector we also need
\begin{align}\label{Pofrhomodes}
 P[\phi_{m}](k,\varepsilon) \cdot S(k) =&~ S(k)^T \cdot {\bf P}^{\rm n.} \left(\lambda_{m}(k^2),k, \varepsilon \right) \cdot N_{m} \cr
\to&~  \frac{2 i \sgn(\mu) \ell^{|\mu|-1}}{g_b k z_{0}^{|\mu|-1}} N_{(w,m)} k^\nu \cT_{\nu\mu}(k) + \frac{2}{g_{5}} N_{(v,m)} V_{\mu}^0(k)~,
\end{align}
as $\varepsilon \to 0$.

We can now collect results and evaluate the on-shell action, \eqref{Sbndrymodeexp}.  We find that the leading terms as $\varepsilon \to 0$ are finite; naively there are no divergences.  One may wonder how this is consistent with the analysis in section \ref{Section:nonnormmodes}, where renormalization was necessary.  However in this section we have been analyzing the limit mode by mode, tacitly assuming that it commutes with the sum over modes.  This is clearly false, taking into consideration \eqref{btbG}: this expresses something that finite in the limit, the bulk-to-boundary propagator, as an infinite sum of terms that vanish in the limit.  In the following we assume that commuting the limit and the sum is equivalent to passing to the renormalized action, at least for computing the physical quantities of interest---masses and decay constants.  We have checked numerically that the formalism of the previous section and this one give the same results for these quantities.  Thus we find
\begin{align}\label{Srenmode}
& \lim_{\varepsilon \to 0} S_{\rm ren}^{(2)} = - \frac{1}{z_{0}^4} \int \frac{d^4 k}{(2\pi)^4} \tr \displaystyle\biggl\{ k^2 \bar{\pi} \pi \sum_m \frac{ N_{(\pi,m)}^2}{k^2 - m_{(\pi,m)}^2} +  \frac{4}{g_{5}^2 k^2} \bar{A}_{\mu}^0 \cP_{\perp}^{\mu\nu} A_{\nu}^0 \sum_{m} \frac{ N_{(a,m)}^2}{k^2 - m_{(a_1,m)}^2} + \cr
& \quad + \frac{2 \ell^{2|\mu|-2}}{g_{b}^2 k^2 z_{0}^{2|\mu|-2}} \sum_m \bar{\cT}^{\mu\nu} \left[ \cP_{\mu\nu,\rho\sigma}^{\parallel} \frac{N_{(w,m)}^2}{k^2 - \lambda_{(\rho,m)}(k^2)}  - \cP_{\mu\nu,\rho\sigma}^{\perp} \frac{N_{(b,m)}^2}{k^2 - m_{(b_1,m)}^2}  \right] \cT^{\rho\sigma} + \cr
& \quad + \frac{4}{g_{5}^2 k^2} \bar{V}_{\mu}^0 \cP_{\perp}^{\mu\nu} V_{\nu}^0 \sum_{m} \frac{ N_{(v,m)}^2}{k^2 - \lambda_{(\rho,m)}(k^2)} + \frac{8 i \sgn(\mu) \ell^{|\mu|-1}}{g_5 g_b k z_{0}^{|\mu|-1}} k_\mu \bar{V}_{\nu}^0 \cT^{\mu\nu} \sum_m  \frac{ N_{(v,m)} N_{(w,m)}}{k^2 - \lambda_{(\rho,m)}(k^2)} \displaystyle\biggr\}~,
\end{align}
where we have introduced the additional notation $\cP_{\perp}^{\mu\nu} = k^2 \eta^{\mu\nu} - k^\mu k^\nu$.

Varying with respect to the sources, applying the holographic prescription, and taking into account the matrix element definitions \eqref{medefs} leads to
\begin{align}\label{2pointers2}
& \Pi^{\pi,\pi} = - \frac{1}{z_{0}^4} \sum_m \frac{N_{(\pi,m)}^2}{k^2 - m_{(\pi,m)}^2}~,  &\Pi^{A,A} = 
\frac{4}{g_{5}^2 k^2 z_{0}^4} \sum_m \frac{N_{(a,m)}^2}{k^2 - m_{(a_1,m)}^2}~, \cr 
& \Pi^{T,T \perp} = -\frac{2 \ell^{2|\mu|-2}}{g_{b}^2 k^2 z_{0}^{2|\mu|+2}} \sum_m \frac{ N_{(b,m)}^2}{k^2 - m_{(b_1,m)}^2}~,  &  \Pi^{T,T \parallel} = \frac{2 \ell^{2|\mu|-2}}{g_{b}^2 k^2 z_{0}^{2|\mu|+2}} \sum_n \frac{ N_{(w,m)}^2}{k^2 - \lambda_{(\rho,m)}(k^2)}~, \cr 
& \Pi^{V,V} = \frac{4}{g_{5}^2 k^2 z_{0}^4} \sum_m \frac{ N_{(v,m)}^2}{k^2 - \lambda_{(\rho,m)}^2}~, & \Pi^{T,V} = -\frac{2 \sgn(\mu) \ell^{|\mu|-1}}{g_b g_5 k z_{0}^{|\mu|+3}} \sum_m  \frac{ N_{(v,m)} N_{(w,m)}}{k^2 - \lambda_{(\rho,m)}(k^2)}~.
\end{align}
These are essentially the sum-over-poles expressions for the matrix elements that we can compare to field theory templates to read off decay constants.  See for example \cite{Domokos:2011dn}.  (To compute residues for the $\Pi$ involving $\rho$ spectra one still needs to use the expansion \eqref{rhoresexp}.  Also the momentum-dependent $N_{(\rho)}$ should be evaluated at the pole.)  We find
\begin{align}\label{decayconstants}
&f_{\pi}^{(m)} = \frac{N_{(\pi,m)}}{m_{(\pi,m)} z_{0}^2}~, \qquad f_{a_1}^{m} = \frac{2 N_{(a,m)}}{g_5 m_{(a_1,m)} z_{0}^2}~, \qquad f_{b_1}^{m} = \frac{\sqrt{2} N_{(b,m)}}{g_b m_{(b_1,m)} z_{0}^2} (\ell/z_0)^{|\mu|-1}~, \cr
& f_{\rho}^{Vm} = \frac{2 N_{(v,m)}(m_{(\rho,m)}) }{g_5 m_{(\rho,m)} z_{0}^2 \sqrt{ 1 - \lambda_{(\rho,m)}'(m_{(\rho,m)}^2) }}~, \cr
& f_{\rho}^{Tm} =  \frac{\sqrt{2} N_{(w,m)}(m_{(\rho,m)}) }{g_b m_{(\rho,m)} z_{0}^2 \sqrt{ 1 - \lambda_{(\rho,m)}'(m_{(\rho,m)}^2) }} (\ell/z_0)^{|\mu|-1}~.
\end{align}
where the prime indicates differentiation with respect to $k^2$.
Let us summarize.  Expressions \eqref{decayconstants}, together with \eqref{evalues} and \eqref{rhoevalues}, constitute our final analytical results for masses and decay constants.  We have boiled them down to quantities that involve simple manipulations with the normalizable-mode wavefunctions, $f^{\rm n.}$.  Masses are obtained by imposing the generalized Neumann condition at the IR boundary, $P[f^{\rm n.}](z_0) = 0$.  Decay constants are expressed in terms of the normalization constants, $N = || f^{\rm n.} ||^{-1}$, where the norm is the natural one on the Hilbert space $\cH$, \eqref{L2}, (for $\varepsilon \to 0$).  The normalizable-mode wavefunctions in turn are easily computed numerically.

\subsection{Three-point couplings}

Having exhausted the quadratic-order Lagrangian, we now predict
three-point couplings that are either new to the hard wall model
(such as the one mediating $b_1\rightarrow \omega\pi$), or modified
due to the mixing between the $b_{MN}$ sector and the vectors
($g_{\rho\pi\pi}$). For the purposes of comparing to experimental
data, we are only interested in tree-level couplings of the
lowest-lying states. 

States in field theory are dual to normalizable modes in the bulk.  We associate to each state $|s_n\rangle$ a 4d field $s_n(k)$, and we identify these fields with Fourier coefficients of corresponding bulk fields expanded in the eigenfunction basis: $\phi^{(s)}(k,z) = \sum_n s_n(k) \phi_{n}^{(s)}(z)$.  The holographic 4d effective action for the fields $s_n(k)$ is obtained by plugging these mode expansions into the bulk action and KK-reducing along $z$.  The mode expansion for each bulk field is summarized in Appendix \ref{app:modeexpansions}.  By design, these expansions diagonalize the quadratic part of our model:
\begin{align}\label{S2modes}
S^{(2)} =&~ \sum_n \int \frac{d^4 k}{(2\pi)^4} \tr \displaystyle\biggl\{ \bar{\pi}^n (k^2 - m_{(\pi,n)}^2) \pi^n - \bar{a}_{\mu}^n (k^2 - m_{(a,n)}^2) a^{n,\mu} + \cr
& \qquad \qquad \qquad \qquad - \bar{b}_{\mu}^n (k^2 - m_{(b,n)}^2) b^{n,\mu} - \bar{\tilde{\rho}}_{\mu}^n (k^2 - \lambda_{(\rho,n)}(k^2)) \tilde{\rho}^{n,\mu} \displaystyle\biggr\}~.
\end{align}
Due to the nontrivial dispersion relation in the $\rho$ sector, the canonically normalized fields are
\begin{equation}\label{rhocannon}
\rho_{\mu}^n(k) =  \left[1 - \lambda_{(\rho,n)}'(m_{(\rho,n)}^2) \right]^{1/2} \tilde{\rho}_{\mu}^n(k)~.
\end{equation}
We now turn to the computation of some three-point couplings.

\subsubsection{$g_{b\omega\pi}$}

Let us begin with the vertex that generates $b_1\rightarrow
\omega\pi$ (or equivalently $h_1\rightarrow\rho\pi$).  $b_1$ modes are contained in 
$b_{MN}$ and $\pi$ modes in $A_M$, while $\rho$ modes can be 
either in $b_{MN}$ or $V_M$.  Thus we are looking
for terms in the bulk action that contain either two factors of
$b_{MN}$ and one $A_M$, or terms that contain one each of $b_{MN}$, $A_M$, and $V_M$. 
In principle, $S_{\rm sd}$, $S_{g_1}$, and $S_{g_3}$
can all contribute, but it turns out that $S_{g_1}$ only contains three
point couplings of the form $b$-$V$-$V$ or $b$-$A$-$A$. Hence the
entire coupling comes from the  gauge covariant derivative in the
$b_{MN}$ kinetic term, or from the $g_3$ interaction term. The relevant contribution from the kinetic term is
\begin{align}\label{Ssd3pt1}
S_{\rm sd} \supset &~ - \frac{i  \sgn(\mu)}{2 g_{b}^2 \ell} \int
\tr \left\{ \bar{b} \wedge Db - b \wedge \overline{Db} \right\} \cr
\supset &~ - \frac{ \sgn(\mu) }{4 g_{b}^2 \ell} \int d^5 x
\epsilon^{MNPQR} \tr \left\{ (b_{MN} \bar{b}_{PQ} + \bar{b}_{MN}
b_{PQ} ) A_{R} \right\}~,
\end{align}
while the relevant portion of the $g_3$ interaction term is
\begin{align}
S_{g_3}&\supset -2g_3\int d^5x\sqrt{g}\tr\left\{  \Im\left[b^P_{\phantom{P}M} \bb^M_{\phantom{M}N}(\d^N A_P-\d_P A^N)\right]\right\}~.
\end{align}
We must take $A_R \to A_{R}^a t^a$ with $t^a \in SU(N_f) \subset U(N_f)$ to
guarantee that we get a pion mode. Meanwhile, one of the $b$ factors
should live in $SU(N_f)$ (to give the $b_1$) and one in $U(1)$ (to
give the $\omega$).

Inserting the mode expansions (see Appendix \ref{app:modeexpansions}), we eventually obtain after a straightforward but tedious calculation the following three-point couplings among excitations in the $b$, $\omega$, and $\pi$ (with momenta $k$,
$p$, and $q$, respectively):
\begin{align}\label{Shrhopi}
S_{b\omega\pi} =&~ \frac{2 z_0}{\sqrt{2 N_f}} \int d^8 Q
\displaystyle\bigg[ \left( a_{mnr} k \cdot p + b_{mnr} p \cdot q +
c_{mnr} q \cdot k \right) b^{a,m}(k) \cdot \omega^{n}(p) +  \cr &
\qquad \qquad \qquad \quad   - (a_{mnr} - b_{mnr} - c_{mnr}) q \cdot
b^{a,m}(k) q \cdot \omega^n(p) \displaystyle\biggr] \pi^{a,r}(q)~.
\end{align}
Here $d^8 Q = (2\pi)^{-8} \delta^{(4)}(k+p+q) d^4 k \cdot d^4 p \cdot d^4 q$ is the measure obtained from Fourier transforming a product of three fields to momentum space, and we have used the notation $\omega$ for the $U(1)$ component of the $\rho$ field, \eqref{rhocannon}.  The dimensionless couplings $a,b,c$, are given by
\begin{equation}\label{renormabc}
(a_{mnr},b_{mnr},c_{mnr}) = \frac{1}{\sqrt{1 - \lambda_{(\rho,n)}'( m_{(\rho,n)}^2) }} \cdot (\tilde{a}_{mnr},
\tilde{b}_{mnr}, \tilde{c}_{mnr})~,
\end{equation}
with
\begin{align}\label{abc}
&  \tilde{a}_{mnr} = \frac{m_{(\pi,r)}}{m_{(b_1,m)} m_{(\omega,n)} z_0}
\int_{0}^{z_0} dz w_{(\pi)}\phi_{r}^{(\pi)} \displaystyle\biggl\{ z_{0}^2 \left( w_{(w)} \phi_{m}^{(b_1)} \d_z \phi_{n}^{(w)} - w_{(b_1)} \phi_{n}^{(w)} \d_z \phi_{m}^{(b_1)}\right) + \cr
& \qquad \qquad \qquad \qquad  - \frac{\sgn(\mu) m_{(\omega,n)} z_0  u_1}{|\mu|
- u_2} \phi_{m}^{(b_1)} \phi_{n}^{(v)} \displaystyle\biggr\} ~, \cr 
&  \tilde{b}_{mnr} = \frac{m_{(b_1,m)}}{m_{(\omega,n)} m_{(\pi,r)} z_0} \int_{0}^{z_0}
dz w_{(b_1)} \phi_{m}^{(b_1)} \displaystyle\biggl\{ z_{0}^2 w_{(\pi)} \phi_{n}^{(w)} \d_z \phi_{r}^{(\pi)} + \cr 
& \qquad \qquad \qquad \qquad  +  g_3g_b^2g_5^2  z w_{(w)} \phi^{(\pi)}_r \left( z\d_z\phi_n^{(w)}-\sgn(\mu) m_{(\omega,n)} z_0u_1\phi_n^{(v)} \right) \displaystyle\biggr\}~,  \cr
& \tilde{c}_{mnr} =
- \frac{m_{(\omega,n)}}{m_{(\pi,r)} m_{(b_1,m)} z_0} \int_{0}^{z_0} dz \displaystyle\biggl\{ w_{(\pi)}  \phi_{m}^{(b_1)} z_0 \d_z \phi_{r}^{(\pi) }  \left[z_0 w_{(w)} \phi_{n}^{(w)} +  \frac{\sgn(\mu) u_1}{m_{(\omega,n)} (|\mu| - u_2)} \d_z \phi_{n}^{(v)} \right] +  \quad \cr
& \qquad \qquad \qquad \qquad   +g_3g_bg_5^2  z w_{(b_1)}w_{(w)}\phi_r^{(\pi)}z\d_z\phi_m^{(b_1)}\left(\phi_n^{(w)} +\frac{\sgn(\mu) u_1}{zm_{(\omega,n)}} z_0 \d_z\phi_n^{(v)} \right) \displaystyle\biggr\}~.
\end{align}
Note that since our framework is insensitive to isospin,
$S_{h\rho\pi}$ is identical to the above with $b^a \to h$, $\omega
\to \rho^a$. 
The first subscript in each of the
above coefficients refers to the $m$th state in the $b_1$ tower, the
second to the $n$th state in the $\omega$ tower, and the third to
the $r$th state in the $\pi$ tower. We are primarily interested in
the first (lightest) state in each tower, but in theory this
framework allows us to compute all of them.

With this 4d action in hand, we now compute the decay rate
$b_1\rightarrow\omega\pi$, as well as the $D/S$ ratio, which
characterizes the angular dependence of the amplitude. We consider
the lowest mode in each tower, defining $a \equiv a_{111}$, and
similarly for $b,c$.

\subsubsection{$b_1$ decay rate}

The matrix element for the decay process, in terms of a basis of orthogonal, transverse
polarizations for the massive vector-like modes, becomes
\begin{align}\label{Mhdecay2}
& \cM^{ij} =  \delta^{ab}  \varepsilon_{(b)}^{(i)\mu}({\bf k})
\left[ A \eta_{\mu\nu} - B \frac{q_\mu q_\nu}{m_{\pi}^2} \right]
\varepsilon_{(\omega)}^{(j)\nu}({\bf p})^\ast~, \quad \textrm{with}
\cr & A = \frac{2 z_0}{\sqrt{2 N_f}} \left(-a k \cdot p + b p \cdot
q - c q \cdot k \right)~, \qquad B =  \frac{2 z_0}{\sqrt{2 N_f}}
m_{\pi}^2 (a - b - c)~.
\end{align}
The indices $i,j=0, \pm$, label all possible 
polarizations for the $b_1$ and for the $\omega$.  From here we find
the $b_1$ decay rate via standard techniques.  The absolute square, averaged over incoming $b_1$ polarizations and isospins and summed over outgoing $\omega$ polarizations and $\pi$ isospins, is found to be
\begin{align}\label{Mspinav}
|\cM|_{av}^2 =&~  \frac{1}{3} \displaystyle\biggl[ A^2 \left( 2 + \frac{(p \cdot k)^2}{m_{b_1}^2 m_{\omega}^2} \right)  + B^2 \left( 1 - \frac{(k \cdot q)^2}{m_{b_1}^2 m_{\pi}^2} - \frac{(p \cdot q)^2}{m_{\omega}^2 m_{\pi}^2} + \frac{(k \cdot q)^2 (p \cdot q)^2}{m_{b_1}^2 m_{\omega}^2 m_{\pi}^4} \right) + \cr
&  \qquad \qquad - 2 A B \left( 1 - \frac{(k \cdot q)^2}{m_{b_1}^2 m_{\pi}^2} - \frac{(p \cdot q)^2}{m_{\omega}^2 m_{\pi}^2} + \frac{(k \cdot q)(p \cdot q)(k \cdot p)}{m_{b_1}^2 m_{\omega}^2 m_{\pi}^2} \right)  \displaystyle\biggr]~.
\end{align}

Working in the lab frame, we consider momenta of the
form
\begin{equation}\label{4momenta}
k_\mu = (m_{b_1}, {\bf 0})~, \qquad p_\mu = (\sqrt{m_{\omega}^2 + {\bf
q}^2}, -{\bf q})~, \qquad q_\mu = (\sqrt{ m_{\pi}^2 + {\bf q}^2},
{\bf q} )~.
\end{equation}
The decay rate is
\begin{align}\label{decayrate2}
\Gamma_{b_1 \to \omega\pi} =&~ \frac{1}{8\pi m_{b_1}} \int_{0}^\infty
d|{\bf q}| \frac{ |{\bf q}|^2 \delta\left( m_{b_1} - \sqrt{m_{\omega}^2 +
|{\bf q}|^2} - \sqrt{m_{\pi}^2 + |{\bf q}|^2} \right)
}{\sqrt{m_{\omega}^2 + |{\bf q}|^2} \sqrt{m_{\pi}^2 + |{\bf q}|^2}}
|\cM(|{\bf q}|) |_{av}^2 \cr =&~ \frac{q_\ast}{8\pi m_{b_1}^2}
|\cM(|{\bf q}| = q_\ast) |_{av}^2~,
\end{align}
where $q_\ast$ is the root of the argument of the $\delta$-function,
\begin{align}\label{qstar}
q_{\ast}^2 =&~ \frac{1}{4 m_{b_1}^2} (m_{b_1}^2 - m_{\omega}^2 - m_{\pi}^2
)^2 - \frac{m_{\omega}^2 m_{\pi}^2}{m_{b_1}^2}~.
\end{align}
The decay rate for the $h_1$ is identical except that we don't need to average over initial state flavors; thus $\Gamma_{h_1 \to \rho\pi} = (N_{f}^2 -1) \Gamma_{b_1 \to \omega\pi}$.

\subsubsection{D/S ratio}

In addition to the polarization-averaged decay rate, we can predict
the angular profile of the final state. This information is
encapsulated in the $D/S$ ratio, or the amplitude ratio for $D$-wave
versus $S$-wave decays.  The ratio is defined from the amplitudes
via some basic quantum mechanics. The initial $b_1$ state has spin
1, and no angular momentum. The final state must thus have $J =1$.
Since the $\pi$ has $S=0$ and the $\omega$ has $S=1$, they may have
some relative angular momentum in the final state, either $L = 2$ or $L = 0$.  Parity forbids $L = 1$.
 The relation between the initial and final states is given by the Clebsch-Gordan
coefficients. For angular momentum $L = 2$, we have
\begin{align}\label{L2cbs}
| 1,1 \rangle_J \rightarrow &~ \sqrt{ \frac{3}{5} } Y_{2,2} | 1, -1 \rangle - \sqrt{ \frac{3}{10}} Y_{2,1} |1, 0 \rangle + \sqrt{ \frac{1}{10} } Y_{2,0} | 1, 1 \rangle~, \cr
|1,0 \rangle_{J} \rightarrow &~ \sqrt{ \frac{3}{10}} Y_{2,1} |1, -1 \rangle - \sqrt{ \frac{2}{5} } Y_{2,0} |1,0 \rangle + \sqrt{ \frac{3}{10}} Y_{2,-1} |1,1 \rangle~,\cr
|1,-1\rangle_J \rightarrow &~ \sqrt{ \frac{3}{5} } Y_{2,-2} |1,1 \rangle - \sqrt{ \frac{3}{10}} Y_{2,-1} | 1,0 \rangle + \sqrt{ \frac{1}{10}} Y_{2,0} |1,-1 \rangle~,
\end{align}
where the orbital angular momentum of the final state has been written in terms of spherical harmonics and the spin has been written as $|s,m_s\rangle$.  If $L=0$, then the $S$ quantum  numbers must be the same as the $J$'s:
\begin{equation}\label{L0cbs}
|1,m \rangle_J \rightarrow Y_{0,0} |1, m \rangle~, \qquad (m=1,0,-1)~.
\end{equation}
The general final state produced in $b_1$ decay can be written as a linear combination of these possibilities that depends on the initial spin of the $b_1$:
\begin{align}\label{ampform}
| 1,1\rangle_{b_1} =&~ S Y_{0,0} |1,1\rangle_\omega + D \left[  \sqrt{ \frac{3}{5} } Y_{2,2} | 1, -1 \rangle_\omega - \sqrt{ \frac{3}{10}} Y_{2,1} |1, 0 \rangle_\omega + \sqrt{ \frac{1}{10} } Y_{2,0} | 1, 1 \rangle_\omega \right]~,\cr
|1,0 \rangle_{b_1} =&~ S Y_{0,0} |1,0 \rangle_\omega + D \left[  \sqrt{ \frac{3}{10}} Y_{2,1} |1, -1 \rangle_\omega - \sqrt{ \frac{2}{5} } Y_{2,0} |1,0 \rangle_\omega + \sqrt{ \frac{3}{10}} Y_{2,-1} |1,1 \rangle_\omega \right]~, \cr
|1,-1 \rangle_{b_1} =&~ S Y_{0,0} |1,-1\rangle_\omega + D \left[ \sqrt{ \frac{3}{5} } Y_{2,-2} |1,1 \rangle_\omega - \sqrt{ \frac{3}{10}} Y_{2,-1} | 1,0 \rangle_\omega + \sqrt{ \frac{1}{10}} Y_{2,0} |1,-1 \rangle_\omega \right]~.
\end{align}

By squaring the above eigenstates, we can see that the coefficients
$S,D$ should satisfy $|S|^2 + |D|^2 = 1$, but their ratio is
determined by the dynamics of the theory via the matrix element
\eqref{Mhdecay2}.

We define the polarization-dependent amplitude $\mathcal{M}^{ij}$
generically as
\begin{equation}\label{Minnerproduct}
\mathcal{M}^{ij} \equiv \langle 1,j |_{\omega} \mathbf{T} | 1,i \rangle_{b_1}~.
\end{equation}
So
\begin{align}\label{Mquantum}
\mathbf{T} |1,1\rangle_{b_1} =&~ \mathcal{M}^{++} |1,1 \rangle_\omega + \mathcal{M}^{+0} |1,0 \rangle_\omega + \mathcal{M}^{+ -} |1,-1 \rangle_\omega~, \cr
\mathbf{T} |1,0 \rangle_{b_1} =&~ \mathcal{M}^{0+} |1,1 \rangle_\omega + \mathcal{M}^{00} |1,0 \rangle_\omega + \mathcal{M}^{0 -} |1,-1 \rangle_\omega~, \cr
\mathbf{T} |1,-1\rangle_{b_1} =&~ \mathcal{M}^{-+} |1,1 \rangle_\omega + \mathcal{M}^{-0} |1,0 \rangle_\omega + \mathcal{M}^{- -} |1,-1 \rangle_\omega~.
\end{align}
Evaluating the amplitude \eqref{Mhdecay2} for various choices of polarization vectors allows us to determine $D$
and $S$.   For this purpose we need an explicit basis of polarization vectors.  
We work in the rest frame of the
$b_1$, so the polarization vectors corresponding to spins $m_s(b_1)
= 0, \pm 1$, are
\begin{equation}\label{h1pols}
\varepsilon_{(b)}^{(0)}({\bf k})_\mu = \left( \begin{array}{c} 0 \\
0 \\ 0 \\ 1 \end{array}\right)~, \qquad
\varepsilon_{(b)}^{(\pm)}({\bf k})_\mu = \frac{\mp 1}{\sqrt{2}}
\left( \begin{array}{c} 0 \\  1 \\   \pm i \\ 0 \end{array}\right)~.
\end{equation}
We can boost an identical set of rest-frame polarizations to arrive
at the polarization of $\omega$ which moves in the lab frame with
three-momentum $-\mathbf{q}$:
\begin{equation}\label{rhopols}
\varepsilon_{(\omega)}^{(0)}(-{\bf q})_\mu = \left( \begin{array}{c}
- \frac{1}{m_\omega} q_z \\ \frac{1}{m_\omega (m_\omega + E_\omega)} q_x q_z
\\ \frac{1}{m_\omega (m_\omega + E_\omega)} q_y q_z \\ 1 + \frac{1}{
m_\omega (m_\omega + E_\omega )}  q_{z}^2  \end{array}\right)~, \quad
\varepsilon_{(\omega)}^{(\pm)}(-{\bf q})_\mu = \frac{\mp
1}{\sqrt{2}} \left( \begin{array}{c} - \frac{1}{m_\omega} q_{\pm} \\
1 + \frac{1}{m_\omega (m_\omega + E_\omega)} q_x q_{\pm} \\ \pm i +
\frac{1}{m_\omega ( m_\omega + E_\omega)}  q_y q_{\pm} \\ \frac{ 1}{m_\omega
(m_\omega + E_\omega)} q_z q_{\pm} \end{array}\right)~,
\end{equation}
where $q_x,q_y,q_y$ are the components of the three-vector
$\mathbf{q}$, and $q_{\pm} \equiv q_x \pm i q_y$.  Note that
$\varepsilon \cdot q =0$, with $q_\mu$ given in \eqref{4momenta}.
Also we have $E_\omega =  \sqrt{m_{\omega}^2 + |{\bf q}|^2}$.

As an example, consider the evaluation of $\cM^{\pm\pm}$.  Taking \eqref{Mhdecay2}, and plugging in the explicit four-momenta and polarizations, leads to
\begin{align}\label{Mpmpm}
\mathcal{M}^{\pm \pm} =&~ A \left[ -1 - \frac{1}{2}\left( \frac{E_\omega}{m_\omega} -1 \right)\sin^2{\theta} \right] - B \left[  \frac{m_{b_1}}{2 m_\omega m_{\pi}^2} \mathbf{q}^2 \sin^2{\theta} \right]  \cr
=&~ -A - \frac{1}{2} \left\{ A \left( \sqrt{1 + x^2} -1 \right) + B \frac{m_{b_1}}{m_\pi} x y \right\}( 1 - \cos^2{\theta} ) \cr
=&~ - \frac{ \sqrt{4\pi} }{3} \left\{ A \left( \sqrt{1+ x^2} +2 \right) + B \frac{m_b}{m_\pi} xy \right\} Y_{0,0} + \cr
& + \frac{1}{6} \sqrt{ \frac{16\pi}{5} }
\left\{ A \left( \sqrt{1 + x^2} -1 \right) + B \frac{m_{b_1}}{m_\pi} x
y \right\} Y_{2,0}~.
\end{align}
In the first line we introduced standard spherical coordinates in ${\bf q}$-space, and in the second line we defined
\begin{equation}\label{xydef}
x \equiv  | {\bf q}|/m_{\omega}~, \qquad y \equiv | {\bf q} |/m_{\pi}~.
\end{equation}
Analogous computations yield the following set of matrix elements:
\begin{align}\label{MDSrelations}
\mathcal{M}^{\pm \pm} =&~ S \ Y_{0,0} + \sqrt{ \frac{1}{10}} D \ Y_{2,0}~, \quad \mathcal{M}^{\pm 0} = - \sqrt{ \frac{3}{10}} D \ Y_{2,\pm1} ~, \quad \mathcal{M}^{\pm \mp} =  \sqrt{ \frac{3}{5} } D \ Y_{2,\pm 2}~, \cr
\mathcal{M}^{0 \pm} =&~ \sqrt{ \frac{3}{10} } D \ Y_{2,\mp 1}~, \quad \mathcal{M}^{00} = S \ Y_{0,0} - \sqrt{ \frac{2}{5} } D \ Y_{2,0}~,
\end{align}
where  $S$ and $D$ are given by
\begin{align}\label{SandD}
S =&~ - \frac{2\sqrt{\pi}}{3} \left[ A \left( \sqrt{1+x^2} +2 \right) + B \frac{m_{b_1}}{m_\pi} x y \right]~, \cr
D =&~ \frac{2 \sqrt{2\pi}}{3} \left[ A \left( \sqrt{1+x^2} -1\right) + B \frac{ m_{b_1}}{m_\pi} x y \right]~.
\end{align}
Note that we should not expect these $D,S$ to satisfy $|S|^2 + |D|^2 =
1$, as we computed the matrix element only to leading order
in perturbation theory.  At this order,
our prediction for $D/S$ is
\begin{equation}\label{DoverS}
(D/S)_{b_1 \to \omega \pi} = - \sqrt{2} \frac{ A \left( \sqrt{1 + x^2} -1 \right) + B \frac{m_{b_1}}{m_\pi} x y }{  A \left( \sqrt{1 + x^2} + 2 \right) + B \frac{m_{b_1}}{m_\pi} x y} ~.
\end{equation}
%

\subsubsection{$g_{\rho\pi\pi}$}

The mixing of the $b_{MN}$ field with the vector field implies
that not only the mass spectrum, but also the interactions of the 4d
$\rho$ meson are modified. We must therefore re-compute the
$\rho$-$\pi$-$\pi$ coupling which appears in \cite{Erlich:2005qh}, and
include the revised version when we fit the undetermined parameters
of the extended model.

Since the pions are contained entirely in $A_M$ while the $\rho$'s
are contained in both $V_M$ and $b_{MN}$, we are interested in bulk
three-point couplings of the form $V$-$A$-$A$ and $b$-$A$-$A$. It
turns out that only $S_{\rm hw}$ and $S_{g_1}$ contribute to this
coupling, with
\begin{align}\label{rhopipi3pt}
S_{\rm hw} \supset &~ \frac{ f^{abc}}{2 g_{5}^2} \int d^4 x \int
\frac{dz}{z} \left\{ 2 V_{\mu}^a A_{z}^b \d^\mu (A_{z}^c - \d_z
\varphi^c) +  (\d^2 V_{\mu}^a) \varphi^b \d^\mu \varphi^c + 2 (\d_z
V_{\mu}^a) A_{z}^b \d^\mu \varphi^c \right\}~, \cr
S_{g_1} \supset &~ \frac{g_1 f^{abc}}{\sqrt{\ell}} \int d^4 x \int
\frac{dz}{z} v(z) \left\{ - (\d^\nu h_{\nu\mu}^a)  \varphi^b \d^\mu
\varphi^c + 2 h_{\mu z}^a A_{z}^b \d^\mu \varphi^c \right\}~,
\end{align}
where $h_{MN} = \Re{b_{MN}}$.  

We can now sum these two contributions, go to momentum space, and plug in the mode expansions of Appendix \ref{app:modeexpansions}.  The result should be compared with the standard definition,
\begin{equation}\label{grhopipidef}
S_{\rho\pi\pi} = g_{\rho\pi\pi} f^{abc} \int d^4 x \rho_{\mu}^a(x) \pi^b(x)
\d^\mu \pi^c(x)~.
\end{equation}
We find the following on-shell value for the coupling:
\begin{align}\label{grhopipi2}
g_{\rho\pi\pi}^{mnr} =&~ \frac{g_5}{ \sqrt{1 - \lambda_{(\rho,n)}'(m_{(\rho,n)}^2)}}  \int_{0}^{z_0} d\xi w_{(\pi)}
\displaystyle\bigg\{ \phi_{m}^{(v)} \phi_{n}^{(\pi)} \phi_{r}^{(\pi)}  + \cr
&  \qquad \qquad -  \left[ w_{(v)} \d_z \phi_{m}^{(v)} - \frac{ \sgn(\mu)
m_{(\rho,m)} u_1}{z_0(|\mu|-u_2)} \phi_{m}^{(w)} \right] \frac{\phi_{n}^{(\pi)} 
z \d_z \phi_{r}^{(\pi)}}{m_{(\pi,n)} m_{(\pi,r)}}
\displaystyle\bigg\}~.
\end{align}
One can verify that this reduces to the result of
\cite{Erlich:2005qh} in the $g_1,g_2 \to 0$ limit.

\section{Results}\label{sec:results}

Based on the above computations and those of the original hard wall
model, we have made predictions for the following physical observables:
$m_\rho$, $m_b$, $m_a$, $m_\pi$, $f_{\rho}^V$, $f_{\rho}^T$,
$f_b$, $f_a$, $f_\pi$, $g_{\rho\pi\pi}$, and the $D/S$ ratio and decay rate for
$b_1\rightarrow\omega\pi$.  In principle we have KK towers associated to each of these, but to compare with data we restrict $m_\rho$ and $g_{\rho\pi\pi}$ to the first three $\rho$ states, $m_{\pi}$ to the first two pion states, and everything else to the lowest KK mode.  These observables depend on the free parameters
of the original hard wall model, $z_0$, $g_5$, $m_q$, $\sigma$, and $\Delta$, and on
the new parameters we introduced by including the AdS two-tensor field $b_{MN}$ and its interactions: $\mu$, $g_b$, $g_1$, $g_2$, $g_3$.\footnote{The AdS radius 
$\ell$ does not play a role for these observables but will for the magnetic susceptibility.  See section \ref{chiresults}.}

We allow all parameters to vary, instead of fixing  them to values calculated in perturbative QCD. This stands in contrast to  the original hard model \cite{Erlich:2005qh} and to   \cite{Alvares:2011wb} where the interactions proposed in \cite{Domokos:2011dn} were also included: these works fix as many observables
as possible to values computed in perturbative QCD.
Of course, since the vector flavor symmetry is preserved, the vector current acquires no anomalous dimension. This implies that we can fix the 5d masses of the corresponding gauge fields to zero.
All other parameters \textit{do} vary between the UV and IR, however. It is  more accurate (though admittedly less predictive) to fit all parameters 
by fitting to  IR data. As we have a large number of free parameters,  our main purpose will be to
explore how well AdS/QCD models conform to hadronic data when we push them beyond $\alpha'=0$ and $N_c\rightarrow\infty$.

As described in detail above, the tensor field $b_{MN}$ and its chiral-symmetry-breaking interaction terms induce
mixing between the vector mesons generated by the QCD two-tensor $O^T$, and the vector mesons generated by the QCD vector current. These terms also
 lift the degeneracy between the  vector- and pseudo-vector states generated by $b_{MN}$.  In other words, coupling this
new sector to the original hard wall model modifies  predictions for observables like $m_{\rho}$ and $g_{\rho\pi\pi}$;
furthermore, the predictions we make for the previously neglected $b_1$ mesons depend critically on the parameters of the original model. In order to
make consistent and accurate predictions, therefore, we perform a comprehensive fit to the experimental data including {\it all}
free parameters. 

The full parameter space is subject to some restrictions, discussed  in subsection \ref{subsec:eomandbcs}, and summarized here.
\begin{itemize}
\item We restrict $2<\Delta<4$, where $\Delta$ is the scaling dimension of $\bar{q}{q}$ dual to the scalar $X$.  Values of $\Delta$ less than $2$ violate the
Breitenlohner-Freedman bound, while it was shown in \cite{de Haro:2000xn} that $\Delta>4$ for the classical solution of $X$ ruins the
asymptotically AdS behavior of the geometry. 
\item  We restrict $|\mu|<6-\Delta$. This guarantees that the $V_\mu$-type solution induced by turning on a source for $b_{\mu\nu}$   does not dominate
the $V_\mu$ mode induced by turning on a source for $V_\mu$ itself.  Violating this restriction implies UV scaling behavior for $V_\mu$
at odds with the scaling dimension of the conserved vector current $J_V$, which is protected from RG flow.  $|\mu| > 6-\Delta$ 
also leads to divergences in the on-shell action that cannot be removed by local counterterms.
\item We must also impose some constraints guaranteeing that the Sturm-Liouville problems defining the masses and decay constants
of each sector remain well-defined. For certain values of the free parameters, the weight functions appearing in the Sturm-Liouville inner products pass through a pole and switch sign.
To guarantee that this does not happen, we require 
constraints
\begin{align}
|\mu|>|u_2(z_0)|\qquad\qquad\text{and}\qquad\qquad 1>\frac{u_1(z_0)^2}{|\mu|-u_2(z_0)}~.
\end{align}
\end{itemize}
Finally we note that the observables we have computed only depend on the relative sign of $\mu$ and $g_1 g_b g_5$.  The fit prefers them to have the same sign, but it does not determine whether this sign is positive or negative.

\subsection{The global fit}

  \begin{table}[t!]
  \centering
  \begin{tabular}{@{} |c|ccc| @{}}
    \hline
    quantity &  fit  & experimental (lattice)  result  & reference \\
    \hline\hline

                        $m_{\rho^0}$ &  $777.1 \pm5.0$ & $775.49 \pm 0.34$ &  \cite{PDG}\\
    $m_{\rho'}$ &   $1433 \pm 6 $ & $1465 \pm 25$& \cite{PDG} \\
    $m_{\rho''}$ &  $1783\pm 11$ & $1720 \pm 20$ &\cite{PDG}\\
         $f_{\rho}$&   $141.7 \pm 0.8$&$153 \pm 7$& \cite{Donoghue:1992dd} \\
      $[f_{\rho}^{T}$  &  184* & $(184 \pm 15)$& 	\cite{Jansen:2009hr}] \\
          \hline

                          $m_{b_1}$ &  $1227\pm4$ & $1229.5 \pm 3.2$& \cite{PDG}\\
                              $[f_{b_1}$ & 3177 & $(236 \pm 23)$& \cite{Jansen:2009yh}]\\
                          \hline
                         $m_{\pi^0}$ &  $138.3\pm 1.2$ & $134.9766\pm 0.0006$ & \cite{PDG}\\
                           $m_{\pi^0}'$ &   $1891\pm15$  & $1300\pm 100$ & \cite{PDG}\\
                         $f_\pi$   &  $75.03\pm0.38$ & $92.4\pm 0.35$ & \cite{PDG} \\
                         $m_{a_1}$ & $1114\pm8$ & $1320\pm 40$& \cite{PDG} \\
                         $f_{a_1}$ &    $410.9\pm2.4$  & $433\pm 13$ & \cite{PDG} \\
\hline\hline 
$g_{\rho^0\pi\pi}$  & $5.54\pm0.02$ & $6.03\pm0.07$& \cite{PDG}\\
$g_{\rho'\pi\pi}/g_{\rho^0\pi\pi}$   & $-0.128\pm 0.002$ & $-0.13\pm 0.02$& \cite{Fujikawa:2008ma}\\
$g_{\rho''\pi\pi}/g_{\rho^0\pi\pi}$ & $0.029 \pm 0.006$& $0.028\pm0.02$& \cite{Fujikawa:2008ma}\\
$[\Gamma(b_1\rightarrow\omega\pi)$  & $108^*$& $108\pm 9$&  \cite{PDG}]\\
$[D/S(b_1\rightarrow\omega\pi)$  & $0.15$ &  $0.277\pm 0.027$ & \cite{PDG}]\\
\hline
  \end{tabular}
  \caption{Masses, decay constants, couplings, and decay rates used in our fit. All units (when relevant) are MeV.  The fit includes 
  all observables listed in the table \textit{with the exception of} those marked with square brackets: data from the  $b_1\rightarrow\omega\pi$ decay ($\Gamma$ and $D/S$), and the $b_1$ and $\rho^T$ decay constants. The decay constants $f_{b_1}$, $f_\rho^T$ are  evaluated at a scale of $2 ~ \textrm{GeV}$. Data from the lattice is shown in parentheses. Fit error bars refer to the average degree of accuracy permitted by our numerical analysis, and do not include any theory error. }
  \label{table:data}
\end{table}

The experimental and lattice data we use for the fit are summarized in Table  \ref{table:data}. We have chosen the data which we expect to be
most accurately modeled by our results. For instance, isospin 0 and 1 states are degenerate
in our framework, so we  have chosen to fit our predicted masses to the measured masses of isospin 1 states (i.e. the $\rho$, the $b_1$ and the $a_1$), as the isospin 0 modes may mix with glueballs. Furthermore, the error bars in the $D/S$ ratio are given by the average PDG estimate, but these errors may in fact
be much larger when systematic effects are included. For example, the $D/S$ ratio is found, experimentally, to have a non-trivial phase -- something we cannot reproduce in our framework. It is suggested in \cite{Nozar:2002br} that the phase may be due in part to final state interactions between the $\pi$ and the $\omega$, which we do not take into account. Similarly,  the $g_{\rho\pi\pi}$ couplings generically assume complex values \cite{Fujikawa:2008ma}, though in our mode they are real by definition. However,
the experimental data indicates that the phase is close to either $1$ or $-1$ (within the error bars), so we fit to a single positive (or negative) number, instead of considering phase and magnitude separately.

Values for the parameters which participate in the fit are given in Table \ref{table:fitparams}. The overall rms value of the best fit result is $0.296\pm 0.026$.
The quantities marked with square brackets in Table \ref{table:data} are not included in the global fit. $f_{b_1}$ and $f_{\rho}^T$ are the only two
  observables which depend on $g_b$ alone (not in combination with $g_1$ or $g_2$), while $\Gamma (b_1\rightarrow\omega\pi)$ and the $D/S$ ratio are the only two quantities which depend on $g_3$. 
  The values quoted in the table are found by using $f_{\rho}^T$ to fix $g_b$, which we then use to determine $f_{b1}$. These quantities are clearly incompatible in the existing model.\footnote{If we were to perform the reverse operation, fixing $g_b$ using
  the $f_{b1}$ we would find $f_{\rho}^T\sim 13.7$ MeV.} Meanwhile, we fix $g_3$ using the value of $\Gamma (b_1\rightarrow\omega\pi)$ and use it to compute a prediction for the $D/S$ ratio.
  
In order to facilitate comparison to the hard wall model, (Model B of reference \cite{Erlich:2005qh}), we define  combinations of certain parameters  that are dimensionless in both frameworks.  The parameters that naturally appear in the numerical fit of our model are
\begin{align}\label{eq:tildeparams}
\tm_{q}= 2m_qz_0\left(\frac{\ell}{z_0} \right)^{\Delta-3}\qquad \tsigma=\frac{\sigma z_0^3}{4(\Delta-2)}\left(\frac{\ell}{z_0} \right)^{\Delta-3}\qquad\tilde{g}_b=g_b\left(\frac{z_0}{\ell}\right)^{|\mu|-1}~,
\end{align}
along with the remaining independent parameters $g_5$, $\Delta$, $z_0$, $\mu$, $g_1 g_b g_5$, $g_{2} g_{b}^2$, and $g_3 g_{b}^2 g_5$.
Five of these, $g_5$, $\tilde{m}_q$, $\tilde{\sigma}$, $\Delta$, and $z_0$, can be compared with analogous parameters in the hard wall model:
\begin{align}
\tm_{q{\rm H}} = m_{q{\rm H}} z_{0{\rm H}} \qquad \tsigma_{\rm H} = \sigma_{\rm H} z_{0{\rm H}}^3~,
\end{align}
together with $g_{5{\rm H}}$, $\Delta_{\rm H}$, $z_{0{\rm H}}$.  (The pairs $(\tilde{m}_q,\tilde{\sigma})$ and $(\tilde{m}_{q{\rm H}}, \tilde{\sigma}_{\rm H})$ are analogous in the sense that they are the coefficients of the two different powers of the dimensionless parameter $\xi = z/z_0$ in the function $v(z)$ controlling the tachyon vev, \eqref{Xvev}.)  In the hard wall, the parameter values $g_{5{\rm H}} = 2\pi$, $\Delta_{\rm H} = 3$ were fixed by UV matching, while the global fit of the remaining parameters yielded $m_{q{\rm H}} = 2.3  \MeV$, $\sigma_{\rm H} = (327 \MeV)^3$, and $z_{0{\rm H}} = (346  \MeV)^{-1}$.  These correspond to $\tilde{m}_{q{\rm H}} = 6.65 \times 10^{-3}$ and $\tilde{\sigma}_{\rm H} = 0.844$.

It is interesting to note that the parameters of the original hard wall framework tend to change very little, even those fixed using perturbative QCD. Meanwhile, the parameter $\mu$, related to the scaling dimension of $O^T$ runs significantly from its 
perturbative value of $|\mu_{\rm pert}|=1$.

\begin{table}[t!]
\renewcommand{\arraystretch}{1.2}
\begin{minipage}[b]{0.45\linewidth}\flushright
\begin{tabular}{| c | c |}
\hline
$g_5/g_{5 {\rm H}}$ & $0.984\pm0.008$ \\
$\tilde{m}_q/\tilde{m}_{q {\rm H}}$ & $0.74\pm 0.04$ \\
$\tilde{\sigma}/\tilde{\sigma}_{\rm H}$ & $1.047\pm 0.008$ \\
$\Delta/\Delta_{\rm H}$ & $1.113\pm 0.009$ \\
$z_0/z_{0 {\rm H}}$ & $1.081\pm 0.006$ \\
\hline
\end{tabular}
\end{minipage}
\hspace{0.5cm}
\begin{minipage}[b]{0.45\linewidth}\flushleft
\begin{tabular}{| c | c |}
\hline
$\pm \mu$ & $2.44\pm0.03$  \\
$\tilde{g}_b$ & $0.31$\\
$\pm g_1 g_b g_5$ & $0.381\pm 0.006$\\
$g_2 g_{b}^2$ & $-3.23\pm 0.04$ \\
$g_3g_b^2g_5$ & $77.3$ \\
\hline
\end{tabular}
\end{minipage}
\caption{Values of free parameters determined by fit to hadronic data. When relevant, the numbers quoted are the ratio between the quantity determined from our fit to that determined in the original hard wall model, Model B of reference \cite{Erlich:2005qh}, denoted by the subscript ``H''.  The signs of $\mu$ and $g_1 g_b g_5$ are correlated, but the overall sign is not determined.  As noted in the text, the quantities $g_b$  and $g_3$ were not included in the fit, as they only appears in $f_{b1}$ and $f_{\rho}^T$, and in $\Gamma$ and $D/S$, respectively.  The value listed
in the table is obtained by matching directly to $f_{\rho}^T$ and to $\Gamma$.}
\label{table:fitparams}
\end{table}

Overall, the results are quite good for the masses and decay constants in the $\rho$ sector. Note that the matching of radial $\rho$ excitations has improved
significantly compared to the hard wall predictions.  The $b_1$ meson is estimated to be far too broad, however. The value of $g_3g_b^2g_5$ is also unusually large.
This indicates that the model  has some fundamental failure.

\subsection{Results for the magnetic susceptibility of the quark condensate}\label{chiresults}

We can use the best-fit parameter values in Table \ref{table:fitparams} to estimate the value of the magnetic susceptibility \eqref{chiuptheory}.  Notice however that the magnetic susceptibility depends on $\sigma$; using \eqref{eq:tildeparams} to convert this to the numerical quantity $\tilde{\sigma}$ leaves us with a dependence on the undetermined ratio $z_0/\ell$.  The theoretical prediction also depends on the sign of $\mu$, which is not uniquely determined by the best fit.  We have
\begin{equation}
\chi_u = \frac{\sgn(\mu) z_{0}^2}{(\Delta -2) g_5 \tilde{g_b} \tilde{\sigma}} \left( \frac{z_0}{\ell} \right)^{\Delta - 3} \cdot \lim_{k \to 0} \frac{1}{k z_0} \left( C_{wv}(k) +C_{vw}(k) \right)~.
\end{equation}
The quantity $\lim_{k \to 0} \frac{1}{k z_0} \left( C_{wv}(k) +C_{vw}(k) \right)$ can be evaluated numerically for the best-fit parameter values and we find its value to be $\approx 0.35$.  Plugging in the remaining quantities yields
\begin{equation}
\chi_u \approx 0.185 \times \sgn(\mu) z_{0}^2 \times \left( \frac{z_0}{\ell} \right)^{0.34} \approx 1.8 \, \textrm{GeV}^{-2} \times \sgn(\mu) \left( \frac{z_0}{\ell} \right)^{0.34}~.
\end{equation}

The magnetic susceptibility is a quantity of growing relevance for a number of experimental results. In particular it enters into computations of
radiative meson decays \cite{Colangelo:2005hv, Rohrwild:2007yt}, the muon anomalous magnetic moment \cite{Czarnecki:2002nt,Nyffeler:2009uw},  and chiral-odd proton distribution functions \cite{Braun:2002en,Pire:2009nn}.  It  has been calculated in QCD using
a variety of theoretical tools \cite{Belyaev:1984ic, Balitsky:1985aq, Vainshtein:2002nv, Dorokhov:2005pg} and most recently has been studied using holographic methods \cite{Gorsky:2009ma, Gorsky:2012ui, Son:2010vc}. It has also been the subject of several lattice
QCD computations \cite{Buividovich:2009ih, Bali:2012jv, Braguta:2010ej}. 

In comparing these results to each other and to our result we find it useful, following \cite{Buividovich:2009ih}, to write the magnetic susceptibility in
terms of a parameter $c_u$  as
\begin{equation}
\chi_u = -c_u \frac{N_c}{16 \pi^2 f_\pi^2} = - c_u~ (2.22~ {\rm GeV}^{-2})
\end{equation}
for $N_c=3$.
Values of $c_u$ in the literature vary  from $0.96$ to $1.04$ for lattice QCD computations \cite{Buividovich:2009ih, Bali:2012jv, Braguta:2010ej}, while $c_u \simeq 1.4$ is a typical value
obtained from radiative meson decays \cite{Colangelo:2005hv, Rohrwild:2007yt}. Sum rule approaches in QCD which can also be applied in conjunction with ideas of holography lead to
$c_u \simeq 2-2.15$ \cite{Vainshtein:2002nv, Gorsky:2009ma} while the resonance sum rules of \cite{Son:2010vc} require $c_u=4$. The only other fully holographic computation of $c_u$
uses the model of  \cite{Domokos:2011dn} but with parameter values fixed to their UV values as in
in \cite{Alvares:2011wb} and leads to an unacceptably small result, $c_u \simeq 0.06$ \cite{Gorsky:2012ui}. 

The first thing we learn from comparing our result to the literature is that the diamagnetic nature of $\chi_u$ requires that we choose
the sign of $\mu$ to be negative.  We can then use the above results for $\chi_u$ to determine a possible range of values for  $z_0/\ell$ in our model.
We find that this ratio varies from roughly $1.8$ if we fit to  the lattice results to $5$ for the radiative decay results to $14$ and $110$ if we try
to fit to the QCD sum rule results of \cite{Vainshtein:2002nv} or the sum rule results of \cite{Son:2010vc} respectively. It is natural to view $1/z_0$
as the QCD scale $ \Lambda_{QCD}$ and $1/\ell$ as a renormalization scale $M_r$ in the ``conformal window" of this class of AdS/QCD models. A ratio $M_r/ \Lambda_{QCD}$ of $1.8$
to $5$ is compatible with this point of view  while the larger values of $M_r/\Lambda_{QCD} \simeq 14-110$ seem implausibly
large, since they extend the conformal window much too far into the ultraviolet.  

Note also that, taking reasonable values for $(z_0/\ell)$  we find a value of $\chi_u$ in good agreement with other determinations, while  \cite{Gorsky:2012ui} do not.
This can be taken as evidence for our approach, in which we fit the model parameters to data rather than demanding that they maintain their
naive UV values \`{a} la \cite{Alvares:2011wb,Gorsky:2012ui}.

\section{Conclusion}\label{sec:conclusions}

We have fully analyzed the extended hard wall framework laid out in \cite{Domokos:2011dn} with one additional interaction term. This extended model incorporates the $b_1$ mesons and a new tower of $\rho$ states.
We included all naive dimension $3$ QCD operators on the field theory side.
We performed a new fit for all free parameters in the model, including those of  the original framework.
The free Lagrangian and boundary conditions we used for introducing these new states are essentially unique. The interaction terms we
included render the model complete up to dimension six (according to 5d dimension counting). We  succeeded in 
reproducing much of the observed data, including the masses of higher radial $\rho$ resonances, and an appropriate estimate for the magnetic susceptibility. 

However, our rms error is $29.6\%\pm 2.6\%$, and
our predictions are quite poor for data depending most heavily on the $b_1$ sector, such as $f_b$, $\Gamma(b_1\to\omega\pi)$ and $D/S$. 
The model prefers parameter values such that the Sturm-Liouville weight function in the $b_1$ sector 
varies strongly near the IR boundary relative to 
weight functions in other sectors.  This is because when $g_2 g_{b}^2$ is large and negative as in Table \ref{table:fitparams}, the pole in $w_{(b_1)}$ is at $z/z_0 \sim 1.05$, close to the 
IR boundary.  $g_2g_b^2$ attains a  value ten times greater in magnitude than $g_1g_bg_5$, though $g_1g_bg_5$ should dominate by a factor of $N_c^{1/2}$. This
indicates a significant departure from the ``conformal'' values of $g_2=0$, $|\mu|=1$.  Though the $\rho$ 
sector masses are affected by $g_2$ as well, their weight functions depend on it only very mildly, and the original hard wall model attained remarkably good fits. 

Our choice of gravity background may underlie this particular failure of the model. As discussed in the introduction, the hard wall framework models
low-energy QCD with a ``conformal window'' in which the coupling does not run. This is clearly a crude approximation, though one which has proved 
 successful until now. One could check the effect of the background quantitatively by comparing our results to the equivalent
 soft wall model prediction, or by introducing more generic (asymptotically AdS) warp factors as polynomials in $z$, and fitting the 
 coefficients to the data.  In the end, however, the model reduces to a set of Sturm-Liouville problems. The functions appearing in
these eigenvalue equations depend on the geometry: both directly through the metric, and indirectly through the form of the $X_0$ vev. Since the Sturm-Lioville equations conflate these two effects,
even our crude analysis indicates that the IR geometry is important.

One should note also that
the results we find for the new couplings are not consistent with $N_c$ counting. The $N_c$ order of each coupling constant is reproduced in Table \ref{table:Nc}.
Even taking into account that $N_c=3$, and that each coupling may also contain some order 1 coefficient, the hierarchy of the couplings is inconsistent with
$N_c$ counting. One should \textit{not} necessarily take this as an indication of a breakdown in the large $N_c$ expansion, but rather a problem with 5d dimension counting. For instance, 
one can increase the 5d dimension, but not the $N_c$-order of any of the interaction terms we have chosen, by sprinkling in appropriately-contracted derivatives. We have
neglected such terms here.  The results indicate that such terms may be important, and that 5d dimension counting breaks down as an organizing principle for
interaction terms.

\begin{table}[th]
\begin{center}
\begin{tabular}{| c | c | c |}
\hline
$g_5$ & $6.28$ & $N_c^{-1/2}$  \\
$\tilde{g}_b$ & $0.31$ & $N_c^{-1/2}$ \\
$\pm g_1 g_b g_5$ & $0.38$ & $N_c^{-1/2}$ \\
$g_3g_b^2g_5$ & $77.3$  & $N_c^{-1/2}$  \\
$g_2 g_{b}^2$ & $-3.23$ & $N_c^{-1}$ \\
\hline
\end{tabular}
\end{center}
\caption{$N_c$ counting and fit values of coupling constants.}
\label{table:Nc}
\end{table}

Though we work in a toy model which does 
not descend from string theory directly, it is nevertheless important to note that we have
not taken into account all corrections that arise at next-to-leading order in $\alpha'$. We have simply included the first state with mass of
order $1/\sqrt{\alpha'}$ on the $\pi$ Regge trajectory. We made this choice because the operator that generates this state effectively renders the model 
complete (a) up to leading
order in perturbative QCD dimension and (b) up to the mass of the $a_1$. Yet the results of our fit seem to indicate that this next-to-leading in $1/\sqrt{\alpha'}$
operator acquires a very large anomalous dimension, so the perturbative QCD dimension counting becomes essentially irrelevant. 
For instance, one should also include $\alpha'$ contributions to the background, or the first 
excitations in $1/\sqrt{\alpha'}$ on the $\rho$ and $a_1$ trajectories.

 In sum, we have produced a model which partially accounts for the finite slope of QCD Regge
trajectories. The results of our fit (as well as our successful estimate of the magnetic susceptibility) suggest that parameters associated with $\cO(1/\sqrt{\alpha'})$ states
acquire significant anomalous dimensions, while those associated with states which appear from the supergravity limit of top-down duals do not. 
One should explore the refinements to the operator content and geometry outlined above to confirm these findings. It would also be very interesting to 
understand what effects (either in holography or directly from QCD) specifically suppress strong-coupling corrections to these ``supergravity-type'' operators.

\section*{Acknowledgements}

We thank Jon Rosner for helpful conversations on tensor mesons in QCD.  We are also grateful to Dima Kharzeev for pointing out the magnetic susceptibility as an additional observable of our model. 
SKD would like to thank the Particle, Field, and String Theory group at the University of Washington for 
hospitality during the final stages of this work, and Ami Katz for useful discussions.  ABR is supported by the U.S. Department of Energy under grant DE-FG02-96ER40959. JH acknowledges 
the support of NSF grant 0855039.

\bigskip

\appendix
\section{The tachyon vev}\label{Appendix:Xvev}

In this appendix we motivate the precise form of \eqref{Xvev}.  The complex field $X_{ij}$ is dual to the operator $\bar{q}_i \half(1- \gamma_5) q_j = \bar{q}_{L,i} q_{R,j}$, and thus its conjugate,
 $\bar{X}_{ij} = X_{ji}^\ast$, is dual to $\bar{q}_{R,i} q_{L,j}$.  If $\mathcal{X}_{ij}$ is the 4d source associated with $X$,  this means that $\mathcal{X}$ appears in the QCD generating functional as 
\begin{equation}
W  = \displaystyle\bigg\langle \exp{ \left\{ i \int d^4x ( \Re(\mathcal{X}_{ij}) \bar{q}_i q_j + i \Im(\mathcal{X}_{ij}) \bar{q}_i \gamma^5 q_j + \cdots) \right\} } \displaystyle\bigg\rangle~.
\end{equation}
Thus $\Re(X_{ij})$ is dual to the scalar bilinear $O_{ij}^S = \bar{q}_i q_j$ while the imaginary part is dual to the pseudoscalar.  Furthermore we see that $\Re(\mathcal{X})$ is naturally identified with the quark mass matrix, $\Re(\mathcal{X}) \equiv M_q$.

In QCD there is a spacetime-independent nonzero condensate for the scalar, $ \langle \bar{q}_i q_i \rangle \equiv \sigma$.  We describe this in holography by turning on a background value for $X_0 = \langle \Re(X) \rangle$ that is independent of the 4d coordinates.  The background field should solve the linearized equations of motion, $\d^M ( \sqrt{g} \d_M X_0) + \sqrt{g} (m_{X}^2/\ell^2) X_0 = 0$, with its value at $z = \varepsilon$ determined by the source:
\begin{equation}\label{Xsource} 
X_0(\varepsilon) = \frac{1}{\sqrt{\ell}} \cdot \frac{\varepsilon^{4-\Delta}}{\ell^{4-\Delta}} M_q~.
\end{equation}
The power of $\varepsilon$ is determined by the scaling behavior of the non-normalizable mode, and the factors of $\ell$ are inserted on dimensional grounds.  This is completely analogous to the expressions \eqref{sources}.  $\Delta$ is related to the tachyon mass $m_{X}^2$ through $\Delta = 2 + \sqrt{ 4 + m_{X}^2}$.  

It is convenient to introduce our basis $t^a = \half (\mathbbm{1}, \vec{\tau})$ and write $X = X^a t^a$, $M_{q} = M_{q}^a t^a$.  The $x^\mu$-independent solution with boundary behavior \eqref{Xsource} is
\begin{equation}\label{exactXbtb}
X_{0}^a(z) = \frac{1}{\sqrt{\ell}} \cdot \frac{\varepsilon^{4-\Delta}}{\ell^{4-\Delta}} M_{q}^a \cdot \frac{z^{4-\Delta} + C_{X} z^{\Delta}}{\varepsilon^{4-\Delta} + C_{X} \varepsilon^{\Delta}}~.
\end{equation}
$C_X$ is related to the quark vev.  Ordinarily it would be fixed by IR boundary conditions and these would determine the vev.  In the hard wall model we instead treat the vev as input and use holography to determine $C_X$.  More precisely we have
\begin{equation}
\Sigma_{q}^a \equiv  \langle \bar{q} t^a q \rangle = -i \frac{\delta}{\delta M_{q}^a} \displaystyle\bigg\langle \exp{ \left\{ i \int d^4x ( M_{q}^b \ \bar{q} t^b q + \cdots) \right\} } \displaystyle\bigg\rangle  = -i \frac{\delta}{\delta M_{q}^a} Z_{\rm grav}[M_q, \ldots]~.
\end{equation}
Following the philosophy discussed below \eqref{qcdW}, we take $Z_{\rm grav}[M_q] \to e^{i S[M_q]}$, where $S[M_q]$ is the action \eqref{Shw} evaluated on the solution \eqref{exactXbtb}.  The leading contribution from \eqref{Shw} is
\begin{align}\label{SXonshell}
S^{(2)}[M_q] =&~ \half \int d^4 x   \left\{ \frac{\ell^3}{\varepsilon^3} X^a(\varepsilon) (\d_z X^a) |_\varepsilon \right\} \cr 
=&~ \frac{1}{2\ell} \int d^4 x \left[ \frac{ \varepsilon^{5-2\Delta}}{\ell^{5-2\Delta}} M_{q}^a M_{q}^a \left( \frac{ (4-\Delta) \varepsilon^{3-\Delta} + \Delta C_X \varepsilon^{\Delta -1}}{ \varepsilon^{4-\Delta} ( 1 + C_X \varepsilon^{2\Delta -4} ) } \right) \right] \cr
=&~ \int d^4 x \left[ \frac{M_{q}^a M_{q}^a}{2 \ell^{6-2\Delta}}  \left( (4- \Delta) \varepsilon^{4-2\Delta} + \left( \Delta C_X - (4-\Delta) C_X \right) + O(\varepsilon^{2\Delta -4}) \right) \right]~, \quad
\end{align}
which, after subtracting counterterms and taking $\varepsilon \to 0$, leads to\footnote{In the first step of \eqref{SXonshell} one implicitly assumes that there is no contribution from the IR boundary.  Equivalently one assumes that there are IR boundary terms in the action which cancel the contribution from integration by parts in the bulk; these are the same boundary terms that would play a role in determining $C_X$.}
\begin{equation}
S[M_q] = \int d^4 x \left[\frac{(2\Delta -4) C_{X}}{2 \ell^{6-2\Delta}} M_{q}^a M_{q}^a \right] \quad \Rightarrow \quad \Sigma_{q}^a = \frac{(2\Delta -4) C_{X}}{\ell^{6-2\Delta}} M_{q}^a~.
\end{equation}
We can solve this for $C_X$ and plug back into \eqref{exactXbtb}.  As $\varepsilon \to 0$ we find
\begin{equation}\label{X0a}
X_{0}^a(z) = \frac{1}{\ell^{3/2}} \left( M_{q}^a \ell^{\Delta -3} z^{4-\Delta} + \frac{\Sigma_{q}^a}{2(\Delta - 2)} \ell^{3-\Delta} z^{\Delta} \right)~.
\end{equation}

Finally, to arrive at \eqref{Xvev}, we specialize to a quark mass matrix that is proportional to the identity, $M_q = m_q \mathbbm{1} = 2 m_q t^0$, and a quark vev which is also proportional to the identity, $\sigma = 2 \langle \bar{q} t^0 q \rangle = 2 \Sigma^0$, with the other components vanishing.  Then $X_{0} = 2 X_{0}^0 t^0$ is
\begin{equation}
X_0 = \frac{1}{2 \ell^{3/2}} \left( 2 m_{q} \ell^{\Delta -3} z^{4-\Delta} + \frac{\sigma}{4(\Delta -2)} \ell^{3-\Delta} z^{\Delta} \right) \mathbbm{1}~.
\end{equation}
%

\section{Eliminating auxiliary fields}\label{S2simp}

In this appendix we outline the steps that take one from the quadratic part of \eqref{Snot} to \eqref{Snot2}.  We decompose the 5d fields into 4d components.  Some of the 4d fields will be auxiliary; we eliminate them via their equations of motion and plug the result back into the action.

The decomposition of $A_M,V_M$ into 4d fields was discussed around \eqref{Aparam}, \eqref{Vparam}.  The quadratic actions for $A_{\mu}^\perp$ and $(\varphi,A_z)$ originate entirely from \eqref{Shw}.  They decouple from the rest of the fields and are given by
\begin{align}\label{S2A}
& S^{(2)}[A^\perp] = \frac{1}{g_{5}^2} \int \frac{d^4 k}{(2\pi)^4} \int_{\varepsilon}^{z_0} \frac{dz}{z} \tr \left\{ \d_z \bar{A}_{\mu}^\perp \d_z A^{\perp\mu} - k^2 \bar{A}_{\mu}^\perp A^{\perp\mu} + \frac{g_{5}^2 v(z)^2}{z^2} \bar{A}_{\mu}^\perp A^{\perp\mu} \right\}~, \\ \label{S2pi}
& S^{(2)}[\varphi,A_z] =  \frac{1}{g_{5}^2} \int \frac{d^4 k}{(2\pi)^4} \int_{\varepsilon}^{z_0} \frac{dz}{z} \tr \left\{ k^2 |\d_z \varphi - A_z|^2 + \frac{g_{5}^2 v(z)^2}{z^2} \left( k^2 |\varphi|^2 - |A_{z}|^2 \right) \right\}~.
\end{align}
They encode the tower of $J^{PC} = 1^{++}$ $f_1/a_1$ mesons and $J^{P} = 0^{-}$ $\eta/\pi$'s respectively.  Our extension of the hard wall model does not modify these sectors at quadratic order.  The equation of motion and boundary conditions for $A^\perp$ derived from \eqref{S2A} are identical to those in \cite{Erlich:2005qh}.  The equations of motion of for the pion sector of \cite{Erlich:2005qh}, described by 5d scalar fields $(\pi,\phi)$, can also be recovered by making the field redefinition $\d_z \varphi - A_z = \d_z \phi$, $\varphi = \phi - \pi$.  This field redefinition can be realized as a gauge transformation that sets $A_z$ to zero at the price of reintroducing pion fluctuations in the phase of $X$.  We prefer to work in a gauge where $X = X_0$ is frozen and $A_z$ is nontrivial.

Integration by parts puts \eqref{S2A} in the form quoted in the text, with a contribution $\int_{\d \cM} \tr \left[ w_{(a)} \bar{\cA}_\mu \d_z \cA^\mu \right]$ to the boundary action.  For the pion sector
we eliminate $A_z$ via its equation of motion and plug the result back into \eqref{S2pi}, obtaining
\begin{equation}\label{Sphi2}
S^{(2)}[\varphi] = - \int \frac{d^4 k}{(2\pi)^4} \int_{\varepsilon}^{z_0} dz \tilde{w}_{(\pi)} \tr \left\{ |\d_z \varphi|^2 - k^2  |\varphi|^2 + \frac{g_{5}^2 v(z)^2}{z^2} |\varphi|^2 \right\}~,
\end{equation}
where
\begin{equation}\label{phiweight}
\tilde{w}_{(\pi)}(k,z) = \frac{ k^2  v(z)^2}{z (k^2 z^2 - g_{5}^2 v(z)^2)}~.
\end{equation}
The unusual momentum dependence in the weight function makes this action awkward to work with.  It can be shown that solutions are completely regular through the pole in $\tilde{w}_{(\pi)}$, but basic properties we expect, such as a complete set of orthogonal wave functions and positive-definiteness of the spectrum, are obscured.  We can bypass these difficulties by a judicious change of field variable.  Let
\begin{equation}\label{picov}
\varphi(k,z) = - \frac{z^3}{k z_{0}^2 v(z)^2} \d_z \hat{A}(k,z)~,
\end{equation}
and consider the action
\begin{equation}\label{S2Ahat}
S^{(2)}[\hat{A}] =  \int \frac{d^4 k}{(2\pi)^4} \int_{\varepsilon}^{z_0} dz w_{(\pi)} \tr \left\{ |\d_z \hat{A}|^2 - k^2 |\hat{A}|^2+ \frac{g_{5}^2 v(z)^2}{z^2} |\hat{A}|^2 \right\}~.
\end{equation}
On shell these two actions are the same: \eqref{picov} maps stationary points of \eqref{S2Ahat} to stationary points of \eqref{Sphi2}.  The values at stationary points are also the same.  Using the equations of motion one can show $\int_{\d \cM} \tr \left[ w_{(\pi)} \bar{\hat{A}} \d_z \hat{A} \right] = - \int_{\d \cM} \tr \left[ \tilde{w}_{(\pi)} \bar{\varphi} \d_z \varphi \right]$.  This is the contribution to the boundary action in \eqref{Snot2}.  Furthermore \eqref{S2Ahat} has a standard form which allows one to demonstrate positive definiteness of the spectrum.

 The vector gauge field $V_\mu$ mixes with the two-form $b_{MN}$ at quadratic order due to the $g_1$ interaction term.  The 4d vector $b_{\mu z}$ is non-dynamical and can be eliminated via its equation of motion.  To linear order in the fields one finds
\begin{align}\label{bmuz}
 b_{\mu z}(k,z) =&~ - \frac{\sgn(\mu) z}{2[\mu^2- u_{2}^2]}  \epsilon_{\mu}^{\phantom{\mu}\nu\rho\sigma} k_\nu \left[ |\mu| b_{\rho\sigma}(k,z) -  u_2 \bar{b}_{\rho\sigma}(-k,z) \right]  + \cr
 & \qquad \qquad \qquad \qquad \qquad  - \frac{2 g_1 g_{b}^2 \sqrt{\ell}}{ [|\mu| - u_2] } v(z) \d_z V_\mu(k,z) + \cO(\Phi^2)~.
\end{align}
We plug \eqref{bmuz} back in to yield a quadratic action mixing $V_\mu$ and the anti-symmetric two-tensor $b_{\mu\nu}$.

It is convenient to split $b_{\mu\nu}$ into its real and imaginary parts,
\begin{equation}\label{btohw}
b_{\mu\nu} = h_{\mu\nu} + i w_{\mu\nu}~,
\end{equation}
such that $\bar{h}(k,z) = h(-k,z)$ and similarly for $w$, and to further split $h,w$ into components that are transverse and longitudinal to the momentum $k$:
\begin{equation}\label{hwTL}
h_{\mu\nu} = h_{\mu\nu}^\perp + h_{\mu\nu}^\parallel~, \qquad w_{\mu\nu} = w_{\mu\nu}^\perp + w_{\mu\nu}^\parallel~.
\end{equation}
The perpendicular and parallel components can be defined in terms of the projectors $\cP^{\perp,\parallel}$ introduced in \eqref{tensorproj}; the only facts we will require are that $k^\mu b_{\mu\nu}^{\perp} = 0$ and $\epsilon_{\mu}^{\phantom{\mu}\nu\rho\sigma} k_\nu b_{\rho\sigma}^\parallel = 0$.  Then we find that the remaining piece of the quadratic action splits into two sectors.  We have the $(h^\perp,w^\parallel)$ sector with action
 \begin{align}\label{S2hTwL}
 S^{(2)}[h^\perp,w^\parallel] =&~  \frac{1}{4\ell g_{b}^2} \int \frac{d^4 k}{(2\pi)^4} \int_{\varepsilon}^{z_0} dz  \tr \displaystyle\biggl\{ \sgn(\mu) \epsilon^{\mu\nu\rho\sigma} ( \bar{h}_{\mu\nu}^\perp \d_z w_{\rho\sigma}^\parallel - \bar{w}_{\mu\nu}^\parallel \d_z h_{\rho\sigma}^\perp) + \cr
&~ \qquad - \frac{2}{z} (|\mu|+ u_2) \bar{w}^{\parallel \mu\nu} w_{\mu\nu}^\parallel - \frac{2}{z} \left(|\mu|- u_2 - \frac{k^2 z^2}{|\mu| + u_2} \right)  \bar{h}^{\perp \mu\nu} h_{\mu\nu}^\perp \displaystyle\biggr\} + \cr
&~ + S_{\rm AF} |_{(h^\perp,w^\parallel)}~,
 \end{align}
 and the $(w^\perp, h^\parallel, V)$ sector with action
\begin{align}\label{S2wThLV}
S^{(2)}[w^\perp,h^\parallel,V] =&~ \frac{1}{4\ell g_{b}^2}  \int \frac{d^4 k}{(2\pi)^4} \int_{\varepsilon}^{z_0} dz  \tr \displaystyle\biggl\{ \sgn(\mu) \epsilon^{\mu\nu\rho\sigma}  (\bar{h}_{\mu\nu}^\parallel \d_z w_{\rho\sigma}^\perp - \bar{w}_{\mu\nu}^\perp \d_z h_{\rho\sigma}^\parallel) + \cr
&~ \qquad - \frac{2}{z} (|\mu|-u_2) \bar{h}^{\parallel \mu\nu} h_{\mu\nu}^\parallel - \frac{2}{z} \left( |\mu|+u_2 - \frac{ k^2 z^2}{|\mu|-u_2}  \right) \bar{w}^{\perp \mu\nu} w_{\mu\nu}^\perp \displaystyle\biggr\} + \cr
&~ + \frac{1}{g_{5}^2}  \int \frac{d^4 k}{(2\pi)^4} \int_{\varepsilon}^{z_0} \frac{dz}{z} \tr \left\{ \left(1 - \frac{u_{1}^2}{|\mu| - u_2} \right)\d_z \bar{V}^\mu \d_z V_\mu - k^2 \bar{V}^\mu V_\mu \right\}  + \cr
&~ + \frac{2 i g_1 \sgn(\mu)}{\sqrt{\ell}} \int \frac{d^4 k}{(2\pi)^4} \int_{\varepsilon}^{z_0} dz \frac{v(z)}{|\mu|-u_2} \epsilon^{\mu\nu\rho\sigma} k_\mu \tr \left\{ (\d_z \bar{V}_\nu) w_{\rho\sigma}^\perp \right\} + \cr
&~ + \frac{4 i g_1}{\sqrt{\ell}} \int \frac{d^4 k}{(2\pi)^4} \int_{\varepsilon}^{z_0} \frac{dz}{z} v(z) \tr \left\{ h_{\mu\nu}^\parallel q^\mu \bar{V}^{\nu} \right\} + S_{\rm AF} |_{(w^\perp, h^\parallel)}~.
\end{align}
$S_{\rm AF}$ refers to the boundary term of Arutyunov and Frolov in the action \eqref{Ssd} \cite{Arutyunov:1998xt,Domokos:2011dn}:
\begin{align}\label{SAFdef}
S_{\rm AF} =&~ - \frac{1}{4\ell g_{b}^2} \int_{\d \cM} \tr b_{\mu\nu} b^{\mu\nu}  \cr
=&~ - \frac{1}{4\ell g_{b}^2} \int \frac{d^4 k}{(2\pi)^4} \tr \left[ \bar{h}^{\perp \mu\nu} h_{\mu\nu}^\perp +  \bar{h}^{\parallel \mu\nu} h_{\mu\nu}^\parallel +  \bar{w}^{\perp \mu\nu} w_{\mu\nu}^\perp +  \bar{w}^{\parallel \mu\nu} w_{\mu\nu}^\parallel  \right]_{\varepsilon}^{z_0}~,
\end{align}
and in \eqref{S2hTwL} and \eqref{S2wThLV} we have restricted it to terms containing the indicated fields only.

These actions are first order with respect to the $h$ and $w$ fields.  On shell they are equivalent to more standard second order actions that can be obtained by solving the equations of motion for the parallel components (say) in terms of the perpendicular ones.  To linear order in the fields we find
\begin{align}\label{paralleleoms}
& w_{\mu\nu}^{\parallel} = - \frac{ \sgn(\mu)  z}{2 (|\mu| + u_2)} \epsilon_{\mu\nu}^{\phantom{\mu\nu}\rho\sigma} \d_z h_{\rho\sigma}^\perp + \cO(\Phi^2)~, \\
& h_{\mu\nu}^{\parallel} = \frac{\sgn(\mu)  z}{2 (|\mu| - u_2)} \epsilon_{\mu\nu}^{\phantom{\mu\nu}\rho\sigma} \d_z w_{\rho\sigma}^{\perp} - \frac{4 i g_1 g_{b}^2 \sqrt{\ell}}{|\mu| - u_2} v(z) k_{[\mu} V_{\nu]} + \cO(\Phi^2)~,
\end{align}
which can be substituted back into \eqref{S2hTwL}, \eqref{S2wThLV} to get second order actions $S^{(2)}[h^\perp]$ and $S^{(2)}[w^\perp,V]$.

The real transverse anti-symmetric two-tensors $h^\perp,w^\perp$ each contain three independent components which together comprise the six degrees of freedom in $b_{MN}$.  The transformation properties of $b_{MN}$ under charge conjugation and parity identify the normalizable modes of $h^{\perp}$ with the $J^{PC} = 1^{+-}$ $h_1/b_1$ mesons, and the normalizable modes of $w^\perp$ with $J^{PC} = 1^{--}$ $\omega/\rho$-mesons.  More precisely, the normalizable eigenmodes of the coupled $w^T$-$V$ system are dual to the tower of $\omega/\rho$ mesons.  The fact that $V$ couples only to $w^\perp$ and not $h^\perp$ is a consequence of the fact that the interaction terms \eqref{Sint} respect the $\mathbb{Z}_2 \times \mathbb{Z}_2$ symmetry of charge conjugation and parity.  It is convenient to parameterize $h^\perp$ and $w^\perp$ in terms of fields $\cH_\mu(k,z)$, $\cW_\mu(k,z)$ via
\begin{align}\label{cHcWdef}
& h_{\mu\nu}^\perp(k,z) = - \frac{i g_b \sqrt{\ell}}{k z_0} \epsilon_{\mu\nu}^{\phantom{\mu\nu}\rho\sigma} k_\rho \cH_{\sigma}(k,z)~, \qquad  w_{\mu\nu}^\perp(k,z) = - \frac{i g_b \sqrt{\ell}}{k z_0} \epsilon_{\mu\nu}^{\phantom{\mu\nu}\rho\sigma} k_\rho \cW_{\sigma}(k,z)~.
\end{align}
These fields transform respectively as a 4d axial vector and vector which may be taken transverse; the coefficients of their normalizable modes are polarization vectors for the towers of $h_1/b_1$ and $\omega/\rho$ modes.  Plugging \eqref{paralleleoms}-\eqref{cHcWdef} back into \eqref{S2hTwL}, \eqref{S2wThLV} leads to the following quadratic actions describing the $h_1/b_1$ and $\omega/\rho$ sectors:
\begin{align}\label{S2H}
S^{(2)}[\cH] =&~ - \int \frac{d^4 k}{(2\pi)^4} \int_{\varepsilon}^{z_0} \frac{dz}{z_{0}^2}  \tr \bar{\cH}^\mu \left[ \d_z \left( \frac{z}{|\mu| + u_2} \d_z \right) +  \frac{k^2 z}{|\mu| + u_2}- \frac{1}{z} ( |\mu| - u_2 )  \right] \cH_\mu + \cr
&~ + S_{\rm AF} |_{(h^\perp,w^\parallel)}[\cH]~,
\end{align}
and
\begin{align}\label{S2WV}
S^{(2)}[\cW,\cV] =&~  - \int \frac{d^4 k}{(2\pi)^4} \int_{\varepsilon}^{z_0} \frac{dz}{z_{0}^2} \tr \bar{\cW}^{\mu} \displaystyle\biggl[  \d_z \left( \frac{z}{|\mu|-u_2} \d_z \right) + \frac{k^2 z}{|\mu|-u_2} - \frac{1}{z} (|\mu|+u_2) \displaystyle\biggr]\cW_\mu+  \cr
&~ + \int \frac{d^4 k}{(2\pi)^4} \int_{\varepsilon}^{z_0} \frac{dz}{z}  \left(1 - \frac{u_{1}^2 }{|\mu|- u_2} \right) \tr \left\{ \d_z \bar{\cV}^\mu \d_z \bar{\cV}_\mu - k^2 \bar{\cV}^\mu \cV_\mu \right\}  + \cr
&~ + 2 \sgn(\mu) \int \frac{d^4 k}{(2\pi)^4} k \int_{\varepsilon}^{z_0} \frac{dz}{z_0} \d_z \left( \frac{u_1}{|\mu|-u_2} \right) \tr \bar{\cV}^\mu \cW_\mu \cr
&~ -  \sgn(\mu) \int \frac{d^4 k}{(2\pi)^4} \frac{k}{z_0} \left( \frac{u_1}{|\mu|-u_2} \right) \tr \bar{\cV}^\mu \cW_\mu \displaystyle\biggr|_{\varepsilon}^{z_0} +  S_{\rm AF} |_{(w^\perp,h^\parallel)}[\cW,\cV]~.
\end{align}

After some integration by parts, the sum of \eqref{S2A}, \eqref{S2Ahat}, \eqref{S2H}, and \eqref{S2WV} can be put in the form \eqref{Snot2}, where the boundary action is
\begin{align}\label{Sbndryapp}
S_{\rm bndry} =&~ \int \frac{d^4 k}{(2\pi)^4} \tr \displaystyle\biggl\{ w_{(\pi)} \bar{\hat{A}} \d_z \hat{A} + w_{(a)} \bar{\cA}_\mu \d_z \cA^\mu + \frac{1}{2 z_{0}^2} \left( |\cH_\mu|^2 - z_{0}^4 w_{(b)}^2 | \d_z \cH_\mu|^2 \right) + \cr
&~ \qquad \qquad \qquad + \frac{1}{2 z_{0}^2} \left[ | \cW_{\mu} |^2 - \left| z_{0}^2 w_{(w)} \d_z \cW_\mu - \frac{ \sgn(\mu) k z_0 u_1}{|\mu| - u_2} \cV_\mu \right|^2 \right] + \cr
&~ \qquad \qquad \qquad + \bar{\cV}_\mu \left[ w_{(v)} \d_z \cV^\mu - \frac{ \sgn(\mu) k u_1}{z_0 (|\mu| - u_2)} \cW^\mu \right] \displaystyle\biggr\} \displaystyle\bigg|_{\varepsilon}^{z_0} ~.
\end{align}
A nice feature of this boundary action is that the contribution from $z = z_0$ vanishes when the IR boundary conditions \eqref{IRbcs} are imposed.  Thus, on shell, \eqref{Sbndryapp} restricts to the $z = \varepsilon$ terms only.

\section{Near-boundary analysis}\label{asymptotic}

In this appendix we discuss the asymptotic analysis that leads from the on-shell action at finite $\varepsilon$, \eqref{Soss}, \eqref{Sosrho}, to the renormalized action in the $\varepsilon \to 0$ limit, \eqref{Sren}.  Let us begin with the pion sector.  Examination of the equation of motion, $(k^2 - O^{(\pi)}) \hat{A} = 0$, near $z = 0$ shows that the leading behaviors of the non-normalizable and normalizable solution are
\begin{align}\label{piwfs}
& f_{(\pi)}^{\rm n.n.} = \xi^{6-2\Delta} \left( 1 + \cO\left(\xi^{ \min\{2,8-2\Delta \} } \right) \right)~,  \qquad f_{(\pi)}^{\rm n.} = 1 + \cO\left(\xi^{ \min\{2,8-2\Delta\} } \right)~.
\end{align}
We have introduced the dimensionless variable $\xi = z/z_0$ and chosen a convenient normalization convention for the solutions: namely the coefficient of the leading power of $\xi$ is equal to one.  This will be our convention in all sectors.  Note that \eqref{piwfs} does not fully specify the non-normalizable solution since we can add a multiple of the normalizable solution without changing the asymptotics.  This ambiguity would manifest itself in a near boundary analysis as a subleading coefficient that is not determined by the differential equation.  We define the non-normalizable modes so that this coefficient is set to zero.

The $P$'s and $Q$'s are given by $(P_{(\pi)}^{\rm n.n.}, P_{(\pi)}^{\rm n.}) = (f_{(\pi)}^{\rm n.n.}, f_{(\pi)}^{\rm n.})$, and
\begin{align}\label{PQpi}
Q_{(\pi)}^{\rm n.n.} = - \frac{ (6 - 2\Delta) \ell^{6-2\Delta}}{m_{q}^2 z_{0}^{8-2\Delta}} \left(1 + \cO \left( \varepsilon^{ \min \{ 2, 8-2\Delta \} } \right) \right)~, \quad Q_{(\pi)}^{\rm n.n.} = \cO \left( \varepsilon^{ \min \{ 2\Delta -4,2 \} } \right)~.
\end{align}
Hence
\begin{align}\label{pilimit}
& \frac{P_{(\pi)}^{\rm n.n.} - C_{(\pi)} P_{(\pi)}^{\rm n.} }{Q_{(\pi)}^{\rm n.n.} - C_{(\pi)} Q_{(\pi)}^{\rm n.} } = \frac{-m_{q}^2 z_{0}^{8-2\Delta}}{ (6-2\Delta) \ell^{6-2\Delta}} \cdot \frac{ \cO ( \varepsilon^{6-2\Delta}) - C_{(\pi)} (1 + \cO(\varepsilon^{>0}) ) }{ 1 +  \cO \left( \varepsilon^{ \min \{ 2, 8-2\Delta \} } \right) - C_{(\pi)} \cO \left( \varepsilon^{ \min \{ 2\Delta -4,2 \} } \right) }~.
\end{align}
We focus on this factor only since the source $S_{(\pi)}$ is $\cO(\varepsilon^0)$.  Expanding the denominator, we see that there can be divergent terms (when $\Delta > 3$) but that none of these terms involve $C_{(\pi)}(k)$.  In particular, terms coming from the $C_{(\pi)}$ term in the denominator always go as a positive power of $\varepsilon$; while terms coming from the $C_{(\pi)}$ term in the numerator give finite contributions.  Recalling the expression for the bulk-to-boundary propagator, \eqref{btbs}, this means that only the normalizable component of it contributes in \eqref{pilimit}.

The set of divergent terms in \eqref{pilimit} can be determined by a purely asymptotic analysis, solving for the coefficients order by order in an asymptotic expansion.  It follows from the form of the equation of motion that these coefficients will be analytic functions of $k^2$, and therefore the divergences can be eliminated by local counterterms.  This results in the pion contribution to the renormalized action, \eqref{Sren}.

The analysis in the $a_1$ and $b_1$ sectors is similar.  In the $a_1$ sector we have
\begin{align}\label{awfs}
& f_{(a)}^{\rm n.n.} =  1 + \cO\left(\xi^{ \min\{2,8-2\Delta \} } \right)~,  \qquad f_{(a)}^{\rm n.} = \xi^2 \left( 1 + \cO\left(\xi^{ \min\{2,8-2\Delta\} } \right) \right)~,
\end{align}
while the $Q$'s and $P$'s are $(Q_{(a)}^{\rm n.n.}, Q_{(a)}^{\rm n.}) = (f_{(a)}^{\rm n.n.}, f_{(a)}^{\rm n.})$, and
\begin{equation}\label{PQa}
P_{(a)}^{\rm n.n.} = \cO\left( \varepsilon^{ \min \{ 0, 6-2\Delta \} } \right)~, \qquad P_{(a)}^{\rm n.} = 2 \left( 1 + \cO\left(\varepsilon^{ \min\{2,8-2\Delta \} } \right) \right)~.
\end{equation}
Again the potentially divergent terms come entirely from $(Q_{(a)}^{\rm n.n.})^{-1} P_{(a)}^{\rm n.n.}$ and can be eliminated with local counterterms.  There may also be a finite contribution from this piece, but it is scheme dependent and may be set to zero as well.  The physical finite contribution comes from the $(Q_{(a)}^{\rm n.n.})^{-1} P_{(a)}^{\rm n.}$ term.

For the $b_1$ sector we have solutions
\begin{align}\label{bwfs}
& f_{(b)}^{\rm n.n.} =  \xi^{-|\mu|} \left( 1 + \cO\left(\xi^{ \min\{2,8-2\Delta\} } \right) \right)~,  \qquad f_{(b)}^{\rm n.} = \xi^{|\mu|} \left( 1 + \cO\left(\xi^{ \min\{2,8-2\Delta\} } \right) \right)~,
\end{align}
and $Q$'s and $P$'s,
\begin{align}\label{PQb}
& Q_{(b)}^{\rm n.n.} = \frac{z_{0}^{|\mu|}}{\varepsilon^{|\mu|}} \left( 1 + \cO\left(\varepsilon^{ \min\{2,8-2\Delta \} } \right) \right)~, \qquad Q_{(b)}^{\rm n.} =  \cO\left(\varepsilon^{|\mu| + \min\{2,8-2\Delta \} } \right)~, \cr
& P_{(b)}^{\rm n.n.} =  \cO\left(\varepsilon^{-|\mu| + \min\{2,8-2\Delta \} } \right)~, \qquad P_{(b)}^{\rm n.} =   \frac{2 \varepsilon^{|\mu|}}{z_{0}^{|\mu|}} \left( 1 + \cO\left(\varepsilon^{ \min\{2,8-2\Delta \} } \right) \right)~.
\end{align}
In this case one should keep in mind that the source factors also has $\varepsilon$ dependence; explicitly,
\begin{equation}\label{Sbcontraction}
\bar{S}_{(b)\mu} S_{(b)}^\mu = - \frac{z_{0}^2 \ell^{2|\mu| -2}}{4 g_{b}^2 k^2 \varepsilon^{2|\mu|}} \bar{\cT}^{\mu\nu} \cP^{\perp}_{\mu\nu,\rho\sigma} \cT^{\rho\sigma}~,
\end{equation}
so we are looking for $\cO(\varepsilon^{2|\mu|})$ terms in the analogous expression to \eqref{pilimit}.  The story is the same.  Divergent terms come from $(Q_{(b)}^{\rm n.n.})^{-1} P_{(b)}^{\rm n.n.}$ and can be removed by counterterms; finite terms come from $(Q_{(b)}^{\rm n.n.})^{-1} P_{(b)}^{\rm n.}$.

Finally there is the $\rho$ sector.  Let $R^i = (f^i,g^i)^T$ denote the set of four linearly independent solutions to $(k^2 - {\bf O}^{(\rho)}) R = 0$.  The index $i$ runs over four values, $i \in \{ \textrm{n.n.}(w), \textrm{n.n.}(v), \textrm{n.}(w),\textrm{n.}(v) \}$.  The leading small $z$ behavior of these solutions is found to be
\begin{align}\label{wwfs}
& f^{\textrm{n.n.}; \textrm{n.}}_{(w)} = \xi^{\mp |\mu|} \left( 1 + \cO \left( \xi^{ \min \{2,8-2\Delta \} } \right) \right)~, \cr
& g^{\textrm{n.n.}; \textrm{n.}}_{(w)} = \frac{2 \sgn(\mu) (4-\Delta) g_1 g_b g_5 m_q k z_{0}^{5-\Delta}}{|\mu| (\mp |\mu| + 6-\Delta) (\mp |\mu| + 4-\Delta) \ell^{3-\Delta}} \xi^{\mp|\mu| + 6-\Delta} \left( 1 + \cO \left( \xi^{ \min \{ 2\Delta -4, 8 -2\Delta \} } \right) \right)~, \cr
\end{align}
for the $w$-type solutions and
\begin{align}\label{vwfs1}
& f_{(v)}^{\rm n.n.} = \frac{2 \sgn(\mu) (4-\Delta) g_1 g_b g_5 m_q
k z_{0}^{5-\Delta}}{ ( (4-\Delta)^2 - \mu^2) \ell^{3-\Delta} }
\xi^{4-\Delta} \left( 1 + \cO \left( \xi^{ \min \{ 2\Delta -4, 8
-2\Delta \} } \right) \right)~, \cr & g_{(v)}^{\rm n.n.} = 1 + \cO
\left( \xi^{ \min \{2,8-2\Delta \} } \right)~,
\end{align}
and
\begin{align}\label{vwfs2}
& f_{(v)}^{\rm n.} = \frac{2 \sgn(\mu) (4-\Delta)  g_1 g_b g_5 m_q k
z_{0}^{5-\Delta}}{ ( (6-\Delta)^2 - \mu^2) \ell^{3-\Delta} }
\xi^{6-\Delta} \left( 1 + \cO \left( \xi^{ \min \{ 2\Delta -4, 8
-2\Delta \} } \right) \right)~, \cr & g_{(v)}^{\rm n.} = \xi^2
\left( 1 + \cO \left( \xi^{ \min \{2,8-2\Delta \} } \right)
\right)~,
\end{align}
for the $v$-type solutions.  Note that the relative normalization between $f,g$ for each solution is fixed by the equations of motion.

In order to evaluate the on-shell action, \eqref{Sosrho}, we need the ${\bf P}$ and ${\bf Q}$ matrices.  Recall that these have components $({\bf P}_{(\rho)}^{\rm n.n.})_{\alpha\beta} = P_{(\alpha)}[ R^{{\rm n.n.}(\beta)} ]$, for example, where $\alpha,\beta$ run over $w,v$.  The action of the $P$'s and $Q$'s is given in \eqref{Ps}, \eqref{Qs}, and applying them to \eqref{wwfs}-\eqref{vwfs2} we find
\begin{equation}\label{Qnn}
{\bf Q}_{(\rho)}^{\rm n.n.} = \left( \begin{array}{c c}  (z_0/\varepsilon)^{|\mu|} (1 + \cO( \varepsilon^m ) ) & \cO( \varepsilon^{4-\Delta}) \\ \cO(\varepsilon^{-|\mu|+6-\Delta} ) & 1 + \cO(\varepsilon^m) \end{array}\right)~,
\end{equation}
\begin{equation}\label{Qn}
{\bf Q}_{(\rho)}^{\rm n.} = \left( \begin{array}{c c} \cO( \varepsilon^{|\mu| +m} )  & \cO( \varepsilon^{6-\Delta}) \\ \cO(\varepsilon^{|\mu|+6-\Delta} ) & (\varepsilon/z_0)^2 (1 + \cO(\varepsilon^m) ) \end{array}\right)~,
\end{equation}
\begin{equation}\label{Pnn}
{\bf P}_{(\rho)}^{\rm n.n.} = \left( \begin{array}{c c}   \cO( \varepsilon^{-|\mu|+m} )  & \cO( \varepsilon^{4-\Delta}) \\ \cO(\varepsilon^{-|\mu|+4-\Delta} ) & \cO(\varepsilon^{m-2}) \end{array}\right)~,
\end{equation}
and
\begin{equation}\label{Pn}
{\bf P}_{(\rho)}^{\rm n.} = \left( \begin{array}{c c} 2 (\varepsilon/z_0)^{|\mu|} (1 + \cO( \varepsilon^{m} ) ) & \cO( \varepsilon^{6-\Delta}) \\ \cO(\varepsilon^{|\mu|+4-\Delta} ) & 2 (1 + \cO(\varepsilon^m) ) \end{array}\right)~,
\end{equation}
where $m \equiv \min \{2,8-2\Delta \}$.

The matrix sandwiched between the sources in \eqref{Sosrho} can be written as
\begin{align}\label{rhoMat}
{\bf M} \equiv &~ \frac{ ({\bf P}_{(\rho)}^{\rm n.n.})^T - ({\bf C}_{(\rho)})^T ({\bf P}_{(\rho)}^{\rm n.})^T }{\left[ ({\bf Q}_{(\rho)}^{\rm n.n.})^T - {\bf C}_{(\rho)}^T ({\bf Q}_{(\rho)}^{\rm n.})^T \right] } \cr
=&~  \left[ {\bf 1} - \left( {\bf Q}_{(\rho)}^{\rm n.} {\bf C}_{(\rho)}  ({\bf Q}_{(\rho)}^{\rm n.n.})^{-1} \right)^T \right]^{-1} \left[ \left({\bf P}_{(\rho)}^{\rm n.n.} ({\bf Q}_{(\rho)}^{\rm n.n.})^{-1} \right)^T - \left( {\bf P}_{(\rho)}^{\rm n.} {\bf C}_{(\rho)} ({\bf Q}_{(\rho)}^{\rm n.n.})^{-1} \right)^{T}  \right] \cr
\equiv &~ \left[ {\bf 1} - {\bf A} \right]^{-1} \left[ {\bf B} - {\bf D} \right]~,
 \end{align}
and the leftmost factor expanded using $({\bf 1} - {\bf A})^{-1} = {\bf 1} + {\bf A} + \cdots$.  We compute the leading behavior of each of these matrices, finding
\begin{equation}\label{Bmatrix}
{\bf B} = \left( \begin{array}{c c}  \cO(\varepsilon^{m}) & \cO(\varepsilon^{-|\mu| + 4-\Delta}) \\ \cO( \varepsilon^{4-\Delta}) & \cO(\varepsilon^{m-2})  \end{array}\right)~,
\end{equation}
\begin{equation}\label{Dmatrix}
{\bf D} = \left( \begin{array}{c c} 2 (\varepsilon/z_0)^{2|\mu|} C_{ww} & 2 (\varepsilon/z_0)^{|\mu|} C_{vw} \\ 2 (\varepsilon/z_0)^{|\mu|} C_{wv} & 2 C_{vv} \end{array}\right) \left(1 + \cO(\varepsilon^{m}) \right) + \cO(\varepsilon^{>0})~,
\end{equation}
and
\begin{align}\label{Amatrix}
A_{ww} =&~ C_{ww} \cdot \cO(\varepsilon^{2|\mu|+m}) + C_{wv} \cdot \cO(\varepsilon^{|\mu| + 6-\Delta + m}) + C_{vw} \cdot \cO(\varepsilon^{|\mu|+6-\Delta}) + C_{vv} \cdot \cO(\varepsilon^{12-2\Delta})~, \cr
A_{wv} =&~ C_{ww} \cdot \cO(\varepsilon^{2|\mu| + 6-\Delta}) + C_{wv} \cdot \cO(\varepsilon^{|\mu| + 12-2\Delta}) + C_{vw} \cdot \cO(\varepsilon^{|\mu| + 2}) + C_{vv} \cO(\varepsilon^{8-\Delta})~, \cr 
A_{vw} =&~ C_{ww} \cdot \cO(\varepsilon^{2|\mu| + 4-\Delta+m}) + C_{wv} \cdot \cO(\varepsilon^{|\mu| +m}) + C_{vw} \cdot \cO(\varepsilon^{|\mu|+ 10-2\Delta}) + C_{vv} \cO(\varepsilon^{6-\Delta})~, \cr
A_{vv} =&~ C_{ww} \cdot \cO(\varepsilon^{2|\mu| + 10-2\Delta}) + (C_{wv} + C_{vw}) \cdot \cO(\varepsilon^{|\mu| + 6-\Delta}) + C_{vv} \cdot \cO(\varepsilon^{2})~.
\end{align}
Since we assume $\Delta < 4$, every term in ${\bf A}$ goes as some positive power of $\varepsilon$.  Thus we will only need to consider a finite number of terms in the expansion of $({\bf 1} - {\bf A})^{-1}$ -- enough to soak up whatever divergences arise from the rest of the factors in the on-shell action.

Now consider the $\varepsilon$ dependence of the source $S_{(\rho)}$, which should match the leading asymptotic behavior of $R =  (\cW,\cV)^T$.  The leading behavior of the $\cW$ wavefunction is clearly given by the non-normalizable $w$-type mode (since we assume $\Delta < 4$), which behaves as $\varepsilon^{-|\mu|}$.  The leading asymptotic behavior of the $\cV$ wavefunction however depends on the quantity $6-\Delta -|\mu|$.  When this quantity is positive the non-normalizable $v$-type mode dominates with leading behavior $\varepsilon^0$, but when this quantity is negative, the non-normalizable $w$-type mode dominates with behavior $\varepsilon^{-|\mu| + 6-\Delta}$.  If we wish to consider general values of $|\mu|$ then the source $S_{(v)}$ in \eqref{Ss} should be multiplied by $\varepsilon^{\textrm{min}(0,-|\mu| + 6-\Delta)}$.  It follows that finite terms in the on-shell action come from terms in $M_{ww}$ that go as $\varepsilon^{-2|\mu|}$, terms in $M_{wv}$ and $M_{vw}$ that go as $\varepsilon^{\textrm{min}(-|\mu|, -2|\mu| + 6-\Delta)}$, and terms in $M_{vv}$ that go as $\varepsilon^{\textrm{min}(0, -2|\mu| + 12-2\Delta)}$.  Terms in the components of ${\bf M}$ that are dominant to these give  divergences, while terms that are subleading can be neglected in the limit $\varepsilon \to 0$.

Consistency of holographic renormalization requires that any divergences be determined by purely asymptotic analysis.  This will fail if any terms containing components of ${\bf C}$ are divergent.  We quickly run into trouble when $|\mu| > 6-\Delta$ since $M_{wv}$ for example contains a term of the form $\varepsilon^{|\mu|} C_{vw}$, which fails to offset the factor of $\varepsilon^{-2|\mu| + 6-\Delta}$ coming from the product of sources $\bar{S}_{(w)} S_{(v)}$.  When $|\mu| < 6-\Delta$ there are no such divergences present: ${\bf B}$ by itself gives removable divergences while all terms in ${\bf A}^n {\bf B}$, $n \geq 1$, vanish in the $\varepsilon \to 0$ limit.  Similarly the leading order terms in ${\bf D}$ contribute finite pieces to the on-shell action while all higher order terms in ${\bf D}$ and ${\bf A}^n {\bf D}$ vanish.  We conclude that the bound
\begin{equation}\label{mubound}
|\mu| < 6-\Delta
\end{equation}
must be imposed.  When it is, the $\rho$-sector contribution to the quadratic renormalized on-shell action is
\begin{align}\label{S2rhocont}
\lim_{\varepsilon \to 0} S_{\rm ren}^{(2)} \supset & ~ \frac{2}{z_{0}^2} \int \frac{d^4 k}{(2\pi)^4} \tr \displaystyle\biggl\{ \frac{\varepsilon^{2|\mu|}}{z_{0}^{2|\mu|}} |S_{(w)}|^2 C_{ww} + |S_{(v)}|^2 C_{vv} + \cr
& \qquad \qquad \qquad \quad + \frac{\varepsilon^{|\mu|}}{z_{0}^{|\mu|}} \left( \bar{S}_{(w)} S_{(v)} C_{vw} + \bar{S}_{(v)} S_{(w)} C_{wv} \right) \displaystyle\biggr\}~,
\end{align}
which yields the appropriate terms in \eqref{Sren}.

\section{Mode expansions}\label{app:modeexpansions}

The mode expansions in each sector are as follows.  For the 5d fields associated with the pion we have
\begin{align}\label{pionmodeexp}
A_{z}(k,z) =&~ \frac{ z^3}{z_{0}^2 v(z)^2} \sum_{n} \pi^n(k) m_{(\pi,n)}  \phi_{n}^{(\pi)}(z)~, \cr
 \varphi(k,z) =&~ - \frac{z^3}{z_{0}^2 v(z)^2} \sum_n \pi^n(k) \frac{\d_z \phi_{n}^{(\pi)}}{m_{(\pi,n)}}~.
\end{align}
The $\phi_{n}^{(\pi)}$ the eigenfunctions of $O^{(\pi)}$, associated with the field $\hat{A}$, and the remaining factors come from evaluating the relations \eqref{Ahat} mode by mode.  The $a_1$ tower is contained in $A_{\mu}^\perp$ via
\begin{equation}\label{a1modeexp}
A_{\mu}^\perp(k,z) = g_5 \sum_{n} a_{\mu}^n(k) \phi_{n}^{(a_1)}(z)~.
\end{equation}
They, together with the pions, are housed in the 5d axial gauge field, $A_M$, according to \eqref{Aparam}.  In the $b_1$ sector we take
\begin{equation}\label{b1modeexp}
\cH_{\mu}(k,z) = \sum_{n} b_{\mu}^n(k) \phi_{n}^{(b_1)}(z)~,
\end{equation}
and finally in the $\rho$ sector,
\begin{equation}\label{rhomodeexp}
R_{\mu}(k,z) = \left( \begin{array}{c} \cW_{\mu}(k,z) \\ \cV_{\mu}(k,z) \end{array}\right) = \sum_{n} \tilde{\rho}_{\mu}^{n}(k) \left( \begin{array}{c} \phi_{n}^{(\rho;w)}(k,z) \\ \phi_{n}^{(\rho;v)}(k,z) \end{array}\right) = \sum_{n} \tilde{\rho}_{\mu}^{n}(k) \phi_{n}^{(\rho)}(k,z)~.
\end{equation}
The fields $\cH,\cW,\cV$ are embedded into the 5d fields $V_M, b_{MN}$ according to \eqref{Vparam}, \eqref{bmunudecomp}, and \eqref{bmuzdecomp}.  Finally, we recall that we work in a gauge where the remaining 5d field, $X$, is frozen to its vev, $X = X_0(z)$, \eqref{Xvev}.


\end{document}